\magnification=1200
\baselineskip=14pt

\def\a{\alpha}
\def\b{\beta}
\def\G{\Gamma}
\def\p{\partial}
\def\M{{\cal M}}
\def\N{{\cal N}}

\def\g{\gamma}
\def\r{{\rm Res}}
\def\d{{\rm d}}
\def\e{\epsilon}
\def\l{\lambda}
\def\F{{\cal F}}

\centerline{{\bf SYMPLECTIC FORMS IN THE THEORY OF SOLITONS}
\footnote*{Research supported in part
by the National Science Foundation under grant DMS-95-05399.}}

\bigskip
\centerline{${\bf I.M.\ Krichever}^{\dagger}\ {\bf and\ D.H.\ Phong}^{\ddagger}$}

\bigskip
\centerline{$\dagger$ Department of Mathematics} 
\centerline{Columbia University}
\centerline{New York, NY 10027}
\centerline{and}
\centerline{Landau Institute for Theoretical Physics}
\centerline{Moscow 117940, Russia}
\centerline{e-mail: krichev@math.columbia.edu}
\medskip
\centerline{$\ddagger$ Department of Mathematics}
\centerline{Columbia University}
\centerline{New York, NY 10027}
\centerline{$\ddagger$ e-mail: phong@math.columbia.edu}\bigskip

\bigskip
\centerline{\bf Abstract}
\medskip
We develop a Hamiltonian theory for 2D soliton equations.
In particular, we identify the spaces of doubly periodic
operators on which a full hierarchy of commuting flows
can be introduced, and show that these flows are
Hamiltonian with respect to a universal
symplectic form $\omega={1\over 2}\r_{\infty}
<\Psi_0^*\delta L\wedge\delta\Psi_0>\d k$. We also construct
other higher order
symplectic forms and compare our formalism with the case of 1D solitons.
Restricted to spaces of finite-gap solitons, the universal
symplectic form agrees with the symplectic forms
which have recently appeared in non-linear WKB theory,
topological field theory, and Seiberg-Witten theories.
We take the opportunity to survey some
developments in these areas where symplectic forms
have played a major role.

\vfill\break

\centerline{\bf I. INTRODUCTION}

\bigskip
There is increasing evidence that
symplectic structures for solitons
may provide a unifying thread
to many seemingly unrelated developments
in geometry and physics. In soliton theory,
the space of finite-gap solutions to the equation
$[\p_y-L,\p_t-A]=0$ is a space $\M_g(n,m)$ of
punctured Riemann surfaces $\G$ and
pair of Abelian integrals
$E$ and $Q$ with poles of order less than
$n$ and $m$ at the punctures. The fibration 
over $\M_g(n,m)$ with $\G$ as fiber carries a natural
meromorphic one-form, namely $\d\l= Q\d E$. It is
a remarkable and still mysterious fact
that the form $\d\lambda$ is actually central to
several theories with very distinct goals and origins. These 
include the non-linear WKB (or Whitham) theory [22][23][26][36][38],
two-dimensional topological models [11][12][13],
and Seiberg-Witten exact solutions of N=2
supersymmetric gauge theories [18][40][52][53]. The form $\d\lambda$
can be viewed as a precursor of a symplectic structure.
Indeed, it can be extended as a 1-form $\sum_{i=1}^k\d\l(z_i)$ on 
the fibration over
$\M_g(n,m)$ with fiber a symmetric $g$th-power of $\G$.
Its differential $\omega$ becomes single-valued
when restricted to a suitable $g$-dimensional leaf of
a canonical foliation on $\M_g(n,m)$, and defines a symplectic form [39].
Earlier special cases of this type of construction
were pioneered by Novikov and Veselov [50]
in the context of hyperelliptic surfaces and
1D solitons, and by Seiberg, Witten, and Donagi [18][53]
in the context of N=2 SUSY gauge theories.

\medskip
The goal of this paper is twofold.
Our first and primary objective is to construct the foundations of a Hamiltonian
theory of 2D solitons. 
\medskip  
$\bullet$ For this, we provide an improved formulation
of 2D hierarchies, since the classical formulations (e.g. Sato [54])
are less pliable than in the 1D case, and inadequate for our
purposes. In particular, the new formulation allows
us to identify suitable spaces ${\cal L}(b)$ of doubly
periodic operators on which 
a full hierarchy of commuting flows $\p_mL=\p_yA_m+[A_m,L]$
can be introduced;
\medskip 
$\bullet$ We can then define a universal symplectic form $\omega$ 
on these spaces ${\cal L}(b)$ by
$$
\omega={1\over 2}\r_{\infty}<\Psi_0^*\delta L\wedge\delta\Psi_0>\d k\eqno(1.1)
$$
where $\Psi_0$ and $\Psi_0^*$ are the formal Bloch and dual Bloch
functions for $L$. This form had been shown in [39] to restrict to
the geometric symplectic form $\delta\sum_{i=1}^g\d\l(z_i)$
when finite-gap solitons are imbedded in the space
of doubly periodic operators. Here we show that
it is a symplectic form in its own right on ${\cal L}(b)$,
and that with respect to this form, the hierarchy of 2D flows
is Hamiltonian. Their Hamiltonians are shown to be $nH_{n+m}$, where $H_s$
are the coefficients of the expansion of the quasi-momentum 
in terms of the quasi-energy.
\medskip
$\bullet$ Our formalism is powerful enough to
encompass many diverse symplectic structures
for 1D solitons. For example, $\omega$ reduces to the 
Gardner-Faddeev-Zakharov
symplectic structure for KdV, while its natural
modifications for $y$-independent equations
(see (2.71) and (2.73) below), reproduce the infinite set
of Gelfand-Dickey as well as Adler-Magri symplectic structures.

\medskip
$\bullet$ The symplectic form (1.1) is algebraic in
nature. However, it suggests new higher symplectic forms,
$$
\omega_{m_0}={1\over 2}\r_{\infty}<\Psi_0^*(A_{m_0}\delta L-L\delta A_{m_0})
\wedge\delta\Psi_0>_0\d k\eqno(1.2)
$$
which are well-defined only on certain spaces of operators 
with suitable growth or ergodicity conditions.
For Lax equations $\p_{t_{m}}L=[A_{m},L]$, these higher symplectic
forms have a remarkable interpretation: they are forms
with respect to which the eigenvalues of $A_{m_0}$, suitably
averaged, can serve as Hamiltonians, just as the eigenvalues
of $L$ are Hamiltonians with respect to the basic symplectic structure
(1.1). It would be very interesting to understand these new
forms in an {\it analytic} theory of solitons.

\medskip
Our second objective is to take this opportunity to
provide a unified survey of some developments where
the form $\d\l$ (or its associated symplectic form
$\omega$) played a central role.
Thus $\d\lambda$ emerges as the generating function for the Whitham hierarchy,
and its coefficients and 
periods are Whitham times (Chapter IV). The same coefficients
are deformation parameters of topological
Landau-Ginzburg models
in two dimensions (Chapter V), while for N=2 SUSY
four-dimensional gauge theories, the periods of $\d\lambda$ generate the lattice
of Bogomolny-Prasad-Sommerfeld states (Chapter VI).
Together with $\d\lambda$, another notion, that of a prepotential
$\F$, emerges repeatedly, albeit
under different guises. In non-linear WKB methods,
$\F$ is the exponential of the $\tau$-function of the Whitham
hierarchy. In topological Landau-Ginzburg models,
it is the free energy. In N=2 supersymmetric gauge theories,
it is the prepotential of the Wilson effective action.
It is an unsolved, but clearly very important problem, to determine
whether these coincidences can be explained from first principles.

\bigskip

\centerline{\bf II. HAMILTONIAN THEORY OF 2D SOLITON EQUATIONS}

\bigskip

Solitons arose originally in the study of shallow
water waves. Since then, the notion of soliton equations
has widened considerably. It embraces now a wide
class of non-linear partial differential equations,
which all share the characteristic feature
of being expressible as a compatibility condition
for an auxiliary pair of linear differential
equations. This is the
viewpoint we also adopt in this paper.
Thus the equations of interest to us are of the form
$$
[\p_y-L,\p_t-A]=0,
\eqno(2.1)
$$
where the unknown functions $\{u_i(x,y,t)\}_{i=0}^n$,
$\{v_j(x,y,t)\}_{j=0}^m$ are the $N\times N$ matrix coefficients
of the ordinary differential operators
$$
L=\sum_{i=0}^nu_i(x,y,t)\p_x^i,\ A=\sum_{j=0}^mv_j(x,y,t)\p_x^j.
\eqno(2.2)
$$
A preliminary classification of equations of the form (2.1)
is by the orders $n,m$ of the operators $L$ and $A$,
and by the dimension $N$ of the square matrices
$u_i(x,y,t)$, $v_j(x,y,t)$. 
In what follows, we assume that the leading coefficients
of $L$ and $A$ are constant diagonal matrices
$u_n^{\a\b}=u_n^{\a}\delta_{\a\b}$,
$v_m^{\a\b}=v_m^{\a}\delta_{\a\b}$. Under this assumption, the equation (2.1) is invariant
under the gauge transformations
$L,A\rightarrow L'=g(x)^{-1}Lg(x),\ A'=g(x)Ag(x)^{-1}$
where $g(x)$ is a diagonal matrix. We fix the gauge by the condition
$u_n^{\a\b}=\delta^{\a\b}u_{\a},\ u_{n-1}^{\a\a}=0$.
We shall refer to (2.1)
as a {\it zero curvature} or {\it 2D soliton equation}. The 1D 
soliton equation corresponds to the special case of 
$y$-independent operators $L$ and $A$. In this case 
the equation (2.1) reduces to
a {\it Lax equation} $L_t=[A,L]$.

\bigskip
\noindent
{\bf A. Difficulties in a Hamiltonian Theory of 2D Solitons}

The Hamiltonian theory of 1D solitons is a rich subject which
has been developed extensively over the years [10][25]. However,
much less is known about the 2D case. We illustrate the
differences between 1D and 2D equations in the basic
example of the hierarchies for the Korteweg-deVries (KdV)
$$u_t-{3\over 2}u\p_x u+
{3\over 4}\p_x^3 u=0\eqno(2.3)$$ 
and the Kadomtsev-Petviashvili (KP) equations
$${3\over 4}u_{yy}
=\p_x(u_t-{3\over 2}u\p_x u+{3\over 4}\p_x^3 u).\eqno(2.4)
$$
The KP equation arises from the choice
$N=1$, $n=2$, $m=3$, and $L=\p_x^2+u$, $A=\p_x^3+{3\over 2}u\p_x+
v_0$ in (2.1). We obtain in this way the system
$$
\p_x v_0={3\over 4}\p_x^2 u+{3\over 4}u_y,\ \
v_{0,y}={3\over 4} \p_x^3u+{3\over 4}\p_x u_y-{3\over 2}
u\p_x u\eqno(2.5)
$$
which is equivalent to (2.4) (up
to an $(x,y)$-independent additive term in $v_0$,
which does not affect the commutator $[\p_y-L,\p_t-A]$).
Taking $L$ and $A$ independent of $y$ gives the KdV equation.

\medskip 
The basic mechanism behind this construction
is that the zero curvature equation actually
determines $A$ in terms of $L$. This remains the case
for the 1D Lax equation $L_t=[A,L]$ even when $A$ is taken
to be of arbitrarily high order $m$,
but not for the 2D zero curvature equation $L_t-A_y=[A,L]$.
The point is that $[A,L]$ is a differential operator of order
$m+1$. The Lax equation requires that it be in fact of order
$0$, while the 2D zero curvature equation requires only
that it be of order $\leq m-1$. The order 0 constraint is quite
powerful. Expressed as differential constraints
on the coefficients of $A$, it implies readily
that the space of such $A$'s for fixed $L$ is
of dimension $m$. An explicit basis can be obtained by the Gelfand-Dickey
construction [10][27], which we present for a general operator $L$ of order $n$. 
Let a pseudo-differential
operator of order $n$ be a
formal Laurent series $\sum_{i=-\infty}^nw_i\p_x^i$ in $\p_x$, with $\p_x$
and $\p_x^{-1}$ satisfying the identities
$$
\p_x u=u\p_x+u',\ \p_x^{-1}u=\sum_{i=0}^{\infty}(-)^iu^{(i)}\p_x^{-i-1}.
$$
Then there exists a unique pseudo-differential operator
$L^{1/n}$  of order $1$ satisfying
$(L^{1/n})^n=L$. 
Evidently, the coefficients of $L^{1/n}$ are differential
polynomials in the coefficients of $L$.
For example, for $L=\d^2+u$, we find $L^{1/2}=\p_x+{1\over 2}u\p_x^{-1}
-{1\over 4}u'\p_x^{-2}+\cdots.$ We set
$$
L^{i/n}=L^{i/n}_++L^{i/n}_-,$$
where the first term on the right hand side
is the differential part of the pseudo-differential
operator $L^{i/n}$, and the second term
on the right hand side is of order $\leq -1$. Then
$[L,L^{i/n}_+]=[L,L^{i/n}]-[L,L^{i/n}_-]=-[L,L^{i/n}_-].$
Since the commutator $[L,L^{i/n}_-]$
is of order at most $n-2$, this shows that the
differential operators $L^{i/n}_+$ provide
the desired basis.

\medskip

Associated to $L$ are then an infinite hierarchy of flows, obtained 
by introducing
``times" $t_1,\cdots, t_m,\cdots$, and considering the evolutions
of $L=\sum_{i=0}^nu(x;t_1,\cdots,t_n)\p_x^i$ defined by
$$
\partial_{m}L=[L,L^{m/n}_+].\eqno(2.6)
$$ 
where we have denoted by $\p_m$ the partial derivative
with respect to the time $t_m$. A key property of these flows is their
commutativity, i.e.
$$
[\p_{i}-L^{i/n}_+,\p_{j}-L^{j/n}_+]=0.\eqno(2.7)
$$
To see this, we note first that if $L$ evolves according
to a flow $\p_tL=[L,A]$, then $L^{\a}$ evolves
according to $\p_tL^{\a}=[L^{\a},A]$. Thus we have
$\p_{i}L^{j/n}=[L^{i/n}_+,L^{j/n}]$, 
$\p_{j}L^{i/n}=[L^{j/n}_+,L^{i/n}]$,
and the left hand side of (2.7) can be rewritten as
$-[L^{i/n}_+,L^{j/n}]
+[L^{j/n}_+,L^{i/n}]
+[L^{i/n}_+, L^{j/n}_+].$
If we replace $L^{i/n}_+$, $L^{j/n}_+$ by $L^{i/n}-L^{i/n}_-$
and $L^{j/n}-L^{j/n}_-$, all terms cancel, except
for $[L^{i/n}_-,L^{j/n}_-]$. This term is however
pseudo-differential, of order $\leq -3$, and cannot
occur in the left hand side of (2.7), which is
manifestly a differential operator. 

\medskip
The flows (2.6) are known to be Hamiltonian with respect to an infinite
number of symplectic structures with different Hamiltonians.
For example, the KdV equation itself can be
rewritten in two Hamiltonian forms
$$
u_t=\p_x{\delta H\over\delta u},
\ \
u_t=(\p_x^3+2(u\p_x+\p_xu)){\delta H'\over\delta u}
$$
where the skew-symmetric operators $K=\p_x$, $K'=\p_x^3+2(u\p_x+\p_xu)$
correspond to two different symplectic structures
(called respectively the Gardner-Faddeev-Zakharov [10][25] and Adler-Magri
structures [1][41]), and $H={1\over 4}u^3-{1\over 8}u_x^2$, 
$H'={1\over 8}u^2$ are the corresponding
Hamiltonians.
\medskip
The situation for the 2D zero curvature equation is much less simple,
since the arguments narrowing $A$ to an $m$-dimensional space
of operators break down. Although formally, we may still introduce
the KP hierarchy as $\p_mL=\p_yA_m+[A_m,L]$, with $A_m$ an operator
of order $m$ which should also
be viewed as an unknown, this is not a closed system of equations for
the coefficients of $L$, as it was in the case of the Lax equation.
Another way, due to Sato [54], is to introduce the KP hierarchy
as a system of commuting flows 
$$\p_m{\rm L}=[{\rm L}_+^m,{\rm L}]\eqno(2.8)$$
on the coefficients $(v_i(x,t))^{i=\infty}_{i=1}$ of a 
{\it pseudo-differential}
operator ${\rm L}$
$${\rm L}=\p+\sum_{i=1}^{\infty}v_i(x,t)\p^{-i}.$$ 
In this form, the KP hierarchy can be
viewed as a completely integrable Hamiltonian system. However, it now
involves an infinite number of functions $v_i$, and its relation
to the original KP equation (which is an equation for a single
function of two variables $(x,y)$) requires additional
assumptions.

\medskip

\noindent{\bf A. Quasi-Energy and Quasi-Momentum}

Our first main task is then to identify the space of differential
operators with periodic coefficients
on which the KP equation and its higher order analogues
can be considered as completely integrable
Hamiltonian systems. Our approach actually applies
systematically to general 2D soliton equations. We present these results
at the end of this section, and concentrate for the moment on 
the simplest case of a differential operator $L$ of order $n$.
We begin with the construction of the formal Bloch eigenfunction for
two-dimensional linear operators with periodic coefficients.

\medskip

\noindent{\bf Theorem 1}.
{\it Let $L$ be an arbitrary linear differential operator of order $n$ with
doubly periodic coefficients
$$
\eqalignno
{&L=\p_x^n+\sum_{i=0}^{n-2}u_i(x,y)\p_x^i,\cr
&u_i(x+1,y)=u_i(x,y+1)=u_i(x,y)&(2.9)\cr}
$$
Then there exists a unique formal solution $\Psi_0(x,y;k)$ of the equation
$$
(\p_y-L)\Psi_0(x,y;k)=0, \eqno(2.10)
$$
which satisfies the following properties
\item{\rm (i)} $\Psi_0(x,y;k)$ has the form
$\Psi_0(x,y;k)=\left(1+\sum_{s=1}^{\infty}\xi_s(x,y)k^{-s}\right)
e^{\left(kx+k^ny+\sum_{i=0}^{n-2}B_i(y)k^i\right)}$;
\item{\rm (ii)} $\Psi_0(0,0;k)=1$; 
\item{\rm (iii)} $\Psi_0(x,y;k)$ is a Bloch function with respect to
the variable $x$, i.e., 
$$ 
\Psi_0(x+1,y,k)=w_1(k)\Psi_0(x,y,k),\ \
w_1(k)=e^{k}\eqno(2.11)
$$

The formal solution $\Psi_0(x,y,k)$ is then also a Bloch function with respect to 
the variable $y$ with a Bloch multiplier $w_2(k)$}
$$ 
\Psi_0(x,y+1;k)=w_2(k)\Psi_0(x,y;k),\ \
w_2(k)=(1+\sum_{s=1}^{\infty} J_s k^{-s})
e^{\left(k^n+\sum_{i=0}^{n-2}B_i(1))k^i\right)}\eqno(2.12)
$$

\medskip

\noindent{\it Proof.} To simplify the notation, we begin with the proof
in the case of $n=2$, with $L=\p_x^2+u(x,y)$.
The formal solution has then the form
$$
\Psi_0(x,y;k)=\left(1+\sum_{s=1}^{\infty}\xi_s(x,y)k^{-s}\right) 
e^{kx+k^2y+B_0(y)}
$$
Substituting this formal expansion in the equation $(\p_y-L)\Psi_0(x,y;k)=0$
gives the following equations for the coefficients $\xi_s$ 
$$
\p_y\xi_s+\p_yB_0\xi_s=2\xi_{s+1}'+\xi_s''+u\xi_s. \eqno(2.13)
$$
(Here and henceforth, we also denote derivatives in $x$ by primes.)
These equations are solved recursively by the formula:
$$
\eqalignno{\xi_{s+1}(x,y)&=c_{s+1}(y)+\xi_{s+1}^0(x,y),\cr
\xi_{s+1}^0(x,y)&={1\over 2}\int_0^x (\p_y \xi_s(\tilde x,y)-
\xi_s''(\tilde x,y)+(\p_yB_0-u(\tilde x,y))\xi_s(\tilde x,y))
\d\tilde x&(2.14)\cr} 
$$ 
where $c_s(y)$ are {\it arbitrary} functions of the variable $y$ 
with the only requirement that $c_s(0)=0$, which is
dictated by (ii). 

Our next step is to show by induction that the Bloch property (i), which 
is equivalent to the periodicity condition 
$$ 
\xi_s(x+1,y)=\xi_s(x,y),  \eqno(2.15) 
$$ 
uniquely defines the functions $c_s(y)$. Assume then that $\xi_{s-1}(y)$ 
is known  and satisfies the condition that
the corresponding function $\xi_s^0(x,y)$ is periodic. The first step
of the induction, namely the periodicity in $x$ of $\xi_1(x,y)$,
requires that $\p_yB_0(y)=2\xi_1'+u$, in view of (2.13) and the
fact that $\xi_0=1$. Thus we fix $B_0(y)$ to be
$$
B_0(y)=\int_0^y\int_0^{1}u( x', y')\d  x'\d  y'
\eqno(2.16)
$$
The choice of the function $c_s(y)$ does not affect the periodicity property of
$\xi_s(x,y)$, but it does affect the periodicity in $x$ of the function 
$\xi_{s+1}^0(x,y)$. In order to make  $\xi_{s+1}^0(x,y)$ periodic,
the function $c_s(y)$ should satisfy the equation
$$
\p_y c_s=
-\int_0^{1} (\p_y \xi_s^0(x,y)-
(\xi_s^0)''(x,y)+(\p_yB_0-u(x,y))\xi_s^0(x,y))\d x
\eqno(2.17)
$$
where $<\cdot>_x$ denotes averaging in $x$ over the interval
$[0,1]$. Together with the initial condition $c_s(0)=0$, this defines $c_s$
uniquely 
$$
c_s=-\int_0^y\d y'\int_0^{1} (\p_y \xi_s^0(x,y')-
(\xi_s^0)''(x,y')+(\p_y B_0-u(x,y'))\xi_s^0(x,y'))\d x, \eqno(2.18)
$$
and completes the induction step.

We can now establish (iii), also by induction.
Assume that $\xi_{s-1}$ satisfies the relation
$$
\xi_{s-1}(x,y+1)-\xi_{s-1}(x,y)=\sum_{i=1}^{s-1} J_i\xi_{s-i-1}(x,y). 
\eqno(2.19)
$$
where $J_1,\ldots, J_{s-1}$ are constants. Then (2.14) implies that
$$
\xi_{s}(x,y+1)-\xi_{s}(x,y)=\sum_{i=1}^{s-1} J_i\xi_{s-i}(x,y)+J_s=
\sum_{i=1}^{s} J_i\xi_{s-i}(x,y), \eqno(2.20)
$$
with
$$
J_s=c_s(y+1)-c_s(y)-\sum_{i=1}^{s-1}c_{s-i}(y). 
$$
We claim that $J_s$ is actually constant. In fact, it follows from 
(2.14) and (2.19) that
$$
\xi_s^0(x,y+1)-\xi_s^0(x,y)=\sum_{i=1}^{s-1} J_{i}\xi_{s-i}^0(x,y)=
\sum_{i=1}^{s-1} J_{i}(\xi_{s-i}(x,y)-c_{s-i}(y)). 
$$
Thus (2.18) implies that
$$
\eqalign{&\p_yc_s(y+1)-\p_yc_s(y)=
-\sum_{i=1}^{s-1}J_i\int_0^{1} (\p_y \xi_{s-i}^0-
(\xi_s^0)''+(\p_yB_0-u)\xi_s^0)\d x \cr
&=\sum_{i=1}^{s-1}J_i(\p_yc_{s-i}-\xi_{s+1}^0(1,y)-\xi_{s+1}^0(0,y))=
\sum_{i=1}^{s-1}J_i\p_yc_{s-i}.\cr}
$$
In particular the derivative of $J_s$ vanishes. This proves
Theorem 1 when $n=2$. 

The proof can be easily adapted to the case of general $n$. 
Let $\xi_s^0(x,y)$ be 
the coefficients of the formal series
$$
\Psi^{(0)}(x,y,k)={\Psi_0(x,y,k)\over \Psi_0(0,y,k)}=
(1+\sum_{s=1}^{\infty}\xi_s^0(x,y,k)k^{-s})e^{kx}. 
$$
Then $(\p_y-L)\Psi_0=0$ is equivalent to the system of equations
$$
\sum_{l=0}^{n-1}C_n^l\left(\p_x^{n-l}\xi_{s+l}^0\right)+
\sum_{i=0}^{n-2} u_i\sum_{l=0}^{i}C_i^l\left(\p_x^{i-l}\xi_{s+l}^0\right)=
\p_y\xi_s^0+\sum_{j=-s}^{n-2}b_j\xi_{j+s}^0, \ \ s\geq -n+2,\eqno(2.21)
$$
where we have assumed that $\xi_s^0=0$ for $s<0$, and set $b_j=b_j(y)$ to
be the 
coefficients of the series
$$
{\p_y \Psi_0(0,y,k)\over \Psi_0(0,y,k)}=k^n+\sum_{j=-\infty}^{n-2} b_j(y)k^j.
\eqno(2.22)
$$
These equations are of the form
$$
n\p_x\xi_{s+n-1}^0=b_{-s}+F_s(\xi_{s'}^0,b_{s''}), 
\eqno(2.23)
$$
Here $F$ is a linear combination of the $\xi_{s'}^0,\ s'\leq s+n-2$ and 
their derivatives, with coefficients which are themselves linear in $u_i$ and 
$b_{s''}, \ s''< s$ (c.f. (2.13)). 
The equations define recursively $b_{-s}(y)$ and $\xi_{s+n-1}^0$. 
The coefficient 
$b_{-s}$ follows from the periodicity in $x$ of (2.23)
$$
b_{-s}(y)=-\int_0^{1} F_s(\xi_{s'}^0,b_{s''})\d x, \eqno(2.24)
$$
and the coefficient $\xi_{s+n-1}^0$ follows in turn
$$
n\xi_{s+n-1}^0=b_{-s}(y)x+
\int_0^xF_s(\xi_{s'}^0,b_{s''})\d x'. 
\eqno(2.25)
$$
We can now integrate (2.22) and find that $\Psi_0$ has the form 
(i) with 
$$
B_i(y)=\int_0^y b_{i}(y),\ \ i=0,\ldots, n-2. \eqno(2.26)
$$
We obtain at the same time the Bloch property for $\Psi_0$ with respect to $y$
with the Bloch multiplier
$$
w_2(k)=\int_0^{1}(k^n+\sum_{i=-\infty}^{n-2}b_i(y)k^i)\d y. 
\eqno(2.27)
$$
The proof of Theorem 1 is complete.

\bigskip
\noindent{\bf B. Basic Constraints}

\medskip

In Theorem 1, we have chosen the simple form $w_1(k)=e^{k}$ for the
Bloch multiplier in $x$. If we view $x$ as a ``space" variable, this
identifies the spectral parameter $k$ with the ``quasi-momentum" (up to a factor
of $i=\sqrt{-1}$). The variable $y$
can then be interpreted as a time variable in the Schr\"odinger-like
equation $(\p_y-L)\Psi_0=0$. This identifies (again up to
a factor of $i$) the logarithm of the second
Bloch multiplier $w_2(k)$ with the quasi-energy $E(k)$.
Alternatively, we may change spectral parameters, and introduce
the spectral parameter $K$ as well as
the coefficients $({\cal E}_i)_{i=-n+2}^{\infty}$ of the expansion in $k$
of the quasi-energy by
$$
\eqalignno{K^n&=E(k)=\log w_2(k)=k^n+\sum_{i=-n+2}^{\infty}
{\cal E}_i k^{-i}\cr 
K&=k+O(k^{-1})&(2.28)\cr}
$$
We observe that a change of spectral parameter
of the form $k\rightarrow K=k+O(k^{-1})$ transforms a formal 
series in $k$ of the form (i) in Theorem 1 into another formal series
of the same form in $K$. With $K$ as spectral parameter, the second
Bloch multiplier $w_2(K)$ reduces to $w_2(K)={\rm exp}(K^n)$,
but the first Bloch multiplier $w_1(K)$ becomes non-trivial. The coefficients
$(H_s)_{s=1}^{\infty}$ of the expansion in $K$ of its logarithm
$$
w_1(K)=e^{k(K)},\ \ k(K)=K+\sum_{s=1}^{\infty}H_sK^{-s}. \eqno(2.29) 
$$
will play an important role in the sequel.

\medskip

The coefficients ${\cal E}_i$ and $H_s$ of the expansions (2.28) and (2.29)
of the spectral parameters are uniquely defined by the coefficients
$(u_i)_{i=0}^{n-2}$ of the operator $L$, and hence can be considered 
as functionals on the space of periodic operators.
We now restrict ourselves to the subspace ${\cal L}(b)$
of operators $L$ satisfying the constraints
$$
b_i(y)=b_i,\ 0\leq i\leq n-2, \eqno(2.30)
$$
where $b_i(y)=\d B_i(y)/\d y$, with
the functions $B_i(y)$ defined by (i) in Theorem 1,
and $b=(b_0,\cdots,b_{n-2})$ are $(n-1)$ fixed constants. 
On the space ${\cal L}(b)$, the essential singularity
in the second Bloch multiplier $w_2(k)$ simplifies to
$k^n+\sum_{i=0}^{n-2}b_ik^i$ (c.f. (2.12)). Comparing with (2.28), we see that
the constraints (2.30) fix the values of the first $(n-1)$ 
functionals ${\cal E}_{-i}=b_i$. This is in turn clearly equivalent to
fixing the values 
of the first $(n-1)$ functionals $H_1,\ldots, H_{n-1}$ in (2.29).
We claim that these constraints can also be expressed under the form
$$
\int_0^1 h_i(u_{j})\d x = H_s=const,\ \ i=1,\ldots, n-1, \eqno(2.31)
$$
where $h_s(u_j), \ j\geq (n-i-1)$, are universal differential polynomials 
depending only on $n$. In fact,
the constraints (2.30) imply that the essential singularity in the Bloch
function $\Psi_0(x,y;k)$ is of the form
$$
kx+(k^n+\sum_{i=0}^{n-2}b_ik^i)y=Kx+K^ny+O(K^{-1})
$$
Since the expression ${\rm exp}\,\big(O(K^{-1})\big)$ contains no essential singularity
and can be expanded as a formal series in $K^{-1}$,
we have shown that
the formal Bloch solution $\Psi_0(x,y;k)$
can be rewritten in terms of the spectral parameter $K$ under the form
$$
\Psi_0(x,y,K)=\left(1+\sum_{s=1}^{\infty}\zeta_s(x,y)K^{-s}\right)
e^{Kx+K^ny}, \eqno(2.32)
$$
Substituting (2.32) in $(\p_y-L)\Psi_0(x,y;K)=0$ gives for the first $(n-1)$
coefficients $\zeta_1,\ldots, \zeta_{n-1}$ a
system of ordinary differential equations
$$
\sum_{l=0}^{n-1}C_n^l\left(\p_x^{n-l}\zeta_{s+l}^0\right)+
\sum_{i=0}^{n-2} u_i\sum_{l=0}^{i}C_i^l\left(\p_x^{i-l}\zeta_{s+l}^0\right)=0,
\ \ s=-n+2\ldots,0,\eqno(2.33)
$$ 
which just coincides with the first equations defining formal eigenfunctions
for ordinary differential operators (see [kr]). Let $\zeta_i(x,y;x_0)$,
be the solution of (2.33) with the normalization $\zeta_0=1$ and
$\zeta_i(x_0,y;x_0)=0$. Then the equations (2.33) 
define recursively differential polynomials $h_i(u_j(x,y))$ such  that
$$
\p_x\zeta_i(x,y;x_0)|_{x=x_0}=h_i(u_j(x,y))|_{x=x_0}. \eqno(2.34)
$$
The above left hand side is equal to the first coefficients of
the logarithmic derivative of $\Psi_0$ at $x=x_0$, i.e.
$$
K+\sum_{s=1}^{n-1}\left(\p_x\zeta_s(x,y;x_0)|_{x_0=x}\right)K^{-s}=
{\p_x \Psi_0(x,y,K)\over \Psi_0(x,y,K)}+O(K^{-n}). \eqno(2.35)
$$
Integrating gives the first (constant) coefficients
of (2.29) and establishes our claim.
 
Henceforth we will always assume
that $L$ is in the space ${\cal L}(b)$
of operators with periodic coefficients $(u_i)_{i=0}^{n-2}$ 
satisfying either one of the equivalent constraints (2.30) or (2.31).

\bigskip

\noindent
{\bf C. Commuting Flows}

\medskip

On the space ${\cal L}(b)$ we can now define
an infinite set of mutually commuting flows as follows. First, we observe
that for any formal series of the form (i) in Theorem 1,
there exists for each integer $m\geq1$ a unique differential operator
$A_m$ of the form
$$
A_m=\p_x^m+\sum_{j=0}^{m-1}u_{i,m}(x,y)\p_x^i, \eqno(2.36)
$$
which satisfies the condition
$$
(A_m-K^m(k))\Psi_0(x,y;k)=O(k^{-1})\Psi_0(x,y;k),
\eqno(2.37)
$$
where $K(k)$ is defined by (2.28). Indeed, this condition is
equivalent to the following finite system of equations for the
coefficients $(u_{i,m})$ of $A_m$
$$
\sum_{i=0}^m u_{i,m}\sum_{l=0}^i C^l_i(\p_x^{i-l}\zeta_{s+l})=\zeta_{s+m},\ \ 
s=-m+2,\ldots,0.
$$
This system is triangular, and identifies uniquely the coefficients
$u_{i,m}$ as differential polynomials in the first $m-1$ coefficients
$\xi_s$ of $\Psi_0(x,y;k)$. For example, we find
$$
u_{m-2,m}=-m\zeta_1',\ \ \ u_{m-3,m}=-m\zeta_2'-C_m^2\zeta_1'' 
$$

Let $L$ be now an operator in ${\cal L}(b)$, $\Psi_0(x,y;k)$
its Bloch function, and $A_m$ be the operators obtained by the
preceding construction. Then 
$$
[\p_y-L,A_m]\Psi_0(x,y,k)=(\p_y-L)O(k^{-1})\Psi_0(x,y,k)=
O(k^{n-2})e^{Kx+K^ny},\eqno(2.38)
$$
which implies that the operator $[\p_y-L,A_m]$ has order
less or equal to $n-2$. We set
$$
[\p_y-L,A_m]=\p_y A_m-[L,A_m]=\sum_{i=0}^{n-2}F_{i,m}(x,y)\p_x^i \eqno(2.39)
$$
The functions $F_{i,m}(x,y)$ are uniquely defined by $u_j(x,y)$ and can by
expressed in terms of multiple integrals of differential polynomials in 
$u_j$. Thus they can also be expressed as
$$
F_{i,m}(x,y)=\hat F_{i,m}(u(x,y)), \ \ u=(u_0,\ldots, u_{n-2}). \eqno(2.40)
$$
where $\hat F_{i,m}(.)$ is {\it a nonlocal} but exact functional of $u_j(x,y)$.

\medskip
\noindent
{\bf Theorem 2.} {\it The system of equations
$$
\p_m u_i= \hat F_{i,m}(u) \longleftrightarrow \p_m L=\p_y A_m+[A_m,L] \eqno(2.41)
$$
defines an infinite set of commuting flows on the subspace 
${\cal L}(b)$ of doubly periodic operators.}

\medskip
Since $\hat F_{i,m}$ and $A_m$ are well-defined
functions of $L$, we need only check 
the commutativity of the flows. For this, we need
the following version of the uniqueness of Bloch solutions, namely,
that if $\Psi(x,y;k)$ is a Bloch solution
($\Psi(x+1,y;k)=w_1(k)\Psi(x,y;k)$)
of the equation $(\p_y-L)\Psi=0$ having the form
$$
\Psi_0(x,y;k)=\left(\sum_{s=-N}^{\infty}\xi_s(x,y)k^{-s}\right)
e^{kx+(k^n+\sum_{i=0}^{n-2}b_ik^i)y}, 
\eqno(2.42) 
$$ 
then $\Psi(x,y;k)$ must be given by 
$$
\Psi(x,y;k)=a(k)\Psi_0(x,y;k), \eqno(2.43)
$$
where $a(k)$ is a constant Laurent series
$a(k)=\sum_{s=N}^{\infty}a_sk^{-s}$.
This is because the leading coefficient $\xi_N$ in any formal
solution $\Psi$ is a constant. As a consequence, if $\Psi(x,y;k)$ 
is a Bloch solution, then the ratio 
$\Psi(x,y;k)\Psi^{-1}(0,0;k)$ is also a solution. This expression
has all the properties of, and can be identified with $\Psi_0$.
Since $\Psi(0,0;k)$ is a Laurent series, our assertion follows.

Returning to the proof of Theorem 2, we observe that if
$\Psi_0$ is the Bloch solution
to (2.22) then $(\p_m-A_m)\Psi_0$ is also a Bloch solution to the same 
equation. The preceding uniqueness property implies 
$$
(\p_m-A_m)\Psi_0=-\tilde\Omega_m(k) \Psi_0,\ \ 
\tilde\Omega_m(k)=K^m+O(k^{-1}). 
\eqno(2.44)
$$
for a suitable $(x,y)$ independent Laurent series $\Omega_m(k)$. 
In particular, 
$[\p_m-A_m,\p_n-A_n]\Psi_0=O(k^{-1})\Psi_0$.
The last equality implies that  
$[\p_m-A_m,\p_n-A_n]$ is an ordinary differential operator in $x$ of order
less than zero. Therefore, it must vanish identically
$$
[\p_m-A_m,\p_n-A_n]=0 \eqno(2.45)
$$
This establishes the commutativity of the flows.

\medskip

Conversely, let $u(x,y,t,k), \ t=(t_1,t_2,\ldots)$ be a solution to the 
hierarchy (2.41).
Then there exists a unique formal Bloch solution $\Psi(x,y,t;K)$
of the equations
$$
(\p_y-L)\Psi=0,\ \ (\p_m-A_m)\Psi=0 \eqno(2.46)
$$
having the form
$$
\Psi(x,y,t,K)=\left(1+\sum_{s=1}^{\infty}\zeta_s(x,y,t)K^{-s}\right)
e^{Kx+K^ny+\sum_m K^mt_m}. \eqno(2.47)
$$

The above expression identifies the original variables $x$ and $y$ with 
the first and the $n$-th {\it times} of the hierarchy,
$$
x=t_1, \ \  y=t_n, \eqno(2.48)
$$
respectively. More generally, the preceding results show that
for periodic operatos $L$, an equation of the zero curvature form
$[\p_y-L,\p_t-A]=0$ must be equivalent to a pencil of
equations for the coefficients of $L$ only. In other words,
there must exist constants $c_i$ such that
$$A=\sum_{i=1}^mc_iA_i\eqno(2.49)$$
and the flow is along the basic times $t_i=c_it$,
of the hierarchy.

\medskip
Finally, we point out that for all $n$, the equations of the
corresponding hierarchy for $L=L_n$ have the same form
$[\p_n-L_n,\p_m-L_m]=0$, and can be considered as reductions
of this system. Our approach to these reductions
is to select two particular times which we treat as spatial
variables, and to impose periodicity conditions in these variables. 

\bigskip

\noindent {\bf D. Dual Formal Bloch Solutions}

A key ingredient in our construction of symplectic structures
on spaces of periodic operators $L$ is the notion
of dual Bloch functions $\Psi_0^*(x,y;k)$. In the one-dimensional case,
dual Bloch functions were introduced in [7]. In our set-up,
its main properties are as follows:

\medskip
$\bullet$ Let $\Psi_0(x,y,k)$ be a formal series of the form
(i) in Theorem 1. Then there exists a unique formal series $\Psi^*_0(x,y;k)$
of the form
$$
\Psi_0^*(x,y;k)=e^{-kx-(k^n+\sum_{i=0}^{n-2}b_ik^i)y}
\left(1+\sum_{s=1}^{\infty}\xi_s^*(x,y)k^{-s}\right), 
\eqno(2.50) $$ 
such that for all non-negative integers $m$ the equalities
$$
\r_{\infty}\left( \Psi_0^*(x,y;k)\p_x^m \Psi_0(x,y;k)\d k\right)=0, \ \ m=0,1,
\ldots \eqno(2.51)
$$
are fulfilled;

\medskip
$\bullet$ If $\Psi_0$ is a Bloch function with Bloch multipliers $w_i(k)$, then
$\Psi_0^*(x,y;k)$ is a Bloch function as well with inverse Bloch multipliers
$$ 
\Psi_0^*(x+1,y;k)=w_1^{-1}(k)\Psi_0^*(x,y;k),\ \
\Psi_0^*(x,y+1;k)=w_2^{-1}(k)\Psi_0^*(x,y;k), \eqno(2.52) 
$$

$\bullet$ If $\Psi_0(x,y;k)$ is a solution to the equation $(\p_y-L)\Psi_0=0$,
then the series 
$\Psi^*(x,y;k)$ is a solution to the adjoint equation
$$
\Psi_0^*(\p_y-L)=0, \eqno(2.53)
$$
where the action on the left of a differential operator is defined as
a formal adjoint action, i.e. for any function $f^*$
$$
\left(f^*\p_x^i\right)=(-\p_x)^if^*. \eqno(2.54)
$$

\medskip

To see this, we begin by noting that,
although each of the factors in (2.51)
has an essential singularity, their product is a meromorphic differential
and the residue is well-defined. It has the form
$$
\r_{\infty}\left( \Psi_0^*\p_x^m \Psi_0\d k\right)=\xi_m+\xi^*_m+
g_m(\xi_1,\ldots, \xi_{m-1};\xi_1^*,\ldots, \xi_{m-1}^*), 
$$
where $g_m$ is linear in $\xi_s^*$, in $\xi_s$ and their derivatives,
$s<m$. The condition (2.51) defines then $\xi_m$ recursively
as differential polynomials in $\xi_s, \ s=1,\ldots,m$.
For example, we have
$$
\xi_1^*=-\xi_1,\ \ \xi_2^*=-\xi_2+\xi_1^2-\xi_1' . 
$$
This shows the existence and uniqueness of $\Psi_0^*$. Since
the second statement is a direct corollary of the uniqueness of $\Psi_0^*$,
we turn to the proof of the last statement. First, we show that if
$\Psi^*(x,y;k)$ is a formal series
$$
\Psi^*(x,y;k)=e^{-kx-(k^n+\sum_{i=0}^{n-2}b_ik^i)y}
\left(\sum_{s=-N}^{\infty}\xi_s^*(x,y)k^{-s}\right), 
\eqno(2.55) $$
satisfying the equations (2.51),
then there exists a unique degree $N$ ordinary linear differential operator 
$D$ such that
$$
\Psi^*(x,y;k)=\Psi^*_0(x,y;k)D. 
$$
Since $\p_x^i \Psi^*_0$ 
satisfies the equations (2.51), we can find $D$ satisfying
the condition
$$
\Psi^*(x,y;k)-\Psi^*_0(x,y;k)D=O(k^{-1})\Psi_0^*(x,y;k). 
$$
The above right hand side has the form (2.55) with
$N<0$ and satisfies (2.51). Evaluating the leading term, we find 
that it must vanish identically.

Let $\Psi_0$ be a solution of $(\p_y-L)\Psi_0=0$. Then 
$$
\r_{\infty}\left(\p_y \Psi_0^*\p_x^m \Psi_0\d k\right)=
\p_y\r_{\infty}\left(\Psi_0^*\p_x^m \Psi_0\d k\right)-
\r_{\infty}\left(\Psi_0^*\p_x^m L\Psi_0\d k\right)=0
$$
In particular, there exists a differential operator $\tilde L$ such that
$$
\p_y\Psi_0^*=\Psi_0^*\tilde L. 
$$
Let $f(x)$ be an arbitrary periodic function on one variable. We have
$$
\p_y<f(x)\Psi_0^*\Psi_0>_x=<f(x)\Psi_0^*(\tilde L+L)\Psi_0)>_x, 
$$
where we have denoted as usual the average value in
$x$ of any periodic function $g(x)$ by $<g>_x$. 
The above left hand side is of order $-1$ in $k$. On the other hand,
if $L+\tilde L$ is not equal to zero and $g_i(x,y), \ 0\leq i\leq n-2$ are
its leading coefficients, then the right hand side is of the form 
$<f(x)g_i(x,y)>_x k^i+O(k^{i-1})$.
This implies that $<fg_i>_x=0$. Since $f$ was arbitrary,
we conclude that $g_i=0$, establishing the last desired
property of dual Bloch functions.

\medskip

We conclude our discussion of dual Bloch functions with several
useful remarks. The first is that the identity  
$$
<\Psi^*_0(x,y;k)\Psi_0(x,y;k)>_x=1. \eqno(2.56)
$$
holds for any formal series $\Psi_0(x,y;k)$ of the form (i) in Theorem 1
and its dual Bloch series $\Psi_0^*(x,y;k)$. Indeed, just as
in Section II.C, we can show the existence of
a unique pseudo-differential operator
$\Phi=1+\sum_{s=0}^{\infty} w_i(x,y)\p_x^{-s}$
so that 
$$
\Psi_0(x,y;k)=\Phi e^{kx+(k^n+\sum_{i=0}^n b_ik^i)y}. \eqno(2.57)
$$
As in [10], this implies
$$
\Psi_0^*(x,y;k)=\left(e^{-kx-(k^n-\sum_{i=0}^n b_ik^i)y}\right) 
\Phi^{-1}. 
\eqno(2.58)
$$ 
More precisely, let $Q=\sum_{s=N}^{\infty}q_i(x)\p_x^{-s}$
be a pseudo-differential operator. Then we may define its residue
$res_{\p}Q$ by
$$
res_{\p}Q=\r_{\infty}(e^{-kx}(Qe^{kx}))\d k=-q_1.
$$
The point is that, while the ring of pseudo-differential
operators is not commutative, the residue is, after averaging
$$
<res_{\p}(Q_1Q_2)>_x=<res_{\p}(Q_2Q_1)>_x.
$$
This shows that the series defined by the right hand side of (2.58) satisfies 
(2.51), and hence must coincide with $\Psi_0^*$. The desired identity (2.56)
is now a direct
consequence of the two preceding identities and of
the associativity of the left and right actions under
averaging.

Secondly, we would like to stress that, 
although $\Psi_0^*$ is a Bloch solution of
the ajoint equation $\Psi_0^*(\p_y-L)=0$, its normalization is different 
from that used for $\Psi_0$. This symmetry may be restored if we
introduce
$$
\Psi_0^+(x,y;k)={\Psi_0^*(x,y;k)\over \Psi_0^*(0,0;k)}. \eqno(2.59)
$$
The inverse relation is then
$$
\Psi_0^*(x,y;k)={\Psi_0^+(x,y;k)\over <\Psi_0^+\Psi_0>_x}. 
$$

Finally, the definition of the action on the left of a differential operator 
adopted earlier implies that for any degree 
$N$ differential operator
$$
D=\sum_{i=0}^N w_i(x)\p_x^i 
$$
there exist degree $(N-i)$ differential operators
$D^{(i)}$ such that for any pair of functions $f^+$ and $g$ the equality
$$
(f^*D)g=f^*(Dg)+\sum_{i=0}^n \p_x^i(f^*(D^{(i)}g)) 
$$
holds. The set of operators $D^{(i)}$ was introduced in [36]. Of particular 
interest is of course $D^{(0)}=D$, and the ``first descendant" of $D$, namely
$$
D^{(1)}=\sum_{i=0}^n iw_i(x)\p_x^{i-1}. \eqno(2.60)
$$

\medskip

\noindent{\bf E. The Basic Symplectic Structure}

We are now in position to introduce a symplectic structure on the space 
${\cal L}(b)$ of periodic operators
$L$ subject to the 
constraints (2.30), and to show that the 
infinite set of commuting flows constructed in Theorem 2 are Hamiltonian.

The main ingredients are the one-forms $\delta L$ and $\delta\Psi_0$.
The one-form $\delta L$ is given by
$$
\delta L =\sum_{i=0}^{n-2} \delta u_i \p_x^i, 
$$
and can be viewed as an operator-valued one-form on the space of operators $L=
\p_x^n+\sum_{i=0}^{n-2}u_i\p_x^i$.
Similarly, the coefficients of the series $\Psi_0$ are 
explicit integro-differential polynomials in $u_i$. Thus
$\delta \Psi_0$ can be viewed as a one-form on the space of operators 
with values in the space of formal series. More concretely, we can write
$$
\delta \Psi_0=\left(\sum_{s=1}^{\infty}\delta \xi_s k^{-s}\right)
e^{kx+(k^n+\sum_{i=0}^{n-2}b_ik^i)y}=
\left(\sum_{s=1}^{\infty}\delta \zeta_s K^{-s}\right)
e^{Kx+K^ny}.
$$
The coefficients $\delta \xi_s$ (or $\delta \zeta_s$ ) can be found from the 
variations of the formulae (2.24), (2.25) for $\xi_s$, or recursively
from the equation
$$
(\p_y-L)\delta \Psi_0=(\delta L)\Psi_0. \eqno(2.61)
$$
Let $f(x,y)$ be a function of the variables $x$ and $y$. We denote 
its mean value by
$$
<f>=\int_0^{1}\int_0^{1} f(x,y)\d x\d y.
$$
\medskip
\noindent
{\bf Theorem 3.} {\it {\rm (a)} The formula
$$
\omega={1\over 2}\ {\rm Res}_{\infty}
<\Psi_0^*\delta L\wedge \delta\Psi_0>\d k, 
\eqno(2.62)
$$
defines a symplectic form, i.e., a closed non-degenerate
two-form on the space ${\cal L}(b)$
of operators $L$ with doubly periodic coefficients. {\rm (b)}
The form $\omega$ is actually independent of the normalization
point $(x_0=0,y_0=0)$ for the formal Bloch solution
$\Psi_0(x,y;k)$. {\rm (c)} The flows (2.41)
are Hamiltonian with respect to this form, with the
Hamiltonians $nH_{m+n}(u)$ defined by (2.29).}

\medskip
\noindent{\it Proof}. 
We require the following formula, which is a generalization of the
well-known expression for the variation of energy for one-dimensional 
operators. Let $E(k)$ be the quasi-energy which is defined by (2.28).
Its coefficients are nonlocal functionals on the space ${\cal L}
(b)$
of periodic functions
$u_i(x,y)$ subject to the constraints (2.30). Then we have
$$
\delta E(k)=<\Psi_0^+\delta L \Psi_0>\eqno(2.63)
$$
Indeed, from the equation $(\p_y-L)\Psi_0=0$ and (2.53), it follows that 
$$
\p_y<\Psi_0^*\delta \Psi_0>_x=<\Psi_0^*\delta L \Psi_0>_x. 
$$
Taking the integral over $y$ and using the following monodromy property of
$\delta \Psi_0$
$$
\delta \Psi_0(x,y+1;k)=w_2(k)(\delta\Psi_0(x,y;k)+\delta E(k)\Psi_0). 
$$
we obtain (2.63).

We begin by checking that the form 
$<\Psi_0^*\delta L\wedge \delta\Psi_0)>_x$ is  periodic in $y$. 
The shift of the argument $y\to y+1$ gives
$$
<\Psi_0^*\delta L\wedge \delta\Psi_0>_x\longrightarrow
<\Psi_0^*\delta L\wedge \delta\Psi_0>_x+
<\Psi_0^*\delta L\Psi_0>_x\wedge \delta E  
$$
The second term on the right hand side can be 
rewritten as $\delta E\wedge \delta E$
and hence vanishes, due to the skew-symmetry of the wedge product. 

Next, we show that (b) is a consequence of the basic
constraints defining the space ${\cal L}(b)$. Let $\Psi_1$
be the formal Bloch solution with the normalization $\Psi_1(x_1,y_1,k)
 =1$.
Then
$$
\Psi_1(x,y;k)=\Psi_0(x,y;k)\Psi_0^{-1}(x_1,y_1;k) 
$$
and
$$
{\rm Res}_{\infty} 
<\Psi_1^*\delta L\wedge \delta\Psi_1>\d k=
2\omega+{\rm Res}_{\infty}\ \left(\delta E\wedge 
{\delta \Psi_0(x_1,y_1;k) \over \Psi_0(x_1,y_1;k)}\right)\d k.
$$
In view of the constraints (2.30), we have $\delta E= O(k^{-1})$. On the other hand, the
second factor in the last term of the above right hand side also has order
$O(k^{-1})$. The product has therefore order $O(k^{-2})$ and its 
residue equals zero.

To see that $\omega$ is a closed form, we  
express the operator $L$ as
$$
L=\Phi D\Phi^{-1}+O(\p_x^{-1}), \ D=\p_x^n+\sum_{i=0}^{n-2}B_i\p_x^i, 
\eqno(2.64)
$$
which can be done in view of (2.57) and (2.58). Therefore
$$
2\omega=<res_{\p}(D\Phi^{-1}\delta \Phi\Phi^{-1}\wedge\delta \Phi)>=
-\delta <res_{\p}(D\Phi^{-1}\delta \Phi)>. \eqno(2.65)
$$
and $\omega$ is closed.

We turn now to the non-degeneracy of $\omega$ on ${\cal L}(b)$.
Let $V$ be a vector field such that
$\omega(V_1,V)=0$ for all vector fields $V_1$. Let
$\Psi_1=\delta \Psi(V)$ be the evaluation of the one-form 
$\delta \Psi$ on $V$. Then the equality
$$
\omega(V_1,V)={\rm Res}_{\infty}
<\Psi_0^*L_1\Psi_1>\d k=0,\eqno(2.66) 
$$
holds for all degree $n-2$ operators $L_1=\delta L(V_1)$. Since $L_1$ is
arbitrary, it follows that
$\Psi_1=O(k^{-n})\Psi_0$. In view of (2.61) we have then 
$$
\delta L(V)\Psi_0=(\p_y-L)\Psi_1=O(k^{-1})\Psi_0
$$
Hence $\delta L(V)=0$. This means
that $V=0$, and the non-degeneracy of $\omega$ is established.

It remains to exhibit the flows (2.41) as Hamiltonian flows.
We recall the classical definition of the Hamiltonian 
vector field $\p_t$ corresponding
to a Hamiltonian $H$ and a two-form $\omega$. The contraction
$i(\p_t)\omega$ of $\omega$ with the vector field $\p_t$ should be the one-form 
given by the differential of the Hamiltonian, i.e. the equality 
$$
i(\p_t)\omega(X)=\omega(X,\p_t)=\d H(X), \eqno(2.67)
$$
should be fulfilled for all vector-fields $X$.

The contraction of the form $\omega$ defined by (2.62) with the 
vector-field $\p_m$ (2.41) is equal to
$$
2i(\p_m)\omega=
{\rm Res}_{\infty}\ <\Psi_0^*\delta L\p_m\Psi_0>\d k-
{\rm Res}_{\infty}\ 
<\Psi_0^*(\p_m L)\delta\Psi_0>\d k.\eqno(2.68) 
$$
Here we use the fact that the evaluations of the forms $\delta L$ and $\delta \Psi_0$
on the vector field $\p_m$ are equal by definition to
$$
\delta L(\p_m)=\p_m L, \ \ \delta \Psi_0(\p_m)=\p_m \Psi_0. 
$$
From (2.44) it follows that
$$
{\rm Res}_{\infty}<\Psi_0^*\delta L\p_m\Psi_0>\d k=
{\rm Res}_{\infty}\left( <\Psi_0^*\delta LA_m\Psi_0>+ 
\tilde \Omega_m(k)<\Psi_0^*\delta L\Psi_0>\right)\d k.
$$
The first term in the right hand side is zero due 
to the definition of $\Psi_0^*$. The usual formula for 
the implicit derivative
$$
\delta E(k) \d k=- \delta k (K) \d E, \eqno(2.69)
$$
implies that the second term is equal to
$$
-{\rm Res}_{\infty}\tilde\Omega_m(K) \delta k (K)\d E=-
{\rm Res}_{\infty}\left(K^m+O(K^{-1})\right)
\left(\sum_{s=n}^{\infty}\delta H_sk^{-s}\right)dK^n=n\delta H_{n+m}.
\eqno(2.70)
$$
(Recall that $\delta H_s=0, \ s<n$ due to the constraints.)
Consider now the second term in the right hand side of (2.68).
The equation (2.41) for $\p_m L$ and the defining equations
for $\Psi_0$ and $\Psi_0^*$ imply
$$
{\rm Res}_{\infty}<\Psi_0^*(\p_m L)\delta\Psi_0>_x\d k=
{\rm Res}_{\infty }\p_y
\left(<\Psi_0^*A_m\delta\Psi_0>_x\right)\d k. 
$$
Therefore,
$$
\int_{0}^{1}\p_y<\Psi_0^+A_m\delta\Psi_0>_{x}\d y=
\delta E(k)<\Psi_0^+A_m\Psi_0>_x|_{y=0}. 
$$
The equality (2.37) implies
$$
<\Psi_0^*A_m\Psi_0>_x=K^m+O(K^{-1}). 
$$
Hence, the second term in (2.68) is equal to
$$
{\rm Res}_{\infty}\ <\Psi_0^*(\p_m L)\delta\Psi_0>\d k=
-n\delta H_{m+n}. 
$$
and Theorem 3 is proved.

\medskip
{\it Example 1.} For $n=2$, the operator $L$ is the
second order differential operator of the form
$L=\p_x^2+u(x,y)$. The space ${\cal L}(b_0)$ is the
space of periodic functions with fixed mean value
in $x$
$$
<u>_x=H_1\to \ <\delta u>_x=0.
$$
and the symplectic form $\omega$ becomes
$$
\omega=-{1\over 2} <\delta u\wedge \int_{x_0}^x\delta u\, \d x>
$$
This symplectic form reduces to the Gardner-Faddeev-Zakharov symplectic
form when $u(x,y)=u(x)$ is a function of a single variable $x$. In this case
the KP equation reduces to the KdV equation.

\medskip
{\it Example 2.} For $n=3$, the operator $L$ is the third order differential
operator $L=\p_x^3+u\p_x+v$. The space ${\cal L}(b_0,b_1)$ is the
space of  doubly periodic functions
$u=u(x,y),\ v=v(x,y)$ satisfying the constraints
$$
<u>_x=const,\ <v>_x=const.
$$
The symplectic form $\omega$ works out to be
$$
\omega=-{1\over 3}\left( \delta u \wedge \int_{x_0}^x \delta v \d x
+\delta v\wedge \int_{x_0}^x \delta u \d x \right).
$$
In the case where $u$ and $v$ are functions of a single variable $x$, this
form gives a symplectic structure for the Boussinesq equation hierarchy
$$
u_t=2v_x-u_{xx}, \ v_t=v_{xx}-{2\over 3}u_{xxx}-{2\over 3}uu_x.
$$
Note that the usual form of the Boussinesq equation,
$u_{tt}+\left( {4\over 3}uu_x+{1\over 3} u_{xxx}\right)_x=0$,
as an equation in one unknown function $u$, is the result of eliminating $v$
from the above system.

\bigskip
\noindent
{\bf F. Lax Equations}

In this section, we compare the results obtained in our formalism with
the one-dimensional case, where the zero curvature equation
reduces to the Lax equation, and where there is a rich theory
of Hamiltonian structures.
It turns out that the symplectic structure constructed above
reduces then to the so-called first (or generalized) 
Gardner-Faddeev-Zakharov symplectic structure. Thus our approach gives a new
representation for this structure, as well as a new proof of its
well-known properties (c.f. [10][25]). As we shall see below, 
the second (Adler-Magri) symplectic structure requires a slight
modification, which explains why it is special to the
one dimensional case and has no analogue in the proposed 
Hamiltonian theory of two-dimensional systems. 
\medskip
Our construction of the basic symplectic form
$\omega$ easily extends to the construction of
an infinite sequence of symplectic structures:

\medskip
\noindent
{\bf Theorem 4.}
{\it Let $l$ be any integer $\geq0$. Then the formula
$$
\omega^{(l)}={1\over 2}\ {\rm Res}_{\infty}\ 
E^l(k)<\Psi_0^*\delta L\wedge \delta\Psi_0>\d k, 
\eqno(2.71)
$$
defines a closed two-form on the space ${\cal L}(H_1,\cdots,H_{nl-1})$
of doubly periodic operators $L$, subject to
the constraints $H_s=const$, $s=1,\ldots, nl-1$. The equations 
(2.8) are Hamiltonian with respect to this form, with the
Hamiltonians $n H_{m+n(l+1)}(u)$ defined by (2.29).}

\medskip
The proof of the theorem is identical to the proof for the basic structure
$\omega$. Specializing to the subspace of periodic $L$ with
coefficients depending only on $x$, we can easily verify that the
symplectic forms $\omega^{(l)}$ coincide with the
Gardner-Faddeev-Zakharov forms.
\medskip 
The construction of the Adler-Magri Hamiltonian is less obvious, 
although formally it has the form (2.71) with $l=-1$ and the residue at 
infinity is replaced by the residue at $E=0$.
Let $L$ be an ordinary linear differential operator of order $n$
with periodic coefficients. Then for
generic values of the complex number $E$, there exist $n$ linearly
independent Bloch solutions $\Psi_i(x,E)$ of the equation
$$
(L-E)\Psi_i=0,\eqno(2.72)
$$
with different Bloch multipliers $w_i(E)$,
$$
\Psi_i(x+1,E)=w_i(E)\Psi_i(x,E). 
$$
The value $p_i(E)=\log w_i(E)$
is called the quasi-momentum. Its differential $\d p_i(E)$ is well-defined.
(In our previous formal theory of Bloch solutions,
there are also $n$ different solutions corresponding to the same $E$, due
to the relation $E=k^n+O(k^{n-2})=K^n$ which defines $k$ and $K$ only up to a root of 
unity.) We fix the Bloch solutions $\Psi_j^*$ of the adjoint equation
$$
\Psi_i^*(L-E)=0. 
$$
by the condition
$$
<\Psi_i^*(x,E)\Psi_j(x,E)>_x=\delta_{ij}
$$

\medskip
\noindent
{\bf Theorem 5.}
{\it The formula
$$
\omega^{(-1)}={1\over 2}\ \sum_{i=1}^n R_i^{-1}
<\Psi_i^*(x,0)\delta L\wedge \delta\Psi_i(x,0)>, 
\eqno(2.73)
$$
where the constants $R_i$ are given by
$$
R_i =<\Psi_i^* L^{(1)}\Psi_i>,
$$
defines a closed two-form on the space of operators $L$ with coefficients
depending only on the variable $x$, and obeying
the constraints $H_s=const,\ s=1,\ldots, n-1$. The equations 
(2.8) are Hamiltonian with respect to this symplectic form, with the
Hamiltonians $n H_m(u)$ defined by (2.29).}

\medskip
We observe that for $n=2$, the operator $L^{(1)}$ reduces to
$L^{(1)}=2\p_x$. The expression $<\Psi_i^*L^{(1)}\Psi_i>_x$
is then just the Wronskian of two solutions
of the Schr\"odinger equation.

The proof of Theorem 5 is analogous to the proof of Theorem 3.
The formula (2.73) can be rewritten in the form
$$
\omega^{(-1)}=-{1\over 2}\ {\rm Res}_0 \ {dE\over E}\ \sum_{i=1}^n
{<\Psi_i^*(x,E)\delta L\wedge \delta\Psi_i(x,E> \over 
<\Psi_i^*(x,E)L^{(1)} \Psi_i(x,E)>} . \eqno(2.74)
$$
Due to the summation over $i$, this expression is independent
of the labeling of the Bloch functions. Thus on the right hand
side, we have the residue of a well-defined function of $E$.
The formula we need for the differential of the branch of the quasi-momentum corresponding
to the Bloch solution  $\Psi(x,E)$ of (2.72) is the following
$$
i\d p<\Psi_i^*(x,E)L^{(1)} \Psi_i(x,E)>=\d E. \eqno(2.75)
$$
Its proof is identical to the proof in the finite gap theory (see [36]).
Consider the differential $d\Psi$ in the variable $E$ of the Bloch function. 
Then
$$
(L-E)\d\Psi=-\Psi \d E. 
$$
Integrating from $x_0$ to $x_0+1$ the identity
$$
0=(\Psi^*(L-E))\d\Psi=-(\Psi^* \Psi)\d E+\sum_{j=1}^n \p_x^j 
(\Psi^* (L^{j}\d\Psi)).
$$
we obtain
$$
\d E=i\d p\big(\Psi_i^*(x_0,E)L^{(1)} \Psi_i(x_0,E)+
\sum_{j=2}^n \p_x^{j-1} (\Psi^*(x_0,E) (L^{j}\Psi(x_0,E))\big).
$$
The desired formula follows after averaging in $x_0$ this last identity.

With the formula (2.75) for $\d E$ and the analyticity in $E$ for $E\not=0$
of all relevant expressions, we can, in the computation of the contracted form
$i(\p_m) \omega^{(1)}$, reduce the residues at $E=0$ to the residue
at $E=\infty$ and get
the desired result. For example, we have
$$
-{\rm Res}_0 \sum_{i=1}^n
<\Psi_i^*\delta L\p_m\Psi_i> {\d p_i(E)\over E}=
{\rm Res}_{\infty} <\Psi^*(x,K)\delta L\p_m\Psi(x,K)> {\d K\over E}=
n\delta H_m.
$$
\bigskip
\noindent
{\bf G. Higher Symplectic Structures in the Two-Dimensional Case}

In this section, we introduce higher Hamiltonian structures 
which exist in both one and two dimensions. 
We would like to emphasize that, in contrast with
the previous results which have basically an {\it algebraic nature},
the following results require in general some additional assumptions on 
the long-time behavior of the solutions to the hierarchy of 
flows $\p_mL=\p_tA_m+[A_m,L]$. In these higher Hamiltonian structures, 
the role of the quasi-momentum
$k(K)$ for the basic structure is assumed by the quasi-momentums $\Omega_m(K)$
corresponding to the higher times of the hierarchy.

\medskip
Our first step is to study in greater detail the
quasi-momentums $\Omega_m(K)$, corresponding to higher times of
the hierarchy. Their ``densities'' $\tilde\Omega_m(K)$ made their
first appearance in (2.44). They can be re-expressed as
$$
\tilde\Omega_m(K)=A_m\Psi_0(x,y;k)|_{x=y=0}=
K^m+\sum_{s=1}^{\infty}\tilde\Omega_{s,m}K^{-s}
\eqno(2.76)
$$ 
The coefficients $\tilde\Omega_{s,m}$ of $\tilde\Omega_m(k)$ are 
integro-differential polynomials in the coefficients of the operator $L$.
As stated above, they do not depend on $(x,y)$, but they do depend
on the times $t$ if the operator $L$ evolves according to the equation 
(2.41), i.e.
$\tilde\Omega=\tilde\Omega(K,t)$. From (2.44) and (2.45) it follows that
$$\p_i \tilde\Omega_m=\p_m\tilde\Omega_i. \eqno(2.77)$$ 

Since the coefficients $\tilde\Omega_{s,m}$ are independent of the choice
of normalization point, they can
be considered as functionals on the space of periodic operators $L$.

The subsequent arguments are based on the variational formulas for these
functionals, which were found originally in the case of finite-gap 
solutions in [36]. Following [36], we use the identity
$$
\eqalignno{&\sum_{j>1}\p_x^{j-1}\left(\p_m(\Psi_0^*(L^{(j)}\delta \Psi_0)-
\p_y(\Psi_0^*(A_m^{(j)}\delta \Psi_0)\right) \cr
&=\sum_{j\geq 1}\p_x^{j-1}
\left((\Psi_0^*(L^{(j)}(\delta A_m+\delta \tilde\Omega_m)\Psi_0)
-(\Psi_0^*(A^{(j)}\delta L\Psi_0)\right) \cr
&+\sum_{k,j\geq 1} \p_x^{j+k-1}\left(\Psi_0^*(A_m^{(k)}L^{(j)}-
L^{(k)}A^{(j)}_m)\delta \Psi_0\right),&(2.78)\cr}
$$
where $L^{(j)}$ and $A^{(j)}$ are the descendants of the operators $L$ and $A$
defined by (2.60). Note that if $L$ and $A_m$ satisfy (2.44), then
the equality
$$
\p_m L^{(j)}-\p_y A_m^{(j)}+\sum_{k} [L^{(k)},A_m^{(j-k)}]=0 \eqno(2.79)
$$
holds. We now average (2.78) first in $x$ and $y$, and then in
the normalization point $x_0$ (the last averaging eliminates all terms with
$j>1$). The outcome is the equality
$$
\delta \tilde\Omega_m(K,t)<\Psi_0^* L^{(1)}\Psi_0>=
<\Psi_0^*(L^{(1)}\delta A_m-
A_m^{(1)}\delta L)\Psi_0>+\p_m<\Psi_0^*L^{(1)}\delta \Psi_0>. \eqno(2.80)
$$

Let ${\cal D}_{m_0}$, be the space of all periodic operators
$L$ which are stationary under the first $m_0$-th flow, i.e.
which satisfy the condition
$$
\p_y A_{m_0}=[L,A_{m_0}]. \eqno(2.81)
$$
It should be emphasized that due to (2.77) onto this space the corresponding
density of the corresponding quasi-momentum is a constant, i.e. does not
depend on the times
$$\Omega_{m_0}(K)=\tilde\Omega_{m_0}(K).$$

The space carrying a higher symplectic structure is a subspace
${\cal D}_{m_0}^{(I)}$ of ${\cal D}_{m_0}$, consisting
of the stationary operators $L$ satisfying in addition the following 
higher constraints 
$$
\Omega_{m_0}=K^{m_0}+\sum_{s=1}^{\infty}\Omega_{s,m_0}K^{-i},
\qquad\Omega_{s,m_0}=I_s, \eqno(2.82)
$$
for a set $I=(I_1,\cdots,I_{n-1})$ of $(n-1)$ fixed constants.
These constraints replace
the constraints (2.30) of our
previous considerations. The subspace ${\cal D}_{m_0}^{(I)}$
is invariant with respect to all the 
other flows corresponding to times $t_i$.

\medskip
\noindent
{\bf Theorem 6.}
{\it The formula
$$
\omega_{m_0}={1\over 2}{\rm Res}_{\infty}
<\Psi_0^*(A_{m_0}^{(1)}\delta L-L^{(1)}\delta A_{m_0})\wedge \delta\Psi_0>
\d k, 
\eqno(2.83)
$$
defines a closed two-form on ${\cal D}_{m_0}^{(I)}$. The restrictions of 
the equations 
(2.41) to this space are Hamiltonian with respect to this form, 
with Hamiltonians $n\Omega_{m_0,n+m}$.}

\medskip
\noindent{\it Remark 1.} 
This statement has an obvious generalization if we replace
the stationary condition (2.81) by the condition that $L$ be stationary with
respect to a linear combination of the first $m_0$ flows, i.e.
$$
\p_y A=[L,A],\ \ A=\sum_{i=0}^{m_0} c_i A_i. $$

\medskip
\noindent{\it Remark 2.} For $m_0=1$, we have $A_{1}=\p,\ A_{1}^{(1)}=1$, and
the formula (2.83) becomes identical to (2.62), i.e. $\omega_1=\omega$.

\medskip
The proof of Theorem 6 is identical to that of Theorem 3, after replacing 
of (2.63) by the formula
$$
\delta \Omega_{m_0}(K)\d E=<\Psi_0^*(L^{(1)}\delta A_{m_0}-
A_{m_0}^{(1)}\delta L)\Psi_0>\d k, \eqno(2.84)
$$
which is valid on ${\cal D}_{m_0}$. This formula is itself a direct corollary
of (2.80) and (2.75).

\medskip

\noindent
As an example, we consider the case $n=2,m_0=3$. For $n=2$, the equations (2.41) define the 
KP hierarchy on the space of periodic functions $u(x,y)$ of two variables.
For $A=L_3$, the condition (2.81) describes the stationary solutions of
the original KP equation, i.e. the space of 
functions described by the equation
$$
3u_{yy}+(6uu_x+u_{xxx})_x=0. \eqno(2.85) 
$$
Theorem 6 asserts that, besides of the basic Hamiltonian structure,
the restriction of any flow of the KP-hierarchy
to the space of functions $u(x,y)$ subject to (2.85)
is Hamiltonian with respect to the structure
given by (2.83).

\medskip
In the one-dimensional case, the constraint 
(2.81) is 
equivalent to the restriction of the Lax hierarchy on the space of finite-gap
solutions. This space is described by the following
commutativity condition for the ordinary 
differential operators $L$ and $A$ of respective degrees $n$ and $m_0$
$$
[L,A_{m_0}]=0.
\eqno(2.86)
$$
This condition is equivalent to a system of ordinary differential
equations for the coefficients $u_i(x)$ of $L$. Theorem 6 asserts that
the restriction of the Lax hierarchy to the space of
solutions to (2.86) is Hamiltonian with respect to the symplectic form
(2.83). In particular, the first flow (which is just a shift in $x$) is 
Hamiltonian. For the KdV case the corresponding symplectic structure coincides
with the {\it stationary} Hamiltonian structure found in [5].

\medskip
{\it Example 3.} We return to the case $n=3$ 
of Example 2, and consider this time operators
$A$ of order $m=2$. Thus the operators $L$ and $A$ are given by
$L=\p_x^3+u\p_x+v$, $A=\p_x^2+{2\over 3} u$. The space ${\cal D}_c$
is the space of
two quasi-periodic 
functions $u(x,y)$ 
and $v(x,y)$ satisfying the constraints
$$
<u>_x=const, \ <v>_x=const,\ <u^2>_x=const. 
$$ 
The operators $L^{(1)}$ and $A^{(1)}$ are given by
$$
L^{(1)}=3\p_x^2+u,\ \ A^{(1)}=2\p_x
$$
and the symplectic form $\omega_m$ of (2.83) becomes
$$
\omega=<{3\over 4} u \delta u \wedge\int_{x_0}^x\delta u\,\d x+
2 \delta v \wedge \int_{x_0}^x \delta v\,\d x +2 \delta u\wedge\delta v -
{3\over 2}\delta u_x \wedge \delta u >.
$$

\bigskip

\noindent
{\bf H. Symplectic Structures under Ergodicity Assumptions} 

It may be worthwhile to point out that the existence of 
the higher Hamiltonian structures obtained in the previous section
requires less that the stationary condition (2.81). The only item
which was necessary to the argument was
the possibility of dropping the last term (which was
a full derivative in $t_{m_0}$) in the formula (2.80). 

This suggests considering the space ${\cal D}_{m_0}^{erg}$ of 
all operators $L$ with smooth
periodic coefficients, for which the corresponding solution $L(t_{m_0}),\ 
L(0)=L$ of the equation (2.41) for
$m=m_0$ exists for all $t_{m_0}$ with uniformly bounded coefficients.
In this case we may introduce the quasi-momentum
$$
\Omega_{m_0}(K)=\lim_{T\to \infty} {1\over T}\int_0^T 
\tilde\Omega_{m_0}(K,t)dt_{m_0}.
$$
Although in this definition, only the dependence on $t_{m_0}$
is eliminated through averaging, 
the quasi-momentum $\Omega_{m_0}$ is actually also independent
of all the other times $t_i$, in view of (2.77).

For $L\in {\cal D}_{m_0}^{erg}$ the formula
$$
\delta \Omega_{m_0}(K)\d E=<\Psi_0^*(L^{(1)}\delta A_{m_0}-
A_{m_0}^{(1)}\delta L)\Psi_0>_0\d k, \eqno(2.87)
$$
where $<f(x,y,t)>_0$ stands for
$$
<f(x,y,t)>_0=\lim_{T\to \infty}{1\over T}\int_0^T <f(x,y,t>dt,
$$
holds. Here we make use of the fact that if
$\delta L$ is variation of the initial Cauchy data $L(0)$ for (1.47)
then the variation $\delta L(t_{m_0})$ is defined by the 
linearized equation
$$
\p_{m_0} \delta L-\p_y\delta A_{m_0}+[\delta L,A_{m_0}]+[L,\delta A_{m_0}].
$$
With (2.87), it is now easy to establish the following theorem
\medskip
\noindent{\bf Theorem 7.} {\it The formula
$$
\omega_{m_0}={1\over 2}{\rm Res}_{\infty}
<\Psi_0^*(A_{m_0}^{(1)}\delta L-L^{(1)}\delta A_{m_0})\wedge 
\delta\Psi_0>_0\d k,\eqno(2.88)
$$
defines a closed two-form on subspaces of ${\cal D}_{m_0}^{erg}$ subject to 
the constraints (2.82). The restrictions of the 
equations (2.41) to this space are Hamiltonian with respect to this 
form, with the Hamiltonians $n\Omega_{m_0,n+m}$.}

\medskip

The space ${\cal D}_{m_0}^{erg}$ appears to be
a complicated space, and we do not
have at this moment an easier description for it.
As noticed in [36], it contains (for an arbitrary $m_0$) 
all the finite-gap solutions.
There exist a few other cases where we can justify the ergodicity assumption.
For example, for the KdV hierarchy, 
the ergodicity assumption is fulfilled
for smooth periodic functions with sufficiently
rapidly decreasing Fourier coefficients.
Indeed, if $u(x)$ can be extended as an analytic function in a
complex neighborhood of real values for $x$ and $y$,
$$
|u(x)|<U ,\ \ |{\rm Im}\, x|<q \eqno(2.89)
$$
then $u(x,t)$ is bounded by the same constant for all $t$ due
to trace formulae. Using the approximation theorem [37] for all 
periodic solutions to (2.5) (also called the KP-2 equation,
by contrast with the KP-1 equation given by (3.19) below) by 
finite-gap solutions, we can prove the ergodicity assumption 
in the case when the
Fourier coefficients $u_{ij}$ of $u$ satisfy the condition
$|u_{ij}|<Uq^{|i|+|j|}$.  
   
\medskip
\noindent{\it Important Remark.} In order to clarify the meaning of the higher
symplectic forms and the higher Hamiltonians, it is instructive
to explain its analogue for the usual Lax equations.
The Lax equations $\p_mL=[M_m,L]$ obviously imply that the eigenvalues of
$L$ are integrals of motion, and usually they serve as Hamiltonians for 
the basic symplectic structure. The higher Hamiltonians correspond to the 
eigenvalues  of the operator $M_{m_0}$ instead of $L$. 
Of course, they are time dependent,
but after averaging with respect to one of the times, namely $t_{m_0}$,
they become nonlocal integrals of motion and can serve as Hamiltonians for
the corresponding symplectic structures. 
\bigskip
\noindent
{\bf I. The Matrix Case and the 2D Toda Lattice}

Our formalism extends without difficulty to a variety of
more general settings. We shall discuss briefly the
specific cases of matrix equations and of the Toda lattice,
which correspond respectively to the cases where $L$
is matrix-valued, and where the differential operator
$\p_x$ is replaced by a difference operator.
\medskip
Let $L=\sum_{i=0}^nu_i(x,y)\p_x^i$ be then an operator with
matrix coefficients $u_i=(u_i^{\a\b})$ which are smooth
and periodic functions of $x$ and $y$, whose leading term
$u_n^{\a\b}=u_n^{\a}\delta_{\a\b}$ is diagonal
with distinct diagonal elements $u_n^{\a}\not= u_n^{\b}$ for $\a\not=\b$,
and which satisfies $u_{n-1}^{\a\a}=0$. Then, arguing as in Section
II.B, we can show that there exists a unique
matrix formal solution $\Psi_0=(\Psi_0^{\a\b}(x,y;k))$ of the equation
$(\p_y-L)\Psi_0=0$, which has the form
$$
\Psi_0(x,y;k)=(I+\sum_{s=1}^{\infty}\xi_s(x,y)k^{-s}){\rm exp}
(kx+u_nk^ny+\sum_{i=0}^{n-2}B_i(y)k^i)\eqno(2.90)
$$
(where $I$ is the identity matrix, $\xi_s=(\xi_s^{\a\b})$ are
matrix functions, and $B_i(y)=(B_i^{\a\b}(y))=(B_i^{\a}(y)\delta_{\a\b})$
are diagonal matrices), and which has the Bloch property
$$
\Psi_0(x+1,y;k)=\Psi_0(x,y;k)w_1(k),\ w_1(k)=e^{k}.
$$
The formal solution $\Psi_0(x,y;k)$ has the Bloch property
with respect to $y$ as well,
$$
\Psi_0(x,y+1;k)=\Psi_0(x,y;k)w_2(k)
$$
with the Bloch multiplier $w_2(k)$ of the form
$$
w_2(k)=(1+\sum_{s=1}^{\infty}J_sk^{-s})
{\rm exp}(u_nk^n+\sum_{i=0}^{n-2}B_ik^i),
$$
where $J_s$ and $B_i$ are diagonal matrices.

As noted at the end of Section II.B, the second Bloch multiplier defines
the quasi-energy $E(k)$. This defines in turn the functionals
${\cal E}_i$ just as in (2.28), with the only difference
the fact that they are now diagonal matrices. If we introduce
the diagonal matrix $K$ by the equality
$$
u_nK^n=E(k)={\rm log}\,w_2(k)=u_nk^n+
\sum_{i=-n+2}^{\infty}{\cal E}_ik^{-i}\eqno(2.91)
$$
then we may define diagonal matrices $H_s=(H_s^{\a}\delta_{\a\b})$ by
$$
kI=K+\sum_{i=1}^{\infty}H_sK^{-s}\eqno(2.92)
$$

The definition of the commuting flows in the matrix case
is then just the same as in the scalar case. The only difference
is that the number of these flows is now $N$ times larger. The
corresponding times are denoted by $t=(t_{\a,m})$, and the flows
are given by
$$
[\p_y-L,\p_{\a,m}-L_{\a,m}]=0\eqno(2.93)
$$
where $L_{\a,m}$ is the unique operator of the form
$$
L_{\a,m}=v_{\a}\p_x^m+\sum_{i=1}^{m-1}u_{i,(\a,m)}(t)\p_x^i,\ 
v_{\a}^{\b\gamma}=\delta_{\a\b}\delta^{\b\gamma}
$$
which satisfies the condition
$$
L_{\a,m}\Psi_0(x,y;k)=\Psi_0(x,y;k)(v_{\a}K^m+O(K^{-1})).\eqno(2.94)
$$
As before, the dual Bloch formal series $\Psi_0^*(x,y;k)$ is defined as
being of the form (2.58), and satisfying the equation
$$
\r_{\infty}{\rm Tr}(\Psi_0^*v\p_x^m\Psi_0)\d k=0,\ m\geq0\eqno(2.95)$$
for arbitrary matrices $v$. We have then

\medskip

\noindent
{\bf Theorem 8.} {\it The formula
$$
\omega={1\over 2}\r_{\infty}Tr<\Psi_0^*\delta L\wedge\delta\Psi_0>
\d k\eqno(2.96)
$$
defines a closed non-degenerate two-form on the space of periodic operators $L$
subject to the constraints $H_s^{\b}=constant$, $\beta=1,\cdots,l$,
$s=1,\cdots,n-1$. The equations (2.93) are Hamiltonian
with respect to this form, with Hamiltonians $nH_{m+n}^{\a}$ defined
by (2.92).}

\medskip
{\it Example 4.}
Consider the case $n=1$, where the operator $L$ is of the form
$L=A\p_x+u(x,y)$, with
$A$ the $N\times N$ matrix
$A^{\alpha,\beta}=a_{\alpha}\delta_{\alpha,\beta}$
and $u(x,y)$ is an $N\times N$ matrix with zero
diagonal entries
$u^{\alpha,\alpha}=0$. 
In this case the symplectic form (2.96) becomes 
$$
\omega = \sum_{\alpha\neq \beta}{1\over a_{\alpha}-a_{\beta}}
\delta u^{\beta,\alpha}\wedge \delta u^{\alpha,\beta}.
$$

\bigskip
Finally, as a basic example of a system corresponding to an auxiliary linear 
equation where the differential operator $\p_x$ is replaced by a 
difference operator acting on spaces of infinite sequences, we consider
the $2D$ Toda lattice.

The 2D Toda lattice is the system of equations for the unknown functions
$\varphi_n=\varphi_n(t_+,t_-)$
$$
{\p^2\over \p t_+\p t_-}\varphi_n=e^{\varphi_n-\varphi_{n-1}}-
e^{\varphi_{n+1}-\varphi_{n}}. 
$$
It is equivalent to the compatibility conditions for the 
following auxiliary
linear problem
$$
\p_+\psi_n=\psi_{n+1}+v_n \psi_n,\ v_n=\p_+ \varphi_n, \ \
\p_-\psi_n=c_n\psi_{n-1},\ c_n=e^{\varphi_n-\varphi_{n-1}}. 
$$

We consider solutions of this system which are periodic in the variables
$n$ and $y=(t_++t_-)$. The relevant linear operator is the difference operator
$$
L\Psi_n=\Psi_{n+1}+v_n\Psi_n+c_n\Psi_{n-1}
$$
with periodic coefficients $v_n(y)=v_{n+N}(y)=v_n(y+l)$ and
$c_n(y)=c_{n+N}(y)=c_n(y+1)$.
Then, arguing as in Section II.B we can show that there exist unique
formal solutions $\Psi^{(\pm)}=\Psi_n^{(\pm)}(y;k)$ of the equation
$$
(\p_y-L)\Psi^{(\pm)}=0, 
$$
which have the form
$$
\Psi_n^{(\pm)}(y;k)=k^{\pm n}
\left(\sum_{s=0}^{\infty}\xi_s^{(\pm)}(n,y)k^{-s}\right) e^{ky+B(y)},\ \ 
\xi_0^{(+)}=1, \eqno(2.97)
$$
the Bloch property
$$
\Psi_{n+N}^{(\pm)}(y;k)=\Psi_{n}^{(\pm)}(y;k)w_1^{(\pm)}(k),\ \ \
w_1^{(\pm)}(k)=k^{\pm N},
$$
and which are normalized by the condition 
$$
\Psi_0^{(\pm)}(0;k)=1.
$$
The coefficients $\xi_s^{(\pm)}$ can be found recursively. The initial value
$\xi_0^{(+)}=1$ and the condition that $\xi_1^{(+)}$ is periodic in $n$
define the function $B(y)$ in (2.97) 
$$ 
B(y)=N^{-1}\int_0^y \left(\sum_{n=1}^N v_n\right)\d y. 
$$ 
The only difference with the previous differential case 
is the definition of the leading 
term $\xi_0^{(-)}(n,y)$. 
Let us introduce $\varphi_n(y)=\log \xi_0^{(-)}(n,y)$. 
Then we have $$ c_n(y)=e^{\varphi_n(y)-\varphi_{n-1}(y)}.  
$$ 
The periodicity condition for $\xi_1^{(-)}$ requires the equality 
$$ 
\sum_{n=0}^{N-1} \p_y \varphi_n =0, 
$$ 
which allows us to define $\varphi_n$ uniquely through $c_n$:
$$ 
\varphi_n=\sum_{n=1}^{n-1}\log c_n- 
\int_0^y \left(N^{-1}\sum_{n=0}^{N-1}\log c_n\right)\d y 
$$ 
The formal solutions $\Psi_n^{(\pm)}(y,k)$ have the Bloch 
property with respect to $y$ as well, 
$$
\Psi_{n}^{(\pm)}(y+1;k)=\Psi_{n}^{(\pm)}(y;k)w_2^{(\pm)}(k) 
$$ with the Bloch multipliers $w_2^{(\pm)}(k)$ of the form 
$$
w_2^{(\pm)}(k)=e^{k+B}\left(\sum_{s=0}^{\infty} J_s^{(\pm)}k^{-s}\right).  
$$
As noted at the end of Section II.B, the second Bloch multiplier defines the 
quasi-energy $E^{(\pm)}(k)$ and the functionals ${\cal E}_i^{(\pm)}$ just as in 
(2.28). If we introduce the variable 
$$
K=E^{(\pm)}(k)=\log w_2^{(\pm)}(k)=k+\sum_{i=0}^{\infty}{\cal 
E}_i^{(\pm)}k^{-i}
$$
then we may define the functionals
$H_s^{(\pm)}$ by
$$
\log k=\log K\pm\sum_{s=0}H_s^{(\pm)} K^{-s}.\eqno(2.98)
$$
The definition of the commuting flows in the discrete case is then
the same as in the scalar case. 
The basic constraints that specify the space of periodic functions $v_n$ and 
$c_n$ have the form
$$
I_0=\sum_{n=1}^n\log c_n(y)=const,\ \ I_1=\sum_{n=1}^N v_n(y)=const.\eqno(2.99)
$$
The corresponding times are denoted by $t=(t_{\pm,m})$ and 
the flows are given by 
$$ 
[\p_y-L, \p_{\pm, m}-L_{\pm,m}]=0 ,\eqno(2.100)
$$ where $L_{\pm,m}$ is the unique operator of 
the form 
$$ 
L_{\pm,m}\Psi_n=\sum_{i=0}^m u_{i,(\pm,m)}(n,t)\Psi_{n \pm i},\ 
u_{i,(+,m)}=1,\ u_{i,(-,m)}(n,t)=e^{\varphi_n-\varphi_{n-i}} 
$$ 
which satisfies the condition 
$$ 
L_{\pm,m}\Psi_n^{(\pm)}=\Psi_n^{(\pm)}(K^m+O(K^{-1})).  
$$ 
The dual formal series $\Psi_n^{*,(\pm)}$ are defined as the formal series 
of the form 
$$ 
\Psi_n^{*,(\pm)}(y,k)=k^{\mp n}
\left(\sum_{s=0}^{\infty}\xi_s^{*,(\pm)}(n,y)k^{-s}\right) e^{-(ky+B(y))},\ \ 
\xi_0^{*,(+)}=1, 
$$
which satisfy the equations
$$
\r_{\infty}\left(\Psi_n^{*,(+)}\Psi_m^{(+)})-\Psi_n^{*,(-)}\Psi_m^{(-)}
\right){\d k\over k}=0, m\in Z
$$
for all integers $m$.

\medskip
\noindent
{\bf Theorem 9.} {\it The formula 
$$
\omega=\r_{\infty} <\Psi^{*,(+)}\delta L\wedge \delta \Psi^{(+)}
-\Psi^{*,(-)}\delta L\wedge\delta \Psi^{(-)}>{\d k\over k}=
<(2\delta v_n-\delta\p_y \varphi_n)\wedge \delta \varphi_n>
\eqno(2.101)
$$
defines a closed non-degenerate two-form on the space of periodic operators $L$
subject to the constraints (2.99). The equations (2.100) are 
Hamiltonian with respect
to this form, with Hamiltonians $H_{m}^{\pm}$ defined by (2.98). }

\medskip
Finally, we would like to conclude this Chapter by calling the reader's
attention to [59][60], where symplectic structures are discussed
from a group theoretic viewpoint, and where applications
of soliton theory to harmonic maps are given.

\bigskip 
\centerline{\bf III. GEOMETRIC THEORY OF 2D SOLITONS}

\bigskip

In Chapter I, we had developed a general Hamiltonian theory of
2D solitons. The central notion was the symplectic form (1.1), which
was defined on the infinite-dimensional space ${\cal L}(b)$ of doubly
periodic operators obeying suitable constraints. Our main goal in this Chapter
is to present and extend the results of [39]. In this work, 
as described in the Introduction, a natural symplectic form $\omega_{\M}$
was constructed on Jacobian fibrations
over the leaves of moduli spaces $\M_g(n,m)$ of finite-gap solutions
to soliton equations. Imbedded in the space of doubly
periodic functions, the form $\omega_{\M}$ was shown to coincide with
$\omega$ (this was in fact our motivation for constructing a general Hamiltonian
theory based on $\omega$ in this paper). Although the infinite-dimensional
symplectic form $\omega$ and its variants in Chapter II
can be expected to play an important role in an analytic theory of solitons,
it is the geometric and finite-dimensional
form $\omega_{\M}$ which has provided a unifying theme
with topological and supersymmetric field theories. 

\medskip

In Chapter II, we have seen how a differential operator $L$ determined
a Bloch function $\Psi_0$, which was a formal series in 
a spectral parameter $k$ or $K$. 
The key to the construction of finite-gap solutions of soliton
equations is the reverse process, namely the association of an operator
$L$ to a series of the form of $\Psi_0$. To allow for evolutions
in an arbirary time $t_m$, it is convenient for us to 
incorporate a factor $e^{t_mk^m}$ in $\Psi_0$ for each $t_m$, and consider
series $\Psi_0(t;k)$ of the form
$$
\Psi_0(t;k)=(1+\sum_{s=1}^{\infty}\xi_s(t)k^{-s}){\rm exp}(\sum_{i=1}^{\infty}
t_ik^i).\eqno(3.1)
$$
As usual, all the times $t_i$ except for a finite number
have been set to 0. Then the operators $L_m$ are uniquely defined
by the requirement that 
$$(\p_m-L_m)\Psi_0(t;k)=O(k^{-1}){\rm exp}
(\sum_{i=1}^{\infty}t_ik^i)$$ 
(this is equivalent to the
earlier requirement that $(L_m-k^m)\Psi_0=O(k^{-1})\Psi_0$ in the case of 
$\xi_s(t)$ independent of $t_m$). In particular, we have the following
identity between formal power series
$$
[\p_n-L_n,\p_m-L_m]\Psi_0=O(k^{-1}){\rm exp}
(\sum_{i=1}^{\infty}t_ik^i)\eqno(3.2)
$$
This identity assumes its full value when the formal series $\Psi_0(t;k)$
is a genuine convergent function of $k$ and has an analytic
continuation as a meromorphic function with $g$ poles
on a Riemann surface of genus $g$. In this case,
the equation (3.2) with zero right hand side becomes exact.
The null space of $[\p_n-L_n,\p_m-L_m]$ is parametrized
then by $k$ and is infinite-dimensional. Since $[\p_n-L_n,\p_m-L_m]$
is an ordinary differential operator, it must vanish.
Thus a convergent $\Psi_0(t;k)$
gives rise to a solution
of the zero curvature equation $[\p_n-L_n,\p_m-L_m]=0$.
The algebraic-geometric theory of solitons provides precisely
the geometric data which leads to convergent
Bloch functions. These functions
are now known as Baker-Akhiezer functions. 

\medskip

\noindent{\bf A. Geometric Data and Baker-Akhiezer Functions}

In a Baker-Akhiezer function, the spectral parameter $k$ is interpreted as the
inverse $k=z^{-1}$ of a local coordinate $z$ on a Riemann surface. Thus let 
a ``geometric data" $(\G,P,z)$
consist of a Riemann surface $\G$ of some fixed genus $g$, 
a puncture $P$ on $\G$, and
a local coordinate $k^{-1}$ near $P$. Let
$\g_1,\cdots,\g_g$ be $g$ points of $\G$ in general position. 
Then for any $t=(t_i)_{i=1}^{\infty}$,
only a finite number of which are non-zero,
there exists a unique function $\Psi(t;z)$ satisfying

\medskip

\item{(i)} $\Psi$ is a meromorphic on $\G\setminus P$,
with at most simple poles at $\g_1,\cdots,\g_g$;
\item{(ii)} in a neighborhood of $P$, $\Psi$ can
be expressed as a convergent series in $k$ of the form
appearing on the right hand side of (3.1).

The exponential factor in (3.1) describes the essential singularity of $\Psi(t;z)$
near $P$. Alternatively, we can view it
as a transition function (on the overlap
betwwen $\G\setminus P$ and a neighborhood of $P$)
for a line bundle ${\cal L}(t)$ on $\Gamma$. The Baker-Akhiezer
function $\Psi$ is then a section of ${\cal L}(t)$,
meromorphic on the whole of $\G$.

The form of the essential singularity of $\Psi$
implies that $\Psi$ has as many zeroes as it has poles
(equivalently, the line bundle ${\cal L}(t)$ has
vanishing Chern class). Indeed, 
$\d\Psi/\Psi=\d(\sum_{i=1}^{\infty}t_ik^{-i})+{\rm regular}$,
and thus has no residue at $P$. From this, the uniqueness of the 
Baker-Akhiezer function
follows, since the ratio $\Psi/\tilde\Psi$
of two Baker-Akhiezer functions would be a meromorphic function
on the whole of $\G$,
with at most $g$ poles (corresponding to the
zeroes of $\tilde\Psi$). By the Riemann-Roch Theorem, it must
be constant. Finally, the existence of $\Psi$ can be deduced
most readily from an explicit formula. Let $A_1,\cdots,A_g$
$B_1,\cdots,B_g$ be a canonical homology basis for $\G$
$$
A_j\cap A_k=B_j\cap B_k=0, \ A_j\cap B_k=\delta_{jk},\eqno(3.3)
$$   
and let 
\item{(a)} $\d\omega_j$, $\tau_{jk}$ be respectively
the dual basis of holomorphic
abelian differentials and the period matrix
$$
\oint_{A_j}\d\omega_k=\delta_{jk},\ \oint_{B_j}\d\omega_k=\tau_{jk}.
$$
\item{(b)} $\theta(z|\tau)$ the Riemann $\theta$-function;
\item{(c)} $\d\Omega_i^0$ the Abelian differential of the second
kind with unique pole of the form
$$
\d\Omega_i^0=(1+O(k^{-i-1}))\d k^i\eqno(3.4)
$$
normalized to have vanishing $A$-periods
$$
\oint_{A_I}\d\Omega_i^0=0,\ 1\leq I\leq g;\eqno(3.5)
$$
\item{(d)} $P_0$ a fixed reference point, with which we can
define the Abel map $A:z\in\G\rightarrow A(z)\in{\bf C}^g$ and 
Abelian integrals $\Omega_i^0$ by
$$
A_j(z)=\int_{P_0}^z\d\omega_j,\ \Omega_i^0(z)=\int_{P_0}^z\d\Omega_i^0
\eqno(3.6)
$$
Here the Abel map as well as the Abelian integrals are path
dependent, and we need to keep track
of the path, which is taken to be the same in both cases.
\item{(e)} $Z=K-\sum_{s=1}^gA(\g_s)$, where $K$ is the vector
of Riemann constants (c.f. [48]).
\medskip
Using the transformation laws for the $\theta$-function,
it is then easily verified that the following expression
is well defined, and must coincide with $\Psi(t;z)$
$$
\Psi(t;z)
={\theta(A(z)+{1\over 2\pi i}\sum_{i=1}^{\infty}t_i\oint_B\d\Omega_i^0
+Z|\tau)\theta(A(P)+Z)
\over\theta(A(z)+Z)
\theta(A(P)+{1\over 2\pi i}\sum_{i=1}^{\infty}t_i\oint_B\d\Omega_i^0
+Z|\tau)}{\rm exp}\bigg(\sum_{i=1}^{\infty}
t_i\Omega_i^0(z)\bigg).\eqno(3.7)
$$

In this formalism, the role of the quasi-momentum $p$ is now assumed
by the Abelian integral $\Omega_1^0$, and we just define $p=\Omega_1^0$.
With the Baker-Akhiezer function assuming now the former role of
formal Bloch functions, we can construct a hierarchy of operators
$L_n$ as in (2.37). The requirement that
$(\p_m-L_m)\Psi(t;z)=O(z)\Psi(t;z)$ 
determines recursively the coefficients of $L_m$ as differential
polynomials in the $\xi_s$. The crucial improvement over
formal Bloch functions is that here,
this requirement actually implies that 
$$
(\p_n-L_n)\Psi(t;z)=0\eqno(3.8)
$$
identically. In fact, $(\p_n-L_n)\Psi(t;z)$ satisfies
all the conditions for a Baker-Akhiezer function,
except for the fact that the Taylor expansion
of its coefficient in front of the essential singularity
${\rm exp}(\sum_{i=1}^{\infty}t_ik^i)$ starts with $k^{-1}$.
If this function is not identically zero, it can be used to
generate distinct Baker-Akhiezer functions from any
given one, contradicting the uniqueness of Baker-Akhiezer functions.
As noted earlier, (3.8) implies that 
$[\p_n-L_n,\p_m-L_m]\Psi(t;z)=0$, and hence that the zero curvature
equation $[\p_n-L_n,\p_m-L_m]=0$ holds. In summary, we have defined
in this way a ``geometric map' ${\cal G}$ which sends the geometric data
$(\G,P,z;\g_1,\cdots,\g_g)$ to an infinite hierarchy of operators [33][34]
$$
{\cal G}:(\G,P,z;\g_1,\cdots,\g_g)\rightarrow (L_n)_{n=2}^{\infty}\eqno(3.9)  
$$

\medskip
The expression (3.7) for $\Psi(t;z)$ leads immediately to
explicit solutions for a whole hierarchy of soliton
equations. Let $t_1=x$, $t_2=y$, $t_3=t$, and $n=2$. We find
then solutions $u(x,y,t)$ of the KP equation (2.4) expressed as [34]
$$
u(x,y,t)=2\p_x^2{\rm log}\,\theta(x\oint_B\d\Omega_1^0+y\oint_B\d\Omega_2^0
+t\oint_B\d\Omega_3^0+Z|\tau)+const.\eqno(3.10)
$$
This formula is at the origin of a remarkable application
of the theory of non-linear integrable models, namely
to a solution of the famous Riemann-Schottky problem.
According to the Torelli theorem, the period matrix defines
uniquely the algebraic curve. The Riemann-Schottky
problem is to describe the symmetric matrices
with positive imaginary part which are period matrices
of algebraic curves. Novikov conjectured that the
function $u(x,y,t)=2\p_x^2{\rm log}\,\theta(Ux+Vy+Wt|\tau)$
is a solution of the KP equation if and only if
the matrix $\tau$ is the period matrix of an algebraic curve,
and $U,V,W$ are the $B$-periods of the corresponding
normalized meromorphic differentials with
poles only at a fixed point of the curve. This conjecture
was proved in [3][55].

\medskip
\noindent{\it The dual Baker-Akhiezer function}

For later use, we also recall here the main properties
of the dual Baker-Akhiezer function $\Psi^{+}(t;z)$ which 
coincides with the formal dual series defined in Section II.D. To define 
$\Psi^+(t;z)$,
we note that, given $g$ points
$\g_1,\cdots,\g_g$ in general position, the unique meromorphic
differential 
$\d\Omega=\d(z^{-1}+\sum_{s=2}^{\infty}a_sz^s)$
with double pole at $P$ and zeroes at $\g_1,\cdots,\g_g$,
must also have $g$ other zeroes, by the Riemann-Roch theorem. Let
these additional zeroes be denoted by
$\g_1^+,\cdots,\g_g^+$. Then the dual Baker-Akhiezer function $\Psi^+(t;z)$
is the unique function $\Psi^+(t;z)$
which is meromorphic everywhere except at $P$, has at most
simple poles at $\g_1^+,\cdots,\g_g^+$, and admits the
following expansion near $P$
$$
\Psi^+(t;z)
={\rm exp}(-\sum_{n=1}^{\infty}t_nz^{-n})(1+\sum_{s=1}^{\infty}\xi_s^+(t)z^s)
$$
To compare the dual Baker-Akhiezer function with the
formal dual Bloch function $\Psi_0^+$ of Section II.D,
it suffices to observe that
$$
{\rm Res}_P\Psi^+(t;z)\big(\partial_x^m\Psi(t;z)\big)\d\Omega=0,
$$
since the differential on the left hand side
is meromorphic everywhere, and holomorphic away from $P$. Together with the 
normalization $\Psi^+(0;z)=1$, this implies that $\Psi^+$ indeed coincides
with the formal dual function $\Psi_0^+$. An exact formula for
$\Psi^+(t;z)$ can be obtained from (3.7) by changing
signs for $t$ and by replacing the vector $Z$ by $Z^+$. From
the definition of the dual set of zeroes $\g_1^+,\cdots,\g_g^+$,
this vector satisfies the equation
$Z+Z^+=2P+K$, where $K$ is the canonical class.
Recalling that the quasi-momentum
$p$ was defined to be $p=\Omega_1$, we also obtain the following
formula for the differential $\d\Omega$ we introduced earlier 
$$
\d\Omega={\d p\over <\Psi^+\Psi>}.\eqno(3.11)
$$

\medskip
\noindent{\it The Multi-Puncture Case}

The above formalism extends easily to the case of $N$ punctures
$P_{\a}$ (with one marked puncture $P_1$). The Baker-Akhiezer function
$\Psi$ is required then to have the essential singularity
$$
\Psi(t;z)=\big(\sum_{s=0}^{\infty}\xi_{s\a}(t)k_{\a}^{-s}\big)
{\rm exp}(\sum_{i=1}^{\infty}t_{i\a}k_{\a}^i)\eqno(3.12)
$$
where $k_{\a}^{-1}$ are local coordinates near each puncture $P_{\a}$,
$t_{i\a}$ are given ``times", only a finite number of which
are non-zero, and the coefficient $\xi_{s1}$ at $P_1$ is normalized to
be 1 for $s=0$. We can introduce as before $\d\Omega_{i\a}^0$
associated now to each puncture $P_{\a}$ and their Abelian integrals
$\Omega_{i\a}^0$. Then the Baker-Akhiezer function $\Psi(t;z)$
becomes
$$
\eqalignno{\Psi(t;z)
=&{\theta(A(z)+{1\over 2\pi i}\sum_{\a=1}^N\sum_{i=1}^{\infty}t_{i\a}
\oint_B\d\Omega_{i\a}^0
+Z|\tau)\theta(A(P_1)+Z)
\over\theta(A(z)+Z)\theta(A(P_1)+{1\over 2\pi i}
\sum_{\a=1}^N\sum_{i=1}^{\infty}t_{i\a}
\oint_B\d\Omega_{i\a}^0
+Z|\tau)}\cr
&\qquad\times{\rm exp}\bigg(\sum_{\a=1}^N\sum_{i=1}^{\infty}
t_{i\a}\Omega_{i\a}^0(z)\bigg)&(3.13)
\cr}
$$
For each pair $(\a,n)$ there is now a unique operator $L_{\a n}$ of the form
(2.36) so that $(\p_{\a n}-L_{\a n})\psi(t,z)=0$, with 
$\p_{\a n}=\p/\p t_{\a n}$. The operators $L_{\a n}$
satisfy the compatibility condition
$[\p_{\a n}-L_{\a n},\p_{\b m}-L_{\b m}]=0$.

\medskip
\noindent{\it Periodic Solutions}

In general, the finite-gap solutions of soliton equations
obtained by the above construction are meromorphic, quasi-periodic
functions in each of the variables $t_{\a i}$ (a quasi-periodic
function of one variable is the restriction
to a line of a periodic function of several variables).
We would like to single out the geometric data
which leads to periodic solutions. For this we need
the following slightly different formula
for the Baker-Akhiezer function. 

Let $\d\Omega_{i\a}$ be the unique differential with pole of
the form (3.4) near $P_{\a}$, but normalized so that all its
periods be purely imaginary, and define the function
$\Phi(\zeta_1,\cdots,\zeta_{2g};z)$ by the formula
$$
\Phi(\zeta;z)
={\theta(A(z)+\zeta_ke_k+\zeta_{k+g}\tau_k+Z|\tau)\theta(A(P_1)+Z)
\over\theta(A(z)+Z)\theta(A(P_1)+\zeta_ke_k+\zeta_{k+g}\tau_k+Z|\tau)}
{\rm exp}\big(2\pi i\sum_{k=1}^gA_k(z)\zeta_{k+g}\big)\eqno(3.14)
$$
where $e_k=(0,\cdots,0,1,0,\cdots,0)$ are the basis vectors in ${\bf C}^g$,
and $\tau_k$ are the vectors with components $\tau_{jk}$.
We observe that $\Phi$ is periodic of
period 1 in each of the variables $\zeta_1,\cdots,
\zeta_{2g}$. Then the Baker-Akhiezer function can be expressed as
$$
\Psi(t;z)
=\Phi(\sum_{i\a}t_{i\a}U_{i\a};z)
{\rm exp}\big(\sum_{i\a}t_{i\a}\Omega_{i\a}\big)\eqno(3.15)
$$
where we have denoted by $U_{i\a}$ the real, $2g$-vector of periods
of $\d\Omega_{i\a}$
$$
U_{i\a}^k={1\over 2\pi i}\oint_{A_k}\d\Omega_{i\a},\ \
U_{i\a}^{k+g}=-{1\over 2\pi i}\oint_{B_k}\d\Omega_{i\a}
$$ 
In particular, for geometric data $\{\G,P_{\a}, z_{\a}\}$
satisfying the condition
$$
{1\over 2\pi i}\sum_{i\a}a_{i\a}U_{i\a}\in {\bf Z},\eqno(3.16)
$$
the Baker-Akhiezer function is a Bloch function with respect
to the variable $\tilde t$ if we set $t_{i\a}=a_{i\a}\tilde t$, with Bloch
multiplier $w={\rm exp}(\sum_{i\a}a_{i\a}\Omega_{i\a}(z))$.
The coefficients of the operators $L_{\a n}$
are then periodic functions of $\tilde t$. As an example, we
consider the one-puncture case. If we express the
data under the form $U_1^k=2\pi m_k/l_1$, $U_2^k=2\pi n_k/l_2$, with
$m_k,n_k\in {\bf Z}$, then the corresponding solution
of the KP hierarchy is periodic in the variables
$x=t_1$, $y=t_2$, with periods $l_1$ and $l_2$ respectively.
\medskip

\noindent{\it Real and Smooth Solutions}

There are two types of conditions which guarantee that
the solutions obtained by the above geometric
construction are real and smooth for real values of $t_{i\a}$.
We present them in the case of the KP hierarchy.

Assume that the geometric data defining the Baker-Akhiezer
function is real, in the sense that
\item{(a)} the algebraic curve $\G$ admits an anti-holomorphic
involution $\iota:\G\rightarrow\G$;
\item{(b)} the puncture $P_1$ is a fixed point of $\iota$;
\item{(c)} the local coordinate $k^{-1}$ in a neighborhood
of $P_1$ satisfies the condition $k(\iota(z))=\overline {k(z)}$;
\item{(d)} the divisor $(\g_1,\cdots,\g_g)$ is invariant
under $\iota$, i.e., $\iota(\g_s)=\g_{\sigma(s)}$,
where $\sigma$ is a permutation.

Then the Baker-Akhiezer function satisfies the reality condition
$$
\Psi(t;\iota(z))=\overline{\Psi(t;z)}\eqno(3.17)
$$
This is an immediate consequence of the uniqueness
of the Baker-Akhiezer function and the fact
that both sides of the equation have the same analytic 
properties. In particular,
the coefficients of $L_{n}$ and the corresponding
solutions of the KP hierarchy are real.

\medskip
In order to have real and {\it smooth} solutions, it is necessary
to restrict further the geometric data. 
In general, the set of fixed points of any anti-holomorphic involution on a 
smooth Riemann surface is a union of disjoint cycles. 
The number of these cycles 
is less or equal to $g+1$.
The algebraic curves which admit an anti-involution
with exactly $g+1$ fixed cycles are called {\it M-curves}.
We claim that the coefficients of $L_n$
are real and smooth functions
of all variables $t_i$ when $\G$ is an M-curve
with fixed cycles $A_0,A_1,\cdots,A_g$, and
$P\in A_0$, $\g_s\in A_s$, $s=1,\cdots,g$.
To see this, we note that, from
the explicit expression for the Baker-Akhiezer function,
the coefficients of $L_n$ have poles at some value
of $t_i$ if and only if
$$
\theta(A(P_1)+\sum_iU_it_i+Z)=0\eqno(3.18)
$$
The monodromy properties of the $\theta$-function
imply that the zeros of the 
function $\theta(A(z)+\sum_iU_it_i+Z)$
are well-defined on $\G$,
even though the function itself is multi-valued.
The number of these zeroes is $g$. They coincide
with the zeroes of $\Psi(t,z)$. In view of (3.17),
the Baker-Akhiezer function is real on the cycles $A_s$.
On each of the cycles $A_1,\cdots, A_g$, there is one pole
of $\Psi$. There must then be at least
one zero on the same cycle. Hence all zeroes
of $\theta(A(z)+\sum_iU_it_i+Z)$ are located on cycles
$A_s$. Since $P_1\in A_0$, the equation (3.18)
cannot be fulfilled for real values of $t_i$.

We observe that the real and smooth
solutions of the KP hierarchy corresponding
to M-curves with a fixed puncture, are parametrized
by the points of a real $g$-dimensional torus which
is the product of the $g$ cycles $A_s$. If we choose
these cycles (as our notation suggests) as half
of a canonical basis of cycles, then this
torus is the real part of the Jacobian.

In the theory of real and smooth solutions
the equation (2.4) is called the KP-2 equation. The other, so-called,
KP-1 equation is the other real form of the same equation. It can be 
obtained from (2.4) by changing $y$ to $iy$. Thus the KP-1 equation
is given explicitly by
$$
-{3\over 4}u_{yy}=(u_t-{3\over 2}uu_x-{1\over 4}u_{xxx})_x. \eqno(3.19)
$$
As complex equations (2.4) and (3.19) are equivalent. But
the conditions which single out real and smooth solutions are different.
These conditions for the KP-1 equation may be found in [37].
Briefly they are :

Assume that the geometric data defining the Baker-Akhiezer
function is real, in the sense that
\item{(a)} the algebraic curve $\G$ admits an anti-holomorphic
involution $\iota:\G\rightarrow\G$;
\item{(b)} the puncture $P_1$ is a fixed point of $\iota$;
\item{(c)} the local coordinate $k^{-1}$ in a neighborhood
of $P_1$ satisfies the condition $k(\iota(z))=-\overline {k(z)}$;
\item{(d)} the divisor $(\g_1,\cdots,\g_g)$ 
under $\iota$ becomes the dual divisor $\g_1^+,\cdots,\g_g^+$ i.e., 
$\iota(\g_s)=\g_{\sigma(s)}^+$, where $\sigma$ is a permutation.

Then the Baker-Akhiezer function satisfies the reality condition
$$
\Psi^+(t';\iota(z))=\overline{\Psi(t';z)},   \eqno(3.20)
$$
where the new variables $t'=(t'_1, \ldots)$ are equal to
$t_{2m+1}'=t_{2m+1}, \ t_{2m}=it_{2m}$.
As before, this is an immediate consequence of the uniqueness
of the Baker-Akhiezer function and the fact
that both sides of the equation have the same analytic 
properties. In particular,
the coefficients of $L_{n}$ and the corresponding
solutions of the KP-1 hierarchy are real for real values of $t'$.

The further restriction of geometric data corresponding
to real and {\it smooth} solutions of the KP-1 hierarchy is as follows.
The fixed cycles $a_1,\cdots,a_l$ of $\iota$ should divide $\G$ into
two disconnected domaines $\G^{\pm}$. The complex domain $\G^+$ defines
the orientation on the cycles considered as its boundary.
The differential $\d\Omega$ of (3.11) should be positive on $a_s$ with respect
to this orientation.

\bigskip

\noindent{\bf B. Moduli Spaces of Surfaces and Abelian Integrals}

The space $\{\G,P,z,\g_1,\cdots,\g_g\}$ provides
geometric data for solutions of
a complete hierarchy of soliton equations,
and is infinite-dimensional. In the remaining
part of this paper, we concentrate rather on
a single equation of zero curvature form $[\p_y-L,\p_t-A]=0$.
The geometric data associated with the pair $(L,A)$
corresponds to the Jacobian bundle
over a finite-dimensional
moduli space $\M_g(n,m)$ of Riemann surfaces with a pair
of Abelian integrals $(E,Q)$ with poles of order
$n$ and $m$ respectively at the puncture $P$. The associated
operators $(L,A)$ are then operators of order
$n$ and $m$, and are obtained by the basic
construction (3.9), after imbedding $\M_g(n,m)$
in the space $(\G,P,z)$ of geometric data.
Alternatively, we may choose to represent the equation
$[\p_y-L,\p_t-A]=0$ as a dynamical system on a
space of operators $L$, with $t$ as time variable.
In this case, a finite-dimensional and geometric
space of operators $L$ is obtained by the same
construction as just outlined, starting instead
from the Jacobian bundle over the moduli space
$\M_g(n)$ of Riemann surfaces $\G$ with just one
Abelian integral with pole of order $n$ at the puncture.

More precisely, given $(\G, E)$, a geometric data $(\G,P,z)$
is obtained by setting
the local coordinate $z\equiv K^{-1}$ near the puncture $P$ to be
$$
E=z^{-n}+R^E {\rm log}\,z\eqno(3.21)
$$
where $n\geq1$ and $R^E$ are respectively the order of the pole of $E$ and
its residue at $P$. When $n=0$, we set instead
$$
E=R^E{\rm log}\, z.\eqno(3.22)
$$
This gives immediately a map
$$
\eqalignno{(\G,E)&\rightarrow (\G,P,z)\cr
(\G,E,\g_1,\cdots,\g_g)&\rightarrow(\G,P,z,
\g_1,\cdots,\g_g)\rightarrow\ L&(3.23)\cr}
$$
where the operator $L$ is characterized by the condition
$(\p_y-L)\Psi=0$, with $\Psi(x,y;k)$ the Baker-Akhiezer function
having the
essential singularity ${\rm exp}(kx+k^ny)$, $k=z^{-1}$.
In presence of a second Abelian integral $Q$,
we can select a second time $t$, by writing the singular part $Q_+(k)$
of $Q$ as a polynomial in $k$ and setting
$$
\eqalignno{Q_+(k)&=a_1k+\cdots+a_mk^m\cr
t_i&=a_it,\ 1\leq i\leq m.&(3.24)\cr}
$$
This means that we consider the Baker-Akhiezer function $\Psi(x,y,t;k)$
with the essential singularity
${\rm exp}(kx+k^ny+Q_+(k)t)$, and construct the operators $L$ and $A$
by requiring that $(\p_y-L)\Psi=(\p_t-A)\Psi=0$. 
The pair $(L,A)$ provides then a solution
of the zero-curvature equation.
By rescaling $t$,
we can assume that $A$ is monic. Altogether, we have restricted the geometric
map ${\cal G}$ of (3.9) to a map on finite-dimensional spaces,
which we still denote by ${\cal G}$
$$
\eqalignno{
{\cal G}: (\G,E;\g_1,\cdots\g_g)&\rightarrow (\G,P,z)\rightarrow\ (L)\cr
{\cal G}: (\G,E,Q;\g_1,\cdots\g_g)&\rightarrow (\G,P,z,t)\rightarrow\ (L,A)
&(3.25)\cr}
$$
Here we have indicated explicitly the choice of time in the
geometric data. The proper interpretation of the
full geometric data $(\G,E,Q;\g_1,\cdots\g_g)$ is as a point
in the bundle $\N_g^g(n,m)$ over $\M_g(n,m)$, whose fiber
is the $g$-th symmetric power $S^g(\G)$ of the curve.
The $g$-th symmetric power can be identified with the
Jacobian of $\G$ via the Abel map
$$
(\g_1,\cdots,\g_g)\rightarrow\phi_j=\sum_{i=1}^g\int_{P_1}^{\g_i}\d\omega_j
\eqno(3.26)
$$ 
More generally, we can construct the bundles 
$\N_g^k(n,m)$ and $\N_g^k(n)$ with fiber $S^k(\G)$ over the bases
$\M_g(n,m)$ and $\M_g(n)$ respectively
$$
\matrix{S^k(\G)&\longrightarrow&\N_g^k(n,m)\cr
         {}  &{}&\downarrow\cr
         {}  &{}&\M_g(n,m)\cr}
\ \ \ \qquad
\matrix{S^k(\G)&\longrightarrow&\N_g^k(n)\cr
         {}  &{}&\downarrow\cr
         {}  &{}&\M_g(n)\cr}
\eqno(3.27)
$$
Thus the bundles
${\cal N}_g^{k=1}(n,m)\equiv{\cal N}_g(n,m)$,
and ${\cal N}_g^{k=1}(n)\equiv\N_g(n)$ are the analogues
in the our context of the universal curve.
Returning to soliton equations, the geometric map ${\cal G}$
of (3.25) can now be succinctly described as a map 
from the fibrations $\N_g^g(n)$ and
$\N_g^g(n,m)$ into the spaces respectively of operators $L$ and pairs
$(L,A)$ of operators
$$
\eqalignno{
{\cal G}: \N_g^g(n)&\rightarrow (L)\cr
{\cal G}: \N_g^g(n,m)&\rightarrow (L,A)&(3.28)\cr}
$$
We emphasize that, although the operators
in its image are not all periodic operators,
the ones arising later upon restriction of ${\cal G}$
to suitable subvarieties of $\N_g^g(n,m)$ and $\N_g^g(n)$
with integral periods (c.f. Section III.C) will be.

\medskip

We conclude this section
by observing that, in the preceding
construction, $L$ and $A$ depend only the singular part of $Q$,
and hence are unaffected if $\d Q$ is shifted by a holomorphic
differential. As we shall see below, the appropriate
normalization in soliton theory is the {\it real normalization}
by which we require that
$$
{\rm Re}\oint_C\d Q=0\eqno(3.29)
$$
In the study of N=2 supersymmetric gauge theories, holomorphicity
is a prime consideration, and we shall rather adopt
in this context the {\it complex normalization}
$$
\oint_{A_j}dQ=0,\ 1\leq j\leq g\eqno(3.30)
$$ 
We note that each normalization provides an imbedding
of $\M_g(n)$ into $\M_g(n,1)$, by making the choice
$Q_+(K)=K$, with periods satisfying either (3.29)
or (3.30), so that the operator
$A$ is just $A=\p_x$ in either case. The image
of $\M_g(n)$ in $\M_g(n,1)$ does depend on the 
normalization, however.

\bigskip

\noindent{\bf C. Geometric Symplectic Structures}

We begin by discussing the basic local
geometry of the moduli spaces $\M_g(n)$ and
$\M_g(n,m)$. They are complex manifolds
with only orbifold singularities, of dimensions
$$
\eqalignno{
{\rm dim}\,\M_g(n)
&=4g-3+2N+\sum_{\alpha=1}^Nn_{\alpha}\cr
{\rm dim}\,\M_g(n,m)
&=5g-3+3N+\sum_{\alpha=1}^N(n_{\alpha}+m_{\alpha})
&(3.31)\cr}
$$
Indeed, the number of degrees of freedom
of an Abelian integral $E$ with poles of order $n=(n_{\alpha})$
is $1+\sum_{\alpha=1}^N(n_{\alpha}+1)-1+g=N+g+\sum_{\alpha=1}^Nn_{\alpha}$,
where the first 1 corresponds to the additive constant, and the remaining
integer on the left is the dimension of meromorphic
differentials with poles of order $\leq n_{\alpha}+1$ at each $P_{\alpha}$.
For $g>1$, the dimension of the moduli space
of Riemann surfaces with $N$ punctures is $3g-3+N$,
which leads immediately to (3.31). For $g\leq 1$, it is easily
verified that the same formula (3.31) holds, although
the counting has to incorporate holomorphic vector
fields and is slightly different in intermediate stages.

\medskip
We can introduce explicit local coordinates on $\M_g(n,m)$.
To obtain well-defined branches of Abelian integrals, 
we cut apart the Riemann surface $\G$
along a canonical homology basis
$A_i,B_j$, $i,j=1,\cdots,g$, and along cuts from
$P_1$ to $P_{\alpha}$ for each $2\leq\alpha\leq N$.
Locally on $\M_g(n,m)$, this construction can be carried out 
continuously, with paths homotopic by deformations not crossing
any of the poles. Denote the resulting surface by
$\G_{cut}$. On $\G_{cut}$, the Abelian integrals $E$ and
$Q$ become single-valued holomorphic functions,
and we can introduce the one-form $\d\l$ by
$$
\d\l=Q\d E\eqno(3.32)
$$
We observe that $\d\l$ has a singularity of order $n_{\a}+m_{\a}+1$
at each puncture $P_{\a}$. Now the Abelian integral $E$ defines
a coordinate system $z_{\a}$ near each $P_{\a}$ by
$$
E=z_{\alpha}^{-n_{\alpha}}+R_{\alpha}^E{\rm log}\,z_{\alpha}
\eqno(3.33)
$$
when $n_{\a}$ is strictly positive.
When $n_{\alpha}=0$, we write instead
$$
E=R_{\alpha}^E{\rm log}\,z_{\alpha}.\eqno(3.34)
$$ 
The coordinate $z_{\a}$ can be used to fix the additive
normalization
of the Abelian integral $\l$, and to describe its Laurent
expansion near each puncture. Thus we fix the additive
constant in $\l$ by demanding
that its expansion in $z_1$ near $P_1$ have no constant term.
The parameters $T_{\a,i}$, $1\leq i\leq n_{\a}+m_{\a}$,
$R_{\a}^{\l}$, $2\leq\a\leq N$, can then be defined by
$$
\eqalignno{
T_{\a, i}&=-{1\over i}\r_{P_{\a}}(z_{\a}^i\d\l),\ 1\leq\a\leq N,
\ 1\leq i\leq n_{\a}+m_{\a}\cr
R_{\a}^{\l}&=\r_{P_{\a}}(\d\l)&(3.35)\cr}
$$
The parameters $T_{\a,i}$ ($1\leq \a\leq N$) and
$R_{\a}^{\l}$ ($2\leq \a\leq N$) account for 
$\sum_{\a=1}^N(n_{\a}+{m_a})+N-1$ parameters.
The remaining
parameters needed to parametrize $\M_g(n,m)$ consist of the
$2N-2$ residues of $\d E$ and $\d Q$
$$
R_{\alpha}^E={\rm Res}_{P_{\alpha}}\d E,\ 
R_{\alpha}^Q={\rm Res}_{P_{\alpha}}\d Q,\ \alpha=2,\cdots,N
\eqno(3.36)
$$
and the following $5g$ parameters which account for the presence
of non-trivial topology
$$
\eqalignno{
\tau_{A_i,E}&=\oint_{A_i}\d E,\ \tau_{B_i,E}=\oint_{B_i}\d E&(3.37)\cr  
\tau_{A_i,Q}&=\oint_{A_i}\d Q,\ \tau_{B_i,Q}=\oint_{B_i}\d Q&(3.38)\cr  
a_i&=\oint_{A_i}Q\d E,\ i=1,\cdots,g&(3.39)\cr}
$$
\medskip
\noindent
{\bf Theorem 10.} {\it Let ${\cal D}$ be the open set in $\M_g(m,n)$ where
the zero divisors of $\d E$ and $\d Q$, namely the sets
$\{z;\d E(z)=0\}$ and $\{z;\d Q(z)=0\}$, do not intersect.
Then
\item{\rm (a)} Near each point in ${\cal D}$, 
the $5g-3+3N+\sum_{\alpha=1}^N(n_{\alpha}+m_{\alpha})$ parameters
$R_{\alpha}^E$, $R_{\alpha}^Q$, $R_{\a}^{\l}$,
$T_{\alpha,k}$,
$\tau_{A_i,E}$, $\tau_{B_i,E}$, $\tau_{A_i,Q}$,
$\tau_{B_i,Q}$, $a_i$ have linearly independent
differentials, and thus define a local holomorphic
coordinate system for $\M_g(n,m)$; 
\item{\rm (b)} The joint level sets of
the set of all parameters except $a_i$ define a smooth
$g$-dimensional foliation of $\cal D$, independent of the choices
we made to define the coordinates themselves.}
\bigskip
Theorem 10 is proved in [39]. Since $\M_g(n)$
can be imbedded in $\M_g(n,m)$ by choosing 
$Q$ to have Laurent expansion $Q_+(K)=K$ and fixing
its $A_k$ periods, Theorem 1 also provides local coordinates for
$\M_g(n)$. Specifically, the coordinates of $\M_g(n)$ which
arise this way are
\medskip 
\item{$\bullet$} $T_{\a,i}$ with $1\leq i\leq n_{\a}$ for $\a\geq 2$,
$T_{1,i}$ with $1\leq i\leq n_1-1$ (since the normalization
$Q=K^{-1}+O(K)$ fixes the coefficients of the two leading terms
$T_{1,n_1+1}$ and $T_{1,n_1}$ in the singularity expansion of $\l$ near $P_1$
to be $T_{1,n_1+1}={n_1\over n_1+1}$,
$T_{1,n_1}=0$);
\item{$\bullet$}
the $2N-2$ residues of $\d E$ and $\d \l$ at $P_{\a}$ for $\a\geq 2$;
\item{$\bullet$}
the $B_k$ periods of $\d Q$, the $A_k$ and $B_k$ periods of $\d E$,
and the $A_k$ periods of $\d\l$, 

\medskip
\noindent
for a total of $4g-3+2N+\sum_{\a=1}^Nn_{\a}$.

\medskip
The intrinsic foliation obtained in Theorem 10 is central to our
considerations, and we shall refer to it
as the {\it canonical foliation}. 
Our goal is to construct now a symplectic form $\omega$
on the complex $2g$-dimensional space obtained
by restricting the fibration ${\cal N}_g^g(n,m)$
to a $g$-dimensional leaf $\M$ of the canonical
foliation of $\M_g(n,m)$.
For this, we need to extend the differentials
$\d E$ and $\d Q$ to one-forms on the whole fibration,
and distinguish between the two ways this
can be done. 

One way is to consider the 
Abelian integrals $E$ and $Q$ as multi-valued
functions on the fibration. Despite their multivaluedness,
their differentials along any leaf of the canonical fibration
are well-defined. In fact, 
$E$ and $Q$ are well-defined in a small neighborhood of the
puncture $P_1$. The ambiguities in their values
anywhere on each Riemann surface consist only of
integer combinations of their residues or periods
along closed cycles. Thus they are constant
along any leaf of the canonical foliation,
and disappear upon differentiation.
The differentials along the fibrations obtained this way
will be denoted by
$\delta E$ and $\delta Q$. Restricted to vectors tangent to the fiber,
they reduce of course to the differentials
$\d E$ and $\d Q$. 

The other way is to trivialize the fibration by using
an Abelian integral, say $E$, as local coordinate in $\G$ 
(c.f. (3.21) and (3.22)).
Equivalently, we note that
at any point of ${\cal N}$, the varieties $E=constant$
are intrinsic and transversal to the fiber. Thus any one-form $\d Q$
on the fiber can be extended to a one-form 
on the total space of the fibration,
by making it zero on vectors tangent to these
varieties. We still denote this one-form by $\d Q$.
In the case of the Abelian integral $E$, we have
$\delta E=\d E$, but this is not true in general.
To compare $\d Q$ with $\delta Q$, let $a_1,\cdots,a_g$ be local coordinates
for the leaf. Then $a_1,\cdots,a_g, E$ are local coordinates
for the whole fibration, $\d a_1,\cdots,\d a_g,\ d E$
are a basis of one-forms, and $\d Q={\d Q\over\d E}\d E$.
On the other hand,
$$
\delta Q=\d Q+\sum_{i=1}^g{\partial Q\over\partial a_i}da_i
\equiv\d Q+\delta^EQ.
\eqno(3.40)
$$  
\medskip
Similarly, the full differential $\delta(Qd E)$ on the total space
of the fibration is well-defined, despite the multivaluedness
of $Q$. The partial derivatives $\partial_{a_i}(Q\d E)$ 
along the base are all holomorphic, since both the singularities
and the ambiguities in the differential are constant, and
disappear upon differentiation. Recalling that $\d\omega_j$
denotes the
basis of holomorphic differentials dual to
the homology basis $A_k$, $B_k$,
we can then write
$$
{\partial\over\partial a_i}(Q\d E)=\d\omega_i.\eqno(3.41)
$$
By extending this construction to the fibration of Jacobians
over a leaf $\M$ of the canonical foliation, we obtain
the desired geometric symplectic form. Furthermore,
this symplectic form coincides with the symplectic
form constructed in Section II.E, upon
imbedding the fibration of Jacobians
in the space of soliton solutions of the equation
$[\p_y-L,\p_t-A]=0$ by the geometric map ${\cal G}$ of (3.28) [39]:

\medskip

\noindent{\bf Theorem 11.} {\it (a) The following two-form
on the fibration ${\cal N}^g(n,m)$ restricted
to a leaf $\M$ of the canonical foliation of $\M_g(n,m)$
$$
\omega_{\cal M}=\delta\big(\sum_{i=1}^gQ(\g_i)\d E(\g_i)\big)=
\sum_{i=1}^g\delta Q(\g_i)\wedge\d E(\g_i)=
\sum_{i=1}^g\d a_i\wedge\d\omega_i\eqno(3.42)
$$
defines a symplectic form; (b) Under the geometric correspondence
${\cal G}$, we have
$$
\omega_{\M}={\cal G}_*(\omega_m)\eqno(3.43)
$$
where $\omega_m$ is the symplectic form constructed in Theorem 7.}

\medskip

\noindent{\bf D. Generalizations and Extensions}

The above set-up can be easily adapted to a variety of important
equations. 
\medskip
For the 2D Toda lattice, the differential operator
$\p_x$ is replaced by a difference operator. Thus for the geometric data,
we consider the moduli space
of Riemann surfaces $\G$ with two punctures $ P_{\pm}$,
and Abelian differentials $E$, $Q$ which are real-normalized,
with $E$ having a pole of order $1$ at $P_+$, while $\d Q$ has simple poles
at $P_{\pm}$. This is a leaf in the foliation
of $\M_g(1,0;0,0)$. The discrete analogue of the Baker-Akhiezer function
$\Psi(z;t)$ is now given by a sequence $\Psi_n(t_+,t_-;z)$
of Baker-Akhiezer functions
characterized by
$$
\Psi_n(t_+,t_-;z)=z_{\pm}^{\pm n}\big(\sum_{s=0}^{\infty}\xi_s(n,t)z_{\pm}^s)
{\rm exp}(t_+z_+^{-1}+t_-z_-^{-1})\eqno(3.44)
$$
As in Section II.I, we define the operators $L$ and $A$ as difference operators
acting on sequences $\psi_n$ which satisfy 
$(\p_{t_+}-L)\Psi=(\p_{t_-}-A)\Psi=0$. Their coefficients $c_n$, $v_n$,
$\phi_n$
are now functions of the geometric data. In analogy with (3.43), we have
then
$$
\eqalignno{
\omega_{\M}=&{\cal G}^*(\omega_{\M})
=-\sum_{\a=\pm}{\rm Res}_{P_{\pm}}<\Psi_n^*\delta L\wedge\delta\Psi_n>\d p
=\sum_n<\delta\phi_n\wedge\delta v_n>\cr
\d p=&-{\d z\over z},\ z\rightarrow P_{\pm}
&(3.45)\cr}
$$ 
Note that the difference between this formula and the one in Section
II.I is due to the choice $t_+$ instead of $t_++t_-$ as the second
variable.

\medskip

In the case of $N\times N$ matrix equations $[\p_y-L,\p_t-A]=0$,
we take the geometric data to consist of surfaces $\G$ with $N$
punctures $P_{\a}$, and $E$, $Q$ to be real-normalized
Abelian integrals with poles of order $n$ and $m$ at $P_{\a}$.
Let $a_{\a}$ be given coefficients, with $a_1=1$. The leaf $\M$ corresponding
to the above specifications for $E$, $Q$, combines with
the space of parameters $a_{\a}$ to a product space
$\M\times{\bf C}^{N-1}$ of dimension $g+N-1$.
On the fibration ${\cal N}^{g+N-1}$ above the product $\M\times{\bf C}^{N-1}$,
we can construct the symplectic form
$$
\omega_{\M}=\sum_{s=1}^{g+N-1}Q(\g_s)\d E(\g_s)\eqno(3.46)
$$
Now corresponding to the preceding geometric data
are local coordinates $z_{\a}$ near each puncture $P_{\a}$ given in analogy with 
(3.21) and (3.22) by
$E=a_{\a}z_{\a}^{-n}$,
polynomials $Q_{\a,+}(z_{\a}^{-1})$ which are the
singular parts of $Q$ near $P_{\a}$, and thus a vector Baker-Akhiezer
function $\Psi(z;x,y,t)=(\Psi_{\a}(z;x,y,t))_{\a=1}^N$ with the following
essential singularity near $P_{\b}$
$$
\Psi_{\a}(z;x,y,t)
=(\delta_{\a\b}+\sum_{s=1}^{\infty}\xi_s^{\a\b}(x,y,t)z_{\b}^s)
{\rm exp}(z_{\b}^{-1}x+a_{\b}z_{\b}^{-n}y+Q_{\b,+}(z_{\b}^{-1})t)\eqno(3.47)
$$
As in Section II.I, there exist then unique matrix operators
$L,A$ so that $(\p_y-L)\Psi=(\p_t-A)\Psi=0$. They have expressions
of the form in (2.93) and subsequent equations. 
We observe that the case where $Q$ has only
simple poles ($m_i=1$, $Q_{\b,+}=z_{\b}^{-1}$) is the 
matrix generalization of the scalar
case considered earlier, where $Q$ still has the interpretation
of a quasi-momentum.  

On the space of such operators, we had
defined in Theorem 8 a symplectic form $\omega$. Here again, the geometric
and the formal symplectic forms (3.46) and (2.96) correspond to
one another under the basic geometric correspondence ${\cal G}$. More
precisely,
$$
\omega_{\M}
=\sum_{\a=1}^N{\rm Res}_{P_{\a}}<\Psi^*(A_m^{(1)}\delta -L^{(1)}\delta A_m)
\Psi>
\d p
={\cal G}_*(\omega_m)\eqno(3.48)
$$
where
$$
\omega_m=\r_{\infty}{\rm Tr}<\Psi_0^*(A_m^{(1)}\delta L-
L^{(1)}\delta A_m)\wedge\delta\Psi_0>\d k.\eqno(3.49)
$$
In the case $m_i=1$, we have
$A=\p_x$, and this relation reduces to
$$
\omega_{\M}=-\sum_{\a=1}^N\r_{P_{\a}}<\Psi^*\delta L\wedge\delta\Psi>\d p=
{\cal G}_*(\omega)
$$
with $\omega$ given by (2.96).

\medskip

Similarly, our formalism can easily identify the
action coordinates for the elliptic Calogero-Moser system,
an issue which had been resolved only relatively recently [29].
We recall that the elliptic Calogero-Moser system is a system of $N$
identical particles on a line, interacting with each other
via the potential  
$V(x)=\wp(x)$ 
$$
\ddot{x}_i=4\sum_{j\neq i} \wp'(x_i-x_j), \ \
\wp(x)={d\wp(x)\over dx}. \eqno(3.50)
$$
Here $\wp(x)=\wp (x|\omega,\omega')$ is the Weierstrass
elliptic function with periods $2\omega,\ 2\omega'$,
and $\omega,\omega'$ are fixed parameters. The complete solution
of the elliptic Calogero-Moser system was constructed by geometric
methods in [35]. There an explicit Lax pair $(L,M)$ was found,
depending on a spectral parameter $z$ varying on the torus
${\bf C}/(2\omega{\bf Z}+2\omega'{\bf Z})$. Thus the dynamical system
(3.50) is equivalent to the Lax equation $\dot L=[M,L]$,
with $L$ and $M$ $N\times N$ matrices given by
$$
\eqalignno{
L_{ij}(z)&=p_i\delta_{ij}+2(1-\delta_{ij})\Phi(x_i-x_j,z),\ p_i=\dot x_i\cr
M_{ij}(z)&=2\delta_{ij}\sum_{k\not=i}\wp(x_i-x_j)+2(1-\delta_{ij})
\Phi'(x_i-x_j)&(3.51)\cr}
$$
and $\Phi(x,z)$ is the function
$$
\Phi(x,z)={\sigma(z-x)\over \sigma(z)\sigma(x)}e^{\zeta(z)x}.\eqno(3.52)
$$
with $\sigma(z)$, $\zeta(z)$ the usual Weierstrass elliptic
functions. In view of the Lax equation, the characteristic polynomial
$R(k,z)=\det (2k+L(z))$
is time--independent, and defines a time-independent spectral curve 
$\Gamma$ 
$$
R(k,z)\equiv\sum_{i=0}^N r_i(z)k^i=0 \eqno(3.53)
$$
where the $r_i(z)$ are elliptic functions of $z$. The Jacobians of 
the spectral curves $\G$ are levels of the involutive integrals of the system. 
In particular, we obtain {\it angle variables} $\varphi_i$ by choosing
$2\pi$-periodic coordinates on them.
However, as noted earlier, the identification of the canonically
conjugate {\it action}
variables is more difficult and has been carried out only recently [29].
In this work, it was shown
that the Calogero-Moser sytem can be obtained through a Hamiltonian
reduction of a Hitchin system, and as a result, 
the action-variables $a_i$ are the periods of the differential
$k\d z$ along the $A$-cycles of the spectral curve $\Gamma$.

We can derive this result directly from our
approach. From our viewpoint, the leaf $\M$ corresponding
to the elliptic Calogero-Moser system is given by
$(\G,k,z)$, where $\G$ is a Riemann surface
of genus $g=N$, $k$ is a function with simple poles
at $g$ points $P_1,\cdots,P_g$, $\d z$ is a holomorphic
Abelian differential whose periods
as well as integrals $\int_{P_1}^{P_{\a}}\d z$,
$2\leq \a\leq g$, are all on the lattice spanned by $2\omega$ and
$2\omega'$, and the residues of $k\d z$ at $P_{\a}$, $2\leq\a\leq N$
are given by
$$
\r_{P_{\a}}(k\d z)=1,\ 2\leq\a\leq N.\eqno(3.54)
$$
The analogues of the Baker-Akhiezer function
and its dual are respectively the column
vector $C(P)=(c_1,\ldots,c_N)$ 
and the row vector $C^+(P)=(c_1^+,\ldots,c_N^+)$, satisfying
$$
\eqalignno{
(2k+L(z))C=0,\ \ \ \ C^+(2k+L(z))&=0, \cr
\sum_{i}c_i\Phi(-x_i,z)=1,\ \ \ \ \sum_{i=1}c_i^+\Phi(x_i,z)&=1. 
&(3.55)\cr}
$$
Here $P=(k,z)$ is a point of the spectral curve $\Gamma$.
The vectors $C(P)$ and $C^+(P)$ are meromorphic
functions on $\Gamma$ outside the points $P_{\alpha}$ on $\G$ corresponding to
$z=0$, and have each $N$ poles. We denote these poles by $\g_1,\cdots,\g_N$
and $\g_1^+,\ldots,\g_N^+$, respectively.
Near the points $P_{\alpha}$, these vectors have the form
$$
c_i(z)=(c_i^{\alpha}+O(z))e^{x_iz^{-1}}, \
c_i^+(z)=(c_i^{\alpha,+}+O(z))e^{-x_iz^{-1}}\eqno(3.56)
$$
where the coefficients $c_i^{\a}$ satisfy
$$
\eqalignno
{c_i^1&=1;\ \ \ \ \ \sum_i c_i^{\alpha}=0 \ \ {\rm for} \ \alpha>1 \cr
c_i^{1,+}&=1;\ \ \ \ \ \sum_i c_i^{\alpha,+}=0 \ \ {\rm for} \ \alpha>1
&(3.57)\cr}
$$
The geometric symplectic form $\omega_{\M}$ constructed in Theorem 11
becomes in this case
$$
\omega_{\M}=\delta(\sum_{s=1}^N k(\g_s)dz) =
{1\over 2}\sum_{\alpha=1}^N{\rm Res}_{P_{\alpha}}{<\delta C^+\wedge \delta L C>
\over <C^+C>}dz, \eqno(3.57)
$$
where $<f^+ g>$ denotes the usual pairing between column
vectors and row vectors.
Substituting in the expansion (3.56), we obtain
$$
\delta(\sum_{s=1}^N k(\g_s)dz) ={1\over 2} \sum _i^N \delta p_i\wedge \delta x_i.
\eqno(3.59)
$$
This identifies our geometric symplectic form with the
canonical symplectic form for the Calogero-Moser dynamical system.
Since by construction (c.f.(3.42)), the geometric symplectic
form admits the periods $a_i$ of $k\d z$
around $A_i$ cycles as action variables dual to the
angle variables on the Jacobian, our argument is complete.

\bigskip

\centerline{\bf IV. WHITHAM EQUATIONS}

\bigskip

\noindent
{\bf A. Non-linear WKB Methods in Soliton Theory}

We have seen that soliton equations
exhibit a unique wealth of exact solutions. Nevertheless,
it is desirable to enlarge the class of solutions further,
to encompass broader data than just rapidly decreasing or quasi-periodic
functions. Typical situations arising in practice
can involve Heaviside-like boundary conditions in the space variable
$x$, or slowly modulated waves which are not exact solutions,
but can appear as such over a small scale in both
space and time.

\medskip

The non-linear WKB method (or, as it is now also called, the Whitham
method of averaging) is a generalization to the case of 
partial differential equations 
of the classical Bogolyubov-Krylov method of averaging. This method is 
applicable to nonlinear equations which have a moduli space 
of exact solutions of the
form $u_0(Ux+Wt+Z|I)$. 
Here $u_0(z_1,\ldots, z_g|I)$ is a periodic function of the 
variables $z_i$; $U=(U_1,\ldots,U_g), \ W=(W_1,\ldots,W_g)$ are vectors which
like $u$ itself, depend on the parameters $I=(I_1,\ldots,I_N)$, i.e.
$U=U(I), \ V=V(I)$. (A helpful example is provided by the
solutions (3.10) of the KP equation, where $I$ is the moduli of
a Riemann surface, and $U,V,W$ are the $B_k$-periods of its
normalized differentials $\d\Omega_1$, $\d\Omega_2$, and $\d\Omega_3$.)
These exact
solutions can be used as a leading term for the construction of 
asymptotic solutions
$$
u(x,t)=u_0(\varepsilon^{-1} S(X,T)+Z(X,T)|I(X,T))+\e u_1(x,t)+\e^2 u_2(x,t)+
\cdots, \eqno(4.1)
$$
where $I$ depend on the {\it slow} variables $X=\e x, T=\e t$ and 
and $\e$ is a small parameter. If the vector-valued function $S(X,T)$ is
defined by the equations
$$
\p_XS=U(I(X,T))=U(X,T),\ \ \p_TS=W(I(X,T))=W(X,T), \eqno(4.2)
$$
then the leading term of (4.1) satisfies the original equation
up to first order one in $\e$. All the other terms of the asymptotic series
(4.1) are obtained from the non-homogeneous linear equations
with a homogeneous part which is just the linearization of the original
non-linear equation on the background of the exact solution $u_0$.
In general, the asymptotic series becomes unreliable on scales of
the original variables $x$ and $t$ of order $\e^{-1}$. In order to have a
reliable approximation, one needs to require a 
special dependence of the parameters
$I(X,T)$. Geometrically, we note that
$\e^{-1}S(X,T)$ agrees to first order with $Ux+Vy+Wt$, and $x,y,t$
are the fast variables. Thus $u(x,t)$ describes a motion which
is to first order the original {\it fast} periodic motion on the Jacobian,
combined with a slow drift on the moduli space of exact
solutions. The equations which describe this drift
are in general called {\it Whitham equations}, although there is no systematic
scheme to obtain them. 

\medskip
One approach for obtaining these equations
in the case when the original equation is Hamiltonian is to consider
the Whitham equations as also Hamiltonian, with the Hamiltonian function being
defined by the average of the original one. In the case when the 
phase dimension $g$ is bigger than one, this approach does not provide a complete
set of equations. If the original equation has a number of integrals
one may try to get the complete set of equations by averaging all of them.
This approach was used in [62] where Whitham equations were
{\it postulated} for the finite-gap solutions of the KdV equation.
The geometric meaning of these equations was clarified in [26]. 
The Hamiltonian approach for the Whitham equations of (1+1)-dimensional 
systems was developed in [23] where the corresponding
bibliography can also be found.

\medskip
In [36] a general approach for the construction of Whitham 
equations for finite-gap solutions of soliton equations
was proposed. It is instructive enough to present it in case of the 
zero curvature equation (2.1) with scalar operators.

Recall from Sections III.A and III.B that the coefficients $u_i(x,y,t),\ v_j(x,y,t)$
of the finite-gap operators $L_0$ and $A_0$ satisfying (2.1) are of
the form (c.f. (3.10))
$$
u_i=u_{i,0}(Ux+Vt+Wt+Z|I), \ \
v_j=v_{j,0}(Ux+Vt+Wt+Z|I), \eqno(4.3)
$$
where $u_{i,0}$ and $v_{j,0}$ are differential polynomials in $\theta$-functions
and $I$ is any coordinate system on the moduli space ${\cal M}_g(n,m)$.

We would like to construct operator solutions of (2.1) of the
form
$$
L=L_0+\e L_1+\cdots, \ A=A_0+\e A_1+\cdots, \eqno(4.4)
$$
where the coefficients of the leading terms have the form
$$
\eqalignno{u_i&=u_{i,0}(\e^{-1}S(X,Y,T)+Z(X,Y,T)|I(X,Y,T)), \cr
v_j&=v_{j,0}(\e^{-1}S(X,Y,T)+Z(X,Y,T)|I(X,Y,T))&(4.5)\cr}
$$ 
From Section III.B, we also recall that $\N_g^1(n,m)$ 
is the bundle over ${\cal M}_g(n,m)$ with 
the corresponding curve $\G$ as fiber.
If $I$ is a system of coordinates on ${\cal M}_g(n,m)$, then we may introduce
a system of coordinates $(z, \ I)$ on ${\cal N}_g^1(n,m)$ by choosing
a coordinate along the fiber $\G$. The Abelian integrals
$p,E,Q$ are multi-valued functions of $(\lambda, \ I)$, i.e.
$p=p(\lambda, I)$, $E=E(\lambda, I)$, $Q=Q(\lambda, I)$. If 
we describe a drift on the moduli space of exact solutions
by a map $(X,Y,T)\rightarrow I=I(X,Y,T)$, then
the Abelian integrals $p,E,Q$ become functions of $z,X,Y,T$, via
$$
\eqalign
{p(z,X,Y,T)&=p(z,I(X,Y,T)),\cr 
E(z,X,Y,T)&=E(z,I(X,Y,T)),\cr 
Q(z,X,Y,T)&=Q(z,I(X,Y,T)).\cr}
$$ 
The following was established in [36]:

\medskip 
\noindent
{\bf Theorem 12.} {\it A necessary condition for
the existence of the asymptotic solution (4.4)
with leading term (4.3) and bounded terms $L_1$ and $A_1$ is that
the equation
$$
{\p p\over\p z}\left({\p E\over \p T}-{\p Q\over \p Y}\right)-
{\p E\over\p z}\left({\p p\over \p T}-{\p Q\over \p X}\right)+
{\p Q\over\p z}\left({\p p\over \p Y}-{\p E\over \p X}\right)=0 
\eqno(4.6)
$$
be satisfied.}

\medskip
The equation (4.6) is called the Whitham equation for (2.1). It can
be viewed as a generalized dynamical system on $\M_g(n,m)$, i.e., a map 
$(X,Y,T)\rightarrow \M_g(n,m)$. Some of its important features are:
\medskip
$\bullet$ Even though the original two-dimensional system may depend
on $y$, Whitham solutions which are $Y$-independent are still
useful. As we shall see later, this particular case
has deep connections with topological field theories.
If we choose the local coordinate $z$ along
the fiber as $z=E$, then the equation simplifies in this case to
$$
\p_Tp=\p_XQ\eqno(4.7)
$$
We note that it followed immediately from the consistency of (4.2) that
we must have
$$\p_T\oint_{B_k}\d p=\p_X\oint_{B_k}\d Q.$$
Thus (4.7) is a strengthening of this requirement
which encodes also the solvability term by term
of the linearized equations for all the successive terms
of the asymptotic series (4.3).
\medskip
$\bullet$ Naively, the Whitham equation seems to impose an infinite
set of conditions, since it is required to hold at every point
of the fiber $\G$. However, the functions involved are all
Abelian integrals, and their equality over the whole
of $\G$ can actually be reduced to a finite set of conditions.
To illustrate this point, we consider the $Y$-independent Whitham equation
on the moduli space of curves of the form
$$y^2=\prod_{i=1}^3(E-E_i)\equiv R(E)
$$ 
Then the differentials $\d p$ and $\d Q$ are given by
$$
\eqalign
{\d p&={\d E\over \sqrt{R(E)}}(E-{\int_{E_2}^{E_3}{E\d E\over \sqrt R}
\over\int_{E_2}^{E_3}{\d E\over \sqrt R}})\cr
\d Q&={\d E\over \sqrt{R(E)}}\big(E^2-{1\over 2}(E_1+E_2+E_3)E-
{\int_{E_2}^{E_3}{(E-{1\over 2}\sum_{i=1}^3E_i)E\d E\over \sqrt R}
\over\int_{E_2}^{E_3}{\d E\over \sqrt R}}\big)
\cr}
$$
We view $p$ and $Q$ as functions of $(E;E_i)$, with $E$ the coordinate
in the fiber $\G$, and $E_i$ the coordinates on the moduli space
of curves. Near each branch point $E_i$, $\sqrt{E-E_i}$ is a local coordinate
and we may expand
$$
\eqalignno{
p&=p(E_i)+\alpha\sqrt{E-E_i}+O(E-E_i)\cr
Q&=Q(E_i)+\beta\sqrt{E-E_i}+O(E-E_i)&(4.8)\cr}
$$
Differentiating with respect to $X$ and $T$, keeping $E$ fixed, we find
that the leading singularities of $\p_Tp$ and $\p_XQ$ are
respectively $-{\alpha\over 2\sqrt{E-E_i}}\p_TE_i$
and $-{\beta\over 2\sqrt{E-E_i}}\p_XE_i$. 
Since ${\beta\over\alpha}={\d Q\over\d p}$, we see that the equation (4.7)
implies
$$
\p_TE_i=\big({\d Q\over \d p})_{|_{E_i}}\p_XE_i\eqno(4.9)
$$
Conversely, if the equation (4.9) is satisfied, then
$\p_Tp-\p_XQ$ is regular and normalized, and hence must vanish.
Thus the equation (4.7) is actually equivalent to the set of differential
equations (4.9).

\medskip 
$\bullet$ The equation (4.7) can be represented in a
manifestly invariant form, without explicit
reference to any local coordinate system $z$.
Given a map $(X,Y,T)\rightarrow\M_g(n,m)$,
the pull back of the bundle ${\cal N}_g^1(n,m)$ defines
a bundle over a space with coordinates $X,Y,T$. The total space ${\cal N}^4$ 
of the last bundle is $4$-dimensional.
Let us introduce on it the one-form
$$
\alpha =p\d X+E\d Y+Q\d T, \eqno(4.10)
$$
Then (4.7) is equivalent to the condition that the wedge product of
$d\alpha$ with itself be zero (as a $4$-form on ${\cal M}^4$)
$$
\d\alpha\wedge \d\alpha=0. \eqno(4.11)
$$

$\bullet$
It is instructive to present the Whitham equation (4.7) in yet another form.
Because (4.7) is invariant with respect to a change
of local coordinate we may use $p=p(z, I)$ by itself as a local
coordinate. With this choice we may view $E$ and $Q$ as functions
of $p,X,Y$ and $T$, i.e. $E=E(p,X,Y,T)$, $Q=Q(p,X,Y,T)$. With this choice of
local coordinate (4.7) takes the form
$$
\p_T E-\p_Y Q+\{E,Q\}=0, \eqno(4.12)
$$
where $\{\cdot,\cdot\}$ stands for the usual Poisson bracket of two 
functions of the variables $p$ and $X$, i.e. 
$$\{f,g\}=f_pg_X-g_pf_X.$$

\medskip

$\bullet$ In Theorem 12 we had focused on constructing an asymptotic solution 
for a single equation. This corresponds to a choice
of $A$, and thus of an Abelian differential $Q$, and
the Whitham equation is an equation for maps from
$(X,Y,T)$ to $\M_g(n,m)$. As in the case of the KP and other hierarchies,
we can also consider a whole hierarchy
of Whitham equations. This means that the Abelian integral
$Q$ is replaced by the really normalized Abelian integral 
$\Omega_i$ which has the following form
$$
\Omega_i =K^i+O(K^{-1}),\ \ K^n=E, \eqno(4.13)
$$
in a neighborhood of the puncture $P$ (compare with (2.48)).
The result is a hierarchy
of equations on maps of the form 
$$(X=T_1, Y=T_2, T=T_3,T_4,\cdots)\rightarrow
\M_g(n).\eqno(4.13)
$$
The whole hierarchy may be written in the form (4.11) where we set now
$$
\alpha=\sum_i\Omega_i\d T_i. \eqno(4.14)
$$

\medskip
\noindent
{\bf B. Exact Solutions of Whitham Equations}

In [38] a construction of exact solutions to the Whitham equations
(4.7) was proposed. In the following section,
we shall present the most important special
case of this construction, which is also
of interest to topological field
theories and supersymmetric gauge theories. It should be
emphasized that for these applications, the definition of the hierarchy
should be slightly changed. Namely, the Whitham equations
describing modulated waves in soliton theory are 
equations for 
Abelian differentials with a real normalization
(3.29). In what follows we shall consider 
the same equations, but where the real-normalized differentials are
replaced by differentials with the complex normalization (3.30).
As discussed in Section III.A, the two types of normalization coincide on the 
subspace corresponding to $M$-curves, which is essentially the space where all
solutions are regular and where the averaging procedure is easily
implemented. Thus the two forms of the Whitham hierarchy 
can be considered as different extensions
of the same hierarchy. The second one is an analytic theory, and we
shall henceforth concentrate on it.

\medskip

In this subsection and in the rest of the paper, we shall restrict
ourselves to the hierarchy
of ``algebraic geometric
solutions" of Whitham equations, that is, solutions
of the following stronger version of the equations (4.11)
$$
{\p\over \p T_i} E=\{\Omega_i, E\}, \eqno(4.15)
$$
We note that the original Whitham equations
can actually be interpreted as consistency conditions
for the existence of an $E$ satisfying (4.15). Furthermore,
the solutions of (4.15) can be viewed in a sense as ``$Y$-independent"
solutions of Whitham equations, since the equation (4.12)
reduces to $\p_TE+\{E,Q\}=0$ for $Y$-independent solutions.
They play the same role as Lax equations
in the theory of (2+1)-dimensional soliton equations. As stressed earlier,
$Y$-independent solutions of the Whitham hierarchy can be
considered even for two-dimensional systems where the $y$-dependence  is
non-trivial in general.

\medskip
Our first step is to show that (4.15) defines a system of commuting
flows on $\M_g(n)$. For the sake of simplicity,
we assume that there is only one puncture.
Let us start with a more detailed description
of this space which is a complex manifold with only orbifold
singularities. Its complex dimension is equal to (c.f. (3.31))
${\rm dim}\,\M_g(n)=4g+n-1$, and we had constructed a set of
local coordinates for it in Theorem 10 and subsequent discussion.
Here we require the following slightly different set of coordinates
(details can be found in [38]).  
The first $2g$ coordinates are still the periods of $\d E$, 
$$ 
\tau_{A_i,E}=\oint_{A_i}\d E,\ \tau_{B_i,E}=\oint_{B_i}\d E \eqno(4.16) 
$$ 
The differential $\d E$ has $2g+n-1$ zeros (counting multiplicities). When 
all zeroes
are simple, we can supplement
(4.16) by the $2g+n-1$ critical values $E_s$ of the Abelian differential 
$E$, i.e.
$$
E_s=E(q_s),\ \ dE(q_s)=0,\ \ s=1,\ldots, 2g+n-1 ,\eqno(4.17)
$$
In general, $\d E$ may have multiple zeroes, and we let
$D=\sum \mu_s q_s$ be the zero divisor of $\d E$. The degree
of this divisor is equal to $\sum_s \mu_s=2g-1+n$. 
Consider a small neighborhood of $q_s$, viewed as a point of 
the fibration ${\cal N}_g^1(n)$, above the original data point $m_0$
in the moduli space $\M_g(n)$. 
Viewed as a function on the fibration, $E$ is a deformation
of its value $E(z,m_0)$ above the original data point, with multiple critical 
points $q_s$.  Therefore, on each of the corresponding curve, there exists 
a local coordinate $w_s$ such that
$$ 
E=w_s^{\mu_s+1}(z,m)+\sum_{i=0}^{\mu_s-1} E_{s,i}(m)w_s^i(z,m). \eqno(4.18)
$$
The coefficients $E_{s,i}(m)$ of the polynomial (4.18) are 
well-defined functions on $\M_g(n)$. Together with
$\tau_{A_i,E}, \ \tau_{B_i,E}$, they 
define a system of local coordinates on $\M_g(n)$.If $\mu_s=1$, 
then $E_{s,0}$ clearly coincides 
with the critical value $E(q_s)$ from (4.17). 

\medskip
Let ${\cal D'}$ be the open set in $\M_g(n)$ where
the zero divisors of $\d E$ and $\d p$, namely the sets
$\{z;\d E(z)=0\}$ and $\{z;\d p(z)=0\}$, do not intersect and 
let ${\cal D}^0$ be the open set in $\M_g(n)$ where all zeros of $\d E$ are 
simple.

\medskip  
\noindent{\bf Theorem 13.}
{\it The Whitham equations (4.15) define a system of commuting meromorphic
vector fields (flows) on $\M_g(n)$ which are holomorphic on
${\cal D'}\subset {\cal M}_g(n)$. On the open set
${\cal D'}\cap {\cal D^0}$, the equations (4.15) have the form}
$$
{\p\over \p T_j}\tau_{A_i,E}=0,\ \ {\p\over \p T_j}\tau_{B_i,E}=0, \eqno(4.18)
$$

$$
\p_{T_j}E_s={d\Omega_j\over dp}(q_s) \p_X E_s. \eqno(4.19)
$$

\medskip
Indeed, the equations (4.15) are fulfilled at each point of $\G$, and
thus 
$$ 
{\p\over \p T_i} \oint_C \d E={\d\Omega_i\over \d p}(z)\p_X \left(\oint_C 
\d E\right)- {\d E\over \d p}(z) \p_X \left(\oint_C \d\Omega_i\right) 
\eqno(4.20)
$$ 
The functions $\d E/\d p$ and $\d\Omega_i/\d p$ are linearly independent. 
It follows that the periods of $\d E$ are constants.
The equations (4.19) coincide with (4.15) at the point $q_s$ (where 
$dE$ equals zero).
\medskip
In order to complete the proof of Theorem 13, it suffices to show that
(4.18-4.19) imply (4.15). The equation (4.19) implies
that the difference between the left and right hand sides of (4.15)
is a meromorphic function $f(z)$ on $\G$. This function is holomorphic outside
the puncture $P$ and the zeros of $\d p$. At the puncture $P$, 
the function $f(z)$
has a pole of order less or equal to $(n-2)$. However,
$f(z)$ equals zero at zeros of $\d E$. Hence, 
the function $g(z)=f(z){\d p\over \d E}$ is holomorphic on $\G$ and equals zero
at $P$. Therefore, $f(z)=0$ identically. Theorem 13 is proved.
\medskip
An important consequence of Theorem 13 is that the space
$\M_g(n)$ admits a natural foliation, namely by
the joint level sets of the functions $\tau_{A_i,E},\ \tau_{B_i,E}$,
which are smooth $(2g+n-1)$-dimensional submanifolds, and which are invariant 
under the flows of the Whitham hierarchy (4.15). We shall sometimes
refer to the leaves $\hat\M$ of this foliation as {\it large leaves},
to stress their distinction from the $g$-dimensional
leaves $\M$ of the canonical foliation of $\M_g(n,m)$.

\medskip
 
The special case of the construction of exact 
solutions to (4.15) in [38] may now be described as follows: 
the moduli space ${\cal M}_g(n,m)$ provides 
the solutions of the first $n+m$-flows of 
(4.15) parametrized by $3g$ constants, which are the set of the last 
three coordinates (3.38-3.39) on ${\cal M}_g(n,m)$.

\medskip
\noindent{\bf Theorem 14.}
{\it Let $T_i, \ i=1,\ldots, n+m,\ \tau_{A_i,E},\  
\tau_{B_i,E}, \tau_{A_i,Q}, \ \tau_{A_i,Q}, \ a_i$ be the coordinates
on $\M_g(n,m)$ defined in Theorem 10. Then the projection
$$
\eqalignno{
\M_g(n,m)&\to \M_g(n)\cr
(\G,E,Q)&\to (\G,E)&(4.21)\cr}
$$ 
defines $(\G,E)$ as a function of the
coordinates on $\M_g(n,m)$. For each fixed set
of parameters $\tau_{A_i,E},\  
\tau_{B_i,E}, \tau_{A_i,Q}, \ \tau_{A_i,Q}, \ a_i$,
the map $(T_i)_{i=1}^{i= n+m}\rightarrow \M_g(n)$
satisfies the Whitham equations (4.15).}

\medskip

\noindent{\it Proof.} Let us use $E(z)$ as a local coordinate on $\G$.
Then as we saw earlier, the equations (4.15) are equivalent to the equations
$\p_{T_i} p(E,T)=\p_X \Omega_i(E,T)$.
These are the compatibility conditions
for the existence of a generating function for
all the Abelian differentials $\d\Omega_{i}$. In fact,
if we set
$$\d \l=Q\d E\eqno(4.22)$$
then it follows from the definition of the coordinates
that
$$
\p_{T_i} \d\l=\d\Omega_i,\ \p_X\d \l=\d p, \eqno(4.23)
$$
(For the proof of (4.23), it is enough to check that the right and the left
hand sides of it have the same analytical properties.)

\medskip
\noindent{\bf Theorem 15.} {\it We consider the same parametrization
of $\M_g(n,m)$ as in Theorem 14. Then as a function
of the parameters $T_i$, $1\leq i\leq n+m$, the second Abelian integral
$Q(p,T)$ satisfies the
same equations as $E$, i.e.
$$
\p_{T_i}Q=\{\Omega_i,Q\}. \eqno(4.24)
$$
Furthermore}
$$
\{E,Q\}=1. \eqno(4.25)
$$

We note that (4.25) can be viewed as a Whitham version of the 
so-called string equation (or Virasoro constraints) in 
a non-perturbative theory of 2-d gravity [19][66].

\bigskip
\noindent{\bf C. The $\tau$-Function of the Whitham Hierarchy}

The solution of the Whitham hierarchy can be succinctly summarized
in a single $\tau$-function defined as follows. The key underlying
idea is that suitable submanifolds of $\M_g(n,m)$ can
be parametrized by $2g+N-1+\sum_{\a=1}^N(n_{\a}+m_{\a})$ Whitham times $T_A$, to each of which is
associated a ``dual" time $T_{DA}$, and an Abelian differential
$\d \Omega_A$. We begin by discussing the simpler
case of one puncture, $N=1$.
Recall that the coefficients of the pole of $\d\lambda$ 
has provided $n+m$ Whitham times $T_j=-{1\over j}\r(z^{j}\d\lambda)$.
Their ``dual variables" are 
$$
T_{Dj}=\r(z^{-j}\d\lambda),\eqno(4.26)
$$
and the associated Abelian differential are the
familiar $\d\Omega_i$ of (3.4) (complex normalized).
When $g>0$, the moduli space $\M_g(n,m)$ has in addition
$5g$ more parameters. We consider only the foliations
for which the following $3g$ parameters are fixed
$$
\oint_{A_k}\d E,\ \oint_{B_k}\d E,\ \oint_{A_k}\d Q.\eqno(4.27)
$$
Thus the case $g>0$ leads to two more sets of $g$ Whitham times each
$$
a_k=\oint_{A_k}\d \lambda, \ T_k^E=\oint_{B_k}\d Q\eqno(4.28)
$$
Their dual variables are
$$
a_{Dk}=-{1\over 2\pi i}\oint_{B_k}\d\lambda,\ 
T_{Dk}^E={1\over 2\pi i}\oint_{A_k^-}E\d\lambda\eqno(4.29)
$$ 
The corresponding Abelian differentials are respectively
the holomorphic differentials $\d\omega_k$ and the differentials
$\d\Omega_k^E$, defined to be holomorphic
everywhere on $\G$ except along the $A_j$ cycles, where
they have discontinuities
$$
\d\Omega_{k}^{E+}-\d\Omega_{k}^{E-}=\delta_{jk}\d E\eqno(4.30)
$$
We denote the collection of all
$2g+n+m$ times by $T_A=(T_j,a_k,T_{k}^E)$.
In the case of $N>1$ punctures, we have $2g+\sum_{\a}(n_{\a}+m_{\a})$ times
$(T_{\a,j},a_k,T_{k}^E)$ and $3N-3$ additional
parameters for $\M_g(n,m)$, namely the residues
of $\d Q$, $\d E$, and $\d\l$ at the punctures $P_{\a}$,
$2\leq\a\leq N$ (c.f. (3.35-3.36)). For simplicity,
we only consider the leaves of $\M_g(n,m)$
where
$$
\r_{P_{\a}}(\d Q)=0, \ \r_{P_{\a}}(\d E)=fixed,\ \ 2\leq\a\leq N,\eqno(4.31)
$$
and incorporate among the $T_A$ 
the residues of $\d\l$ at $P_{\a}$, $2\leq\a\leq N$,
$R_{\a}^{\l}=\r_{P_{\a}}(\d\l)$
(c.f.(3.35)).
The dual parameters to these $N-1$ additional Whitham times
are then the regularized integrals
$$
R_{D\a}^{\l}=-\int_{P_1}^{P_{\a}}\d\l,\ 2\leq\a\leq N.\eqno(4.32)
$$
More precisely, recall that the Abelian integral
$\l$ has been fixed by the condition that its expansion
near $P_1$, in terms of the local coordinate
$z_{\a}$ defined by $E$, have no constant term. Near each $P_{\a}$,
$2\leq\a\leq N$, in view of (4.31), it admits
an expansion of the form
$$
\l(z_{\a})=
\sum_{j=1}^{n_{\a}+m_{\a}}{T_{\a,j}\over z_{\a}^j}+
\r_{P_{\a}}(\d\l)\,{\rm log}\,z_{\a}+\l_{\a}+O(z_{\a})\eqno(4.33)
$$
For each $\a$, $2\leq \a\leq N$,
we define the right hand side of (4.32) to be $\l_{\a}$.
The Abelian differential $\d\Omega_{\a}^{(3)}$ associated
with $R_{\a}^{\l}$ is the Abelian differential of the third
kind, with simple poles at $P_1$ and $P_{\a}$, and residue
1 at $P_{\a}$. In summary, we have the following table
$$
\matrix{&{\bf Times} & &{\bf Dual\ Times}& &{\bf Differentials}\cr
        &{} & & {}  & & {} \cr
&-{1\over j}\r_{P_{\a}}(z^j\d\l) & & \r_{P_{\a}}(z^{-j}\d\l)& &
\d\Omega_{\a,j}^0\cr
&\oint_{A_k}\d\l & &-{1\over 2\pi i}\oint_{B_k}\d\l& &\d\omega_k\cr
&\r_{P_a}\d\l & & -\int_{P_1}^{P_{\a}}\d\l & &\d\Omega_{\a}^{(3)}\cr
&\oint_{B_k}\d Q & &{1\over 2\pi i}\oint_{A_k} E\d\l& &\d\Omega_{k}^E
\cr}
\eqno(4.34)
$$

We can now
define the $\tau$-function of the Whitham hierarchy by
$$
\eqalignno{
\tau(T)&=e^{\F(T)}\cr
\F(T)&={1\over 2}\sum_A T_AT_{DA}+
{1\over 4\pi i}\sum_{k=1}^g a_kT_{k}^EE(A_k\cap B_k)&(4.35)\cr}
$$
where $A_k\cap B_k$ is the point of intersection
of the $A_k$ and $B_k$ cycles. 
In  the case $\r_{P_{a}}\d E=0$ we have then (see [38])

\medskip 
\noindent {\bf Theorem 16.} {\it The derivatives of $\F$ with
respect to the $2g+\sum_{\a}(n_{\a}+m_{\a})+N-1$ Whitham times $T_A$ are 
given by} 
$$ 
\eqalignno{\p_{T_A}\F&=T_{DA}+{1\over 2\pi i}\sum_{k=1}^g
\delta_{a_k,A}T^E_{k}E(A_k\cap B_k)\cr
\p_{T_{\a,i},T_{\b,j}}^2\F&=\r_{P_{\a}}(z_{\a}^i \d\Omega_{\b,j})\cr
\p_{a_j,A}^2\F&={1\over 2\pi i}\left(E(A_k\cap B_k)\delta_{(E,k),A}-\oint_{B_k}
\d\Omega_A\right) \cr
\p_{(E,k)A}^2 \F&={1\over 2\pi i}\oint_{A_k} E\d \Omega_A \cr
\p_{ABC}^3\F&=\sum_{q_s}\r_{q_s}\left({\d\Omega_A\d\Omega_B\d\Omega_C\over
\d E \d Q} \right)&(4.36) \cr}
$$
\medskip
These formulae require some modifications when $\r_{P_{\a}}\d E\neq 0$,
which is a case of particular interest in supersymmetric QCD 
(see Section VI below).
In particular, the first derivatives with respect to $R_{\a}^{\l}$ become [17] 
$$
\p_{R_{\a}^{\l}}\F=
T_{D\a}^{\l}+
{1\over 2}\pi i\sum_{\b}c_{\a,\b}R_{\b}^{\l}, 
$$
where $c_{\alpha,\beta}$ is an integer which is antisymmetric in 
$\a$ and $\b$. It is then easy to see that the
second derivatives are modified accordingly by constant shifts, while the
third derivatives remain unchanged.
 
\medskip
We would like to point out that the $2g+\sum_{\a}(n_{\a}+m_{\a})+N-1$
submanifolds of $\M_g(n,m)$ defined by fixing (4.27)
as well as the residues of $\d E$ and $\d Q$ are yet another
version of the $2g+n-1$ large leaf $\hat\M$ of the foliation of $\M_g(n)$
encountered earlier in the case of one puncture. Indeed, imbedding
$\M_g(n)$ in $\M_g(n,1)$ by choosing $Q_+=k$ would fix
two Whitham times $T_{n}$ and $T_{n+1}$, as we saw
after Theorem 10. This reduces the dimension $2g+n+1$ to
the desired dimension $2g+n-1$.

\medskip 
We observe that the first derivatives of $\F$ give the coefficients
of the Laurent expansions and the periods of the form $\l$. 
The second derivatives
give the coefficients of the expansions and the periods of the differentials
$\d\Omega_A$. In that sense the function $\F$ encodes all the information
on the Whitham hierarchy. The formulae for the first and the second derivatives
with respect to the variables $T_i$ are the analogues in the case of 
the averaged equations of the corresponding formulae for the $\tau$-function of
the KP hierarchy. The formulae for the third derivatives are 
specific to the Whitham theory. As we shall see later,
they are reminiscent rather
of marginal deformations of topological or conformal field theories
and of special geometry.

\medskip
Finally, it may be worth noting that the expression (4.35) for
$\F$ can be elegantly
summarized as
$$
\F=\int_{\G}\bar{\d} \lambda\wedge\d \lambda.\eqno(4.37)
$$

\bigskip

\centerline{\bf V. TOPOLOGICAL LANDAU-GINZBURG MODELS}
\centerline{\bf ON A RIEMANN SURFACE} 

\bigskip

In this section, we shall show that each $2g+n-1$ leaf of the foliation 
of $\M_g(n)$ (or, equivalently, of the foliation
of $\M_g(n,1)$ upon imbedding $\M_g(n)$
into $\M_g(n,1)$) actually parametrizes the marginal deformations 
of a topological
field theory on a surface of genus $g$. Furthermore, the free energy of
these theories coincides with the restriction to the leaf of
the exponential of the $\tau$-function for the Whitham hierarchy.
We begin with a brief discussion of some key features of topological
field theories [63-65][11-13].

\medskip
\noindent
{\bf A. Topological Field Theories}

In general, a two-dimensional quantum field theory is specified
by the correlation functions $<\phi(z_1)\cdots\phi(z_N)>_g$ 
of its local physical observables $\phi_i(z)$ on any surface
$\G$ of genus $g$.
Here $\phi_i(z)$ are operator-valued tensors on $\G$. The
operators act on a Hilbert space of states with a designated
vacuum state $|\Omega>$. The correlators
$<\phi(z_1)\cdots\phi(z_N)>$ usually depend on the background metric on $\G$
and on the location of the insertion points $z_i$. In particular, they
may develop singularities as $z_i$ approaches $z_j$. Equivalently,
the operator product $\phi_i(z_i)\phi_j(z_j)$ may develop
singularities. For example,
in a conformal field theory, $\phi_i(z_i)\phi_j(z_j)$ will
have an operator product expansion of the form
$$
\phi_i(z_i)\phi_j(z_j)=\sum_kc_{ij}^k(z_i-z_j)^{h_i+h_j-h_k}\phi_k(z_j)+
descendants\eqno(5.1)
$$
where $h_i$ is the conformal dimension of $\phi_i$.
If we let $\phi_0(z)$ be the field corresponding to $|\Omega>$,
under the usual $states\leftrightarrow fields$ correspondence,
then we obtain a metric by setting
$$
\eta_{ij}=c_{ij}^0\eqno(5.2)
$$
Using $\eta_{ij}$ to raise and lower indices, and noting that
$<\phi_0(z)>_0=1$, we can easily recognize $c_{ijk}=c_{ij}^k\eta_{kl}$
as the three-point function on the sphere
$$
c_{ijk}=<\phi_i(z_i)\phi_j(z_j)\phi_k(z_k)>_0,\eqno(5.3)
$$
which is actually independent of the insertion points $z_i$, $z_j$,
$z_k$ by SL(2,{\bf C}) invariance.

\medskip

Topological field theories are theories where the correlation
functions are actually independent of the
insertion points $z_i$. Thus they depend only on
the labels of the fields $\phi_i$ and the genus $g$ of $\G$. This
independence implies that the OPE of (5.1) contains no
singularity, and that for all practical purposes, the operator
product $\phi_i(z_i)\phi_j(z_j)$ can be replaced
by the formal operator algebra
$$
\phi_i\phi_j=\sum_kc_{ij}^k\phi_k\eqno(5.4)
$$
The associativity of operator compositions translates into
the associativity of the operator algebra (5.4). Furthermore,
the operator algebra is commutative, since
factorization of the 4-point function
in the $s$ and $t$ channels must give the same
answer. If we assume that in a topological
field theory, physical correlation functions can
be factored through only physical states (as
is the case, if the Hilbert space of physical states
arises as the cohomology of a nilpotent BRST operator
$Q$ acting on a larger Hilbert space containing
spurious states), then the correlation functions of the theory on surfaces
of all genera can all be expressed explicitly, through factorization,
in terms of the structure constants $c_{ijk}$.

\medskip

\noindent
{\bf B. Deformations of a Topological Field Theory}

We shall be particularly interested in the
the case where the topological field theory
arises as a deformation by parameters $t_i$ of a fixed theory.
In a topological field theory,
the physical fields $\phi_i$ have clearly scaling dimension 0.
However, in presence of a BRST symmetry, they actually fit
in a multiplet $\Phi_i=(\phi_i^{(0)}=\phi_i,\phi_i^{(1)},\phi_i^{(2)})$,
where the ``descendants" $\phi_i^{(d)}$ are tensors of scaling dimension $d$
which satisfy the descent equations $\d\phi^{(d)}=\{Q,\phi^{(d+1)}\}$.
In particular, $\phi_i^{(2)}$ can be integrated on the surface $\G$.
If we let $I(0)$ be the Lagrangian of the
original theory, we can deform the theory to
another theory with Lagrangian 
$$
I(t)=I(0)-\sum_it_i\int_{\G}\phi_i^{(2)}\eqno(5.5)
$$
The structure constants $c_{ijk}(t)$ of the new theory are then given
by
$$
c_{ijk}(t)=<\phi_i\phi_j\phi_k{\rm exp}(\sum_it_i\int_{\G}\phi_i^{(2)})>_0\eqno(5.6)
$$
They define in turn a topological field theory. For
deformations around a topological conformal field theory, i.e.,
a topological field theory whose stress tensor is already
traceless before restricting to physical states,
the structure constants $c_{ijk}(t)$ are known to satisfy
the key compatibility condition
$\p_lc_{ijk}=\p_kc_{ijl}$ [11-13]. This means that we can find a function $\F$,
called the free energy and formally denoted by
$$
{\cal F}(t)=<{\rm exp}(\sum_it_i\int\Phi_i)>_0,\eqno(5.7)
$$
which satisfies the third derivative condition
$$
\p^3_{ijk}{\cal F}(t)=c_{ijk}\eqno(5.8)
$$
In terms of ${\cal F}$, the commutativity and associativity conditions
of the operator algebra with structure constants $c_{ijk}(t)$
become a system of differential equations of third order, called
the Witten-Dijkgraaf-Verlinde-Verlinde (WDVV) equation.

\medskip
\noindent
{\bf C. The Framework of Solitons}

In their original work [11-12], Dijkgraaf, Verlinde, and Verlinde
derived an explicit expression for the free energy $\F(t)$
in the case of topological Landau-Ginzburg theories. These are the topological
theories arising from twisting an N=2 superconformal field theory,
which is itself obtained by following the renormalization group
flow to the fixed point of a Landau-Ginzburg model. Although
the renormalization group flow modifies the kinetic terms
in the Landau-Ginzburg action, the superpotential $W(x)$ remains
unchanged, and thus characterizes both the associated superconformal and
topological models [30][61][67]. Our goal in this section
is to show how this theory can fit in the framework of solitons,
and to exhibit the natural emergence of the differential
$Q\d E$ and Whitham times. 

\medskip
We consider first the case of genus 0, with $\G=\{z\in{\bf C}\cup\infty\}$. 
The role of the superpotential $W(x)$ is
played in our context by the Abelian integral $E$ with
a unique pole of order $n$ at $\infty$. We consider then
a leaf in the space $\M_0(n,1)$, characterized by the condition
that $Q=z$, and $E$ is of the form
$$
E=z^n+\sum_{i=0}^{n-2}u_iz^i+O(z^{-1})\eqno(5.9)
$$
This is of dimension $n-1$, and can be parametrized by
the $n-1$ Whitham times $T_A$, $A=1,\cdots,n-1$, with the
other times fixed to $T_n=0$, $T_{n+1}={n\over n+1}$.
At each point $E$ on the leaf, the primary fields $\phi_i$
can be identified with $\d\Omega_i/\d Q$. The structure constants 
are defined by
$$
<\phi_i\phi_j>=\r({\d\Omega_{(i)}\d\Omega_{(j)}\over\d E}),\
<\phi_i\phi_j\phi_k>=\r
({\d\Omega_{(i)}\d\Omega_{(j)}\d\Omega_{(k)}\over\d E\d Q})\eqno(5.10)
$$
We note that $\d Q$ corresponds to the field defined by the
vacuum. The following can be derived from Theorem 16: 
\medskip

\noindent
{\bf Theorem 17.} {\it Let $\F(T)$ be the $\tau$-function of the (complex
normalized) Whitham hierarchy, restricted to the $(n-1)$
dimensional leaf described above. Then
\item{(i)} the fields $\phi_i$ anti-diagonalize the pairing $\eta_{ij}$,
i.e., $<\phi_i\phi_j>=\delta_{i+j,n}$;
\item{(ii)} $\F(T)$ is the free energy of the theory, i.e.,
$\p^3_{T_iT_jT_k}\F(T)=c_{ijk}$.
In particular, $\F(T)$ satisfies the system of WDVV equations.}

\medskip
More generally, the case of $\G$ of genus $g$ (with one
puncture to simplify the notation) has been treated in [22][38]. In 
this case, the relevant leaf
within $\M_g(n,1)$ is of dimension $n-1+2g$
and is given by the constraints
$$
\eqalignno{
&T_n=0,\ T_{n+1}={n\over n+1}\cr
&\oint_{A_k}\d E=0,\ \oint_{B_k}\d E=fixed,\cr
&\oint_{A_k}\d Q=0 &(5.11)\cr}
$$ 
Thus the leaf is parametrized by the $(n-1)$ Whitham times
$T_A$, $A=1,\cdots,n-1$, and by the periods
$a_j$ and $T_{j}^E$ of (4.28).
The fields $\phi_A$ of the theory need to be augmented accordingly.
We take the $2g$ additional ones to be given by $\d\omega_j/\d Q$ and
$\d \Omega_{j}^E/\d Q$, where the differentials
$\d\omega_j$ and $\d\Omega_{j}^E$ are the ones
associated to $a_j$ and $T_{j}^E$, as described
earlier in (4.30).

\medskip
\noindent
{\bf Theorem 18.} {\it Let $\eta_{AB}$ and $c_{ABC}$ be defined as
$$
\eta_{A,B}=\sum_{q_s}\r_{q_s}{\d\Omega_A\d\Omega_B\over \d E}\eqno(5.12) 
$$
$$
c_{ABC}=\sum_{q_s}\r_{q_s}{\d\Omega_A\d\Omega_B\d\Omega_C\over \d E \d Q},
\eqno(5.13)
$$
with $q_s$ the zeroes of $\d E$, and the indices 
A,B,C running this time through the augmented set
of $n-1+2g$ indices given by $T_A=(T_i,a_j,T_{E,j})$.  
Then 
\item{\rm (i)}  $\eta_{i,j}=\delta_{i+j,n}$, $\eta_{a_j,(E,k)}=\delta_{j,k}$.
All other pairings vanish; \item{\rm (ii)} 
Let $\F(T_i,a_j,T_{E,j})$ be the $\tau$-function of the Whitham hierarchy 
restricted to the leaf (5.11). Then $\p^3_{ABC}\F(T)=c_{ABC}$, where $A$ 
runs through the augmented set of $n-1+2g$ indices.}

\medskip
Note that in the genus 0 case when $Q=z$, the sum in (5.12-5.13) 
over the residues 
at the zeros of $\d E$ reduces to the residue at infinity. 
\medskip
Remarkably, the larger spaces $\M_g(n,m)$ can accommodate the 
gravitational descendants of the fields $\phi_A$. More precisely, consider 
for $g=0$ the leaf of the space $\M_0(n,mn+1)$ given by the 
following evident modification of the earlier normalization (5.11) 
$$ 
T_{in}=0,\ 
i=1,\ldots,m,\ \ \
T_{nm+1}={nm\over nm+1}.\eqno(5.14) 
$$
The 
space of Whitham times is automatically increased to the correct number by 
taking all the coefficients of $Q\d E$. The additional $m(n-1)$ 
fields may be identified with the first $m$ gravitational descendants
of the primary fields. Namely, the $p$-th descendant $\sigma_p(\phi_i)$ of the
primary field $\phi_i$ is
just ${\d \Omega_{pn+i}\over dQ}$. This statement is a direct corollary of
the following theorem.

\medskip
\noindent
{\bf Theorem 19.} {\it The correlation functions given by 
$<\phi_A\phi_B\phi_C>=\p_{ABC}^3\F$ with
$\sigma_p(\phi_i)=\d\Omega_{i+pn}/\d Q$ satisfy 
the factorization properties for descendant fields
$$
<\sigma_p(\phi_i) \phi_B\phi_C>=<\sigma_{p-1}(\phi_i)\phi_j>\eta^{jk}
<\phi_k \phi_B\phi_C>\eqno(5.15)
$$
where $\phi_i, i=1,\ldots,n-1$  are primary fields, 
and $\phi_A$ are all fields
(including descendants).}

\medskip
Factorization properties for descendant fields were derived by 
Witten [63-66]. 
We note that the completeness of the operator algebra
requires a larger set of fields that just $g$, which is
the dimension of the small leaves of the canonical foliation of
$\M_g(n,m)$, and which will be shown in the next section
to be the dimension of the moduli space of vacua of certain 
supersymmetric gauge theories. This is one of the
difficulties in establishing direct contact between topological field
theories and supersymmetric gauge theories, although there
has been progress in this direction [6][44][45].

\bigskip

\centerline{\bf VI. SEIBERG-WITTEN SOLUTIONS OF N=2 SUPERSYMMETRIC}
\centerline{\bf GAUGE THEORIES}

\bigskip
Moduli spaces of geometric structures are appearing
increasingly frequently as the key
to the physics of certain supersymmetric gauge
or string theories. One recurrent feature is a moduli space of
degenerate vacua in the physical theory. The physics of the theory
is then encoded in a K\"ahler geometry on the space of vacua, or,
in presence of powerful constraints such as N=2 supersymmetry,
in an even more restrictive {\it special geometry},
where the K\"ahler potential is dictated by
single holomorphic function $\F$, called the prepotential.
This was the case for Type IIA and Type IIB strings, where
the vacua corresponding to compactifications on Calabi-Yau threefolds [31][68]
produce effective N=2 four dimensional supergravity theories.
The massless scalars of such theories (in this case, the moduli
of the Calabi-Yau threefold) must parametrize a manifold equipped with
special geometry [9][51][57]. More recently, a similar
phenomenon has been brought to light
by Seiberg-Witten [52][53] for N=2 supersymmetric
gauge theories. Remarkably, the space of vacua
of these theories, which is classically just a
space of diagonalizable and traceless matrices,
becomes upon quantization a moduli space of Riemann
surfaces. The prepotential $\F$ for the quantum effective theory can 
then be derived from
a meromorphic one-form $\d\l$ on each surface. A particularly striking
feature of these effective theories, noticed by
many authors [24][28][42-43][49][56], is a strong but as yet
ill-understood similarity with the Whitham theory of solitons.
Indeed, the quantum spaces of vacua for many N=2 SUSY theories
actually coincide with certain leaves of the canonical
foliation of $\M_g(n,m)$ [39],
the Seiberg-Witten form $\d\l$ with the one-form $Q\d E$
central to Whitham theory, and the effective prepotential
of the gauge theories with the exponential of the $\tau$-function
of the Whitham theory! The purpose of this section
is to review some of these developments.

\medskip
\noindent{\bf 
A. N=2 Supersymmetric Gauge Theories}

We begin with a brief account of N=2 SUSY Yang-Mills theories
in four dimensions with gauge group $G$ [2]. The Yang-Mills
gauge field $A=A_{\mu}\d x^{\mu}$
is imbedded in an N=2 gauge multiplet consisting
of $A$, left and right Weyl spinors $\l_L$ and $\l_R$, and
a complex scalar field $\phi$, with all fields valued in the adjoint
representation of $G$. The requirement of N=2 SUSY and renormalizability
fixes uniquely the action
$$
I=\int_{M^4}\d^4x\,Tr\big[{1\over 4g^2}F\wedge F^*
+{\theta\over 8\pi^2}F\wedge F+D\phi^{\dagger}\wedge *D\phi+
[\phi,\phi^{\dagger}]
^2\big]+fermions\eqno(6.1)
$$
where $g$ is the coupling constant,
$\theta$ is the instanton angle, and we have written 
explicitly only the bosonic part of the action.
The classical vacua are given by the critical points of the
action. In this case, they work out to be $A=0$, $\phi$ is constant
(up to a gauge transformation), and
$$
[\phi,\phi^{\dagger}]=0\eqno(6.2)
$$  
Thus $\phi$ must lie in the Cartan subalgebra. For $G$=SU($N_c$) ($N_c$
is commonly referred to as number of ``colors"), we set
$$
\phi=\pmatrix{\bar a_1&{}&{}&{}&{}\cr
{}&\bar a_2&{}&{}&{}\cr
{}&{}&\cdot&{}&{}\cr
{}&{}&{}&{\cdot}&{}\cr
{}&{}&{}&{}&\bar a_{N_c}\cr},\ \sum_{k=1}^{N_c}\bar a_k=0.\eqno(6.3)
$$
Thus the classical space of vacua is parametrized by the $\bar a_k$, up to
a Weyl permutation.
\medskip
For a generic configuration $\bar a_k$, we have $\bar a_j\not=\bar a_k$
for any $j\not=k$, and the gauge group
SU($N_c$) is spontaneously broken down to 
${\rm U(1)}^{N_c-1}$. 
At the quantum level, we expect then the space
of inequivalent vacua to be parametrized by $N_c-1$
parameters $a_k$ (thought of as renormalizations of the $\bar a_k$,
$\sum_{k=1}^{N_c}a_k=0$), with each vacuum corresponding to a theory
of $N_c-1$ interacting U(1) gauge fields
$A_j$, i.e., $N_c-1$ 
copies of electromagnetism. In the weak coupling
regime, we expect singularities at $a_k=a_l$, where
the gauge symmetry is suddenly enhanced. Since $N=2$
SUSY remains unbroken, each gauge field $A_j$ is part of
an N=2 SUSY U(1) gauge multiplet $(A_j, \l_{Lj},\l_{Rj},\phi_j)$,
all in the adjoint representation of U(1). Again, the action for
a theory with such a field content 
is fixed by N=2 supersymmetry. 
To leading order in the low momentum
expansion for these fields, it must be of the form
$$
I_{eff}={1\over 8\pi}
\int_{M^4}\d^4 x\big[({\rm Im}\,\tau^{jk})F_j\wedge *F_k+
({\rm Re}\,\tau^{jk})F_j\wedge F_k
+\d\phi^{\dagger i}\wedge
\d\phi_D^i\big]+fermions\eqno(6.4)
$$
where 
$$\tau^{jk}={\p^2\F\over\p a_j\p a_k},\
\phi_D^j={\p\F\over\p a_j}(\phi)\eqno(6.5)
$$ 
for a suitable complex and analytic function ${\cal F}(a,\Lambda)$, called
the {\it prepotential}. We note that the prepotential
${\cal F}$ is a function not just of the vacua
parameters $a_k$, but also of
a scale $\Lambda$ introduced by renormalization.

\medskip
Thus the physics of the quantum theory
is encoded in a single function ${\cal F}$.
What is known
about ${\cal F}?$ To insure the positivity
of the kinetic energy, we must have
$$
{\rm Im}{\p^2\F\over\p a_j\p a_k}>0.\eqno(6.6)
$$
Geometrically, ${\cal F}$ defines then a K\"ahler metric on
the quantum moduli space by
$$
\d s^2=\sum {\rm Im}({\p^2\F\over\p a_j\p a_k})\d a_j\overline{\d a_k}\eqno(6.7)
$$
Furthermore, at weak-coupling $\Lambda<<1$, 
${\cal F}$ can be evaluated in
perturbation theory. For pure SU($N_c$) Yang-Mills, one finds
$$
{\cal F}(a,\Lambda)
={2N_c\over2\pi i}\sum_{k=1}^{N_c}a_k^2
-{1\over 8\pi i}\sum_{j,k=1}^{N_c}(a_k-a_j)^2{\rm log}
{(a_k-a_j)^2\over\Lambda^2}+\sum_{d=1}^{\infty}{\cal F}_d\Lambda^{2N_cd}
\eqno(6.8)
$$
The first term on the right hand side is the classical prepotential.
The second term is the perturbative one-loop quantum correction.
In view of N=2 non-renormalization theorems,
it is known that higher loops do not contribute.
The third term is 
the instanton contribution, consisting of $d$-instanton processes
for all $d$. 
We observe that the expansion (6.8) implies
in particular that ${\cal F}$ has non-trivial
monodromy around $a_j=a_k$ in the $\Lambda<<1$ regime.
The exact solution of N=2 Yang-Mills theories is reduced in this way
to finding a holomorphic $\F$ satisfying the constraints
(6.6) and (6.8).
\medskip
We have just described the main problem
for N=2 SUSY pure SU($N$) Yang-Mills theories. However, the same
problem should be addressed for general N=2 SUSY
gauge theories with gauge group G, with matter fields (``hypermultiplets")
in a representation $R$ of $G$. As in the case of pure Yang-Mills,
the Wilson effective Lagrangian of these theories is
dictated by a prepotential $\F_{G,R}(a,\Lambda)$,
and the problem is to determine $\F_{G,R}(a,\Lambda)$,
subject to the constraints (6.6) and (6.8), where the right
hand side of (6.8) has been modified to incorporate the
contributions of the hypermultiplets. For example, in presence
of $N_f$ hypermultiplets
in the fundamental representation of bare masses $m_i$,
$1\leq i\leq N_f$,
the one-loop correction to the prepotential for the SU($N_c$) theory
contains the additional term 
$$
\sum_{k=1}^{N_c}\sum_{i=1}^{N_f}
(a_k+m_j)^2{\rm log}{(a_k+m_j)^2\over\Lambda^2}.
$$

\medskip
\noindent{\bf B. The Seiberg-Witten Ansatz}

The requirements that ${\cal F}$ have monodromy
and a Hessian with positive
definite imaginary part,
suggest an underlying non-trivial
geometry on the quantum space of vacua.
In [52][53], Seiberg and Witten made the fundamental
Ansatz that 

\medskip
\item{$\bullet$} For each $\Lambda$, the quantum moduli space should
parametrize a family of Riemann surfaces
$\G(a,\Lambda)$ of genus $g=N_c-1$, now known as the spectral
curves of the theory;
\item{$\bullet$} on each $\G(a,\Lambda)$, there is a meromorphic
one-form $\d\l$;
\item{$\bullet$} ${\cal F}$ is determined by the periods of $\d\l$
\footnote*{In this section, we adopt the normalization (6.9)
for the periods $a_i$ of $\d\l$ rather than (3.39), in keeping
with the literature on Seiberg-Witten theory. Similarly,
the present $\F$ differs from the earlier $\tau$-function 
$\F_{\rm Whitham}$ of soliton theory (c.f. (4.35)) by 
$\F=-{1\over 2\pi i}\F_{\rm Whitham}$.} 
$$
a_k={1\over 2\pi i}\oint_{A_k}\d\l,
\quad
\ a_{D,k}={1\over 2\pi i}\oint_{B_k}\d\l, \quad
{\p\F\over\p a_k}=a_{D,k}\eqno(6.9)
$$ 

The gauge theories under consideration contain dyons, i.e.,
particles which carry both electric and magnetic charges. 
Let $(n,m)\in {\bf Z}^{N_c-1}
\times{\bf Z}^{N_c-1}$ be their charges,
with $(n_i,m_i)$ the charge with respect to the $i$-th U(1) factor.
The N=2 SUSY algebra implies the bound $M^2\geq 2|an+a_Dm|^2$
from below for their masses. Thus the states saturating this bound,
known as Bogomolny-Prasad-Sommerfeld or BPS states, are described
by the lattice spanned by the periods of $\d\l$.
The singular locus of the fibration $\G(a,\Lambda)$, namely 
the points where the curve degenerates and a period
$a_j$ or $a_{D,j}$ vanishes, corresponds
then to vacua where one or several dyons become massless.
\medskip
For pure SU(2) Yang-Mills, the monodromy prescription at $\infty$
is restrictive enough to suggest the identification of 
the quantum moduli space of vacua with $H/\G(2)$ ($H$
denotes the upper half space, and $\G(2)$ the subgroup
of SL(2,{\bf Z}) matrices congruent to 1 mod 2),
assuming the minimal number two of
singularities in the interior of the quantum moduli
space. Since then, spectral curves have been proposed
for a variety of gauge theories with matter, based on
physical considerations such as decoupling, or analogies with 
singularity theory or soliton theory
(see e.g. [40] and references therein). However,
at the present time, we still do not have
a complete correspondence between the group theoretic
characterization of the gauge theory, consisting
of the group $G$ and the representation $R$ for the
hypermultiplets, and the fibration of spectral curves which  
characterizes its geometric and physical content.
\medskip
\noindent{\bf C. The Framework of the Theory of Solitons}

Nevertheless, an intriguing feature of most of the
spectral curves for N=2 SUSY gauge theories known so far,
is that they, together with the one-form $\d\l$, fit exactly in 
the framework of the foliation on $\M_g(n,m)$ with
$\d\l=Q\d E$. In particular,

\medskip
$\bullet$ The spectral curves for SU($N_c$) theories
with $N_f<2N_c$ hypermultiplets in the fundamental
representation of bare masses $m_i$,
$1\leq i\leq N_f$, are given by the leaf $(\G,E,Q)$
with the following properties.

\item{-} $\d E$ has simple poles, at points $P_+$, $P_-$,
$P_i$,with residues
$-N_c$, $N_c-N_f$, and $1$ ($1\leq i\leq N_f$) respectively.
Its periods around homology cycles are integer multiples of $2\pi i$;

\item{-} $Q$ is a well-defined meromorphic $function$,
with simple poles only at $P_+$ and $P_-$;

\item{-} The other parameters of the leaf are fixed
by the following normalization of the one-form $\d\l=Q\d E$
$$
\eqalignno{\ {\rm Res}_{P_i}(\d\l)&=-m_i\cr
\ {\rm Res}_{P_+}(z\d\l)&=-N_c2^{-1/N_c},
\ {\rm Res}_{P_-}(z\d\l)=(N_c-N_f) 
({\Lambda^{2N_c-N_f}\over 2})^{1/(N_c-N_f)}\cr
{\rm Res}_{P_+}(\d\l)&=0&(6.10)\cr}
$$
Here $z=E^{-1/N_c}$ or $z=E^{1/(N_c-N_f)}$ is as usual the holomorphic 
coordinate system near $P_+$ or near $P_-$
adapted to the Abelian integral $E$.
\medskip
These conditions imply that $\G$ is hyperelliptic,
and admits an equation of the form (see [39])
$$
y^2=\prod_{k=1}^{N_c}(Q-\bar a_k)^2-\Lambda^{2N_c-N_f}
\prod_{j=1}^{N_f}(Q+m_j)\equiv A(Q)^2-B(Q)\eqno(6.11)
$$
Strictly speaking, the parameters $\bar a_k$ of (6.11) agree with the
classical vacua in (6.3) only when 
$N_c<N_f$. For $N_f\geq N_c$, there are $O(\Lambda)$ corrections,
which can be absorbed in a reparametrization leaving
the prepotential $\F$ invariant [15]. Thus we may view
the $\bar a_k$ of (6.3) and (6.11) as identical. 
If we represent the Riemann surface (6.11) by a two-sheeted
covering of the complex plane, then the meromorphic function $Q$ on $\G$ 
in $\d\l=Q\d E$ is just the coordinate in each sheet,
while the Abelian integral $E$ is given by $E={\rm log} (y+A(Q))$. 
The points $P_{\pm}$ correspond to the points at infinity, with the two
possible choices of signs $\pm$ for $y=\pm\sqrt{A^2-B}$. To choose a 
canonical homology
basis, we let $x_k^{\pm}$, $1\leq k\leq N_c$, be the branch points 
$A(x_k^{\pm})^2-B(x_k^{\pm})$,
$A_k$ be a simple contour enclosing the slit from $x_k^-$ to $x_k^+$
for $2\leq k\leq N_c$, and $B_k$ be the curve going from
$x_k^-$ to $x_1^-$ on each sheet. We can now give a preliminary and easy check
that the curve (6.11) is consistent with the expected behavior
of the theory at weak-coupling. Consider for simplicity the case of 
pure Yang-Mills,
$N_f=0$. Then as $\Lambda\rightarrow 0$, the discriminant of the curve
behaves as $\Lambda^{N_c}\prod_{j<k}(\bar a_j-\bar a_k)^2$, and the 
singularities are at the expected location. Furthermore, in this limit,
$\d\l\sim Q{A'(Q)\over A(Q)}\d Q+\cdots$, and the cycles $A_k$ are just
contours in the complex plane around a slit which shrinks to a single
point $\bar a_k$. The residue formula gives at once
$$
a_k={1\over 2\pi i}\oint_{A_k}Q{A'(Q)\over A(Q)}\d Q+O(\Lambda^{N_c})=
\bar a_k+O(\Lambda^{N_c}),\eqno(6.12)
$$
identifying $\bar a_k$ as a classical order parameter.
 
\medskip

$\bullet$ The spectral curves for the other classical gauge
groups with matter in the fundamental representation are
restrictions of the ones for SU($N_c$) [4][16];

\medskip
$\bullet$ The SU($N_c$) theory with matter in the adjoint
representation is of particular interest. For massless matter,
the theory has actually an N=4 supersymmetry, and is conformally
invariant. As the hypermultiplet acquires mass,
the N=4 SUSY is broken down to an N=2 SUSY. In [18], Donagi and
Witten argued that the spectral curves for the theory
are then given by Hitchin systems. Expressed in terms of
elliptic Calogero-Moser systems, the curves they proposed are
given precisely by the leaf $(\G,k,z)$ in Section III.D.
Here the hypermultiplet mass has been scaled to 1, and the moduli
$\tau=\omega_2/\omega_1$ of the torus is the microscopic
gauge coupling.

\medskip
Although this suggests a deep relation between N=2 gauge theories
and integrable models, such a relation is still not fully
understood at the present time. Nevertheless, the parallelism
between the two fields allows us to apply to the study
of the prepotential $\F$ of gauge theories the methods
developed in the theory of solitons. Thus Theorem 16 implies readily [17]
\medskip
\noindent
{\bf Theorem 20.} {\it The prepotential $\F$ for 
{\rm SU}($N_c$) gauge theories with $N_f<2N_c$ hypermultiplets
of masses $m_j$ in the fundamental representation,
satisfies the following differential equation}
$$
\eqalignno{
\sum_{j=1}^{N_c}a_j{\p\F\over\p a_j}+\sum_{j=1}^{N_f}m_j{\p\F\over\p m_j}
-2\F=&-{1\over 2\pi i}[\r_{P_+}(z\d\l)\r_{P_+}(z^{-1}\d\l)\cr
&\qquad
+\r_{P_-}(z\d\l)\r_{P_-}(z^{-1}\d\l)]&
(6.13)\cr}
$$

We observe that there is a slight abuse of language here,
since in the case of the effective prepotential of gauge
theories, $\F$ is only fixed up to $a_k$ independent
terms by (6.9). This is consistent with the fact that only derivatives
of $\F$ with respect to $a_k$ occur in the effective action.
Thus the $a_k$ independent terms on the right hand side
of (6.13) can be ignored by adjusting $\F$.
The prepotential (4.35) (restricted to the leaf
corresponding to the SU($N_c$) theory) is one choice of $\F$. 
Another choice suggested by
dimensional analysis (c.f. (6.8)) is the
prepotential $\F$ satisfying the homogeneity condition
$({\p\over\p\Lambda}+{\cal D})\F=0$. In this case, we recognize
(6.13) as a renormalization group equation, with the beta
function given by the right hand side of (6.13). Earlier versions
of (6.13) appear in [24][46][47][56].
\medskip
To illustrate the power of Theorem 20, we shall show how it can
generate explicit expressions for the contributions of instanton
processes to any order. Thus we consider the regime where
$\Lambda$ is small and all the $A_k$-cycles degenerate simultaneously.
A fundamental observation is that in this
regime, the quantum order parameters $a_k$ are perturbations
of their classical
counterparts $\tilde a_k$ which can be determined
explicitly to any order. In fact, as noted in the
arguments leading to (6.12), the $A_k$ cycles are simple contours
shrinking to a point in
one sheet of the Riemann surface and residue
formulae apply. The approximation (6.12) can be improved to
$$
\eqalignno
{&a_k=\bar a_k+\sum_{m=1}^{\infty}{\Lambda^{(2N_c-N_f)m}\over2^{2m}(m!)^2}
({\p\over\p \bar a_k})^{2m-1}\bar S_k(\bar a_k)\cr
&\bar S_k(x)={\prod_{j=1}^{N_f}(x+m_j)\over\prod_{l\not=k}(x-\bar a_l)^2}
&(6.14)\cr}
$$
The evaluation of the dual periods $a_{Dk}$ is of course
more difficult. We need to show that the prepotential $\F$,
as defined by the $B_k$-periods, reproduces the classical
prepotential $\F^{(0)}$ in (6.8) (with hypermultiplets)
and satisfies the non-renormalization
theorem. This requires an analytic continuation
in an auxiliary parameter $\xi$, as explained in [15]. However,
once this is established, the difficult instanton contributions
can be derived from the renormalization group equation.
Setting $\F=\F^{(0)}+\F^{(1)}+\F^{(2)}+\cdots$, we have, say
up to 2-instanton order and 
using Euler's homogeneity relation,
$$
\sum_{j=1}^{N_c}a_j{\p\F\over\p a_j}+\sum_{j=1}^{N_f}m_j{\p\F\over\p m_j}=
(N_f-2N_c)({1\over 4\pi i}\sum_{k=1}^{N_c}a_k^2+\F^{(1)}+2\F^{(2)})
\eqno(6.15)
$$
(We note the overall factor $N_f-2N_c$, which confirms the known
conformal invariance of the theory with $N_f=2N_c$. For the spectral curves
of this theory, we refer to [4]).
On the other hand, dropping all $a_k$-independent terms,
the right hand side of (6.13) is easily found
$$\sum_{j=1}^{N_c}a_j{\p\F\over\p a_j}+\sum_{j=1}^{N_f}m_j{\p\F\over\p m_j}
-2\F={1\over 4\pi i}(N_f-2N_c)\sum_{k=1}^{N_c}\bar a_k^2.
\eqno(6.16)
$$
We can now use (6.14) to reexpress the right hand side of (6.16)
in terms of the quantum order parameters $a_k$. 
The instanton contributions can then be read off
after suitable rearrangements [15][17]
$$
\eqalignno{\F^{(1)}&={1\over 8\pi i}\Lambda^{2N_c-N_f}
\sum_{k=1}^{N_c}S_k(a_k)\cr
\F^{(2)}&={1\over 32\pi i}\Lambda^{2(2N_c-N_f)}\big(
\sum_{k\not=l}{S_k(a_k)S_l(a_l)\over (a_k-a_l)^2}
+{1\over 4}\sum_{k=1}^{N_c}S_k(a_k)\p_{a_k}^2S_k(a_k)\big)&(6.17)\cr}
$$
Here the function $S_k(x)$ is defined in analogy with
(6.14) by $S_k(x)={\prod_{j=1}^{N_f}(x+m_j)
\over\prod_{j\not=k}^{N_c}(x-a_j)^2}$.
\medskip
We turn next to a determination of the effective prepotential at strong
coupling. In general, when a single cycle $A_k$ or $B_k$ degenerates,
we expect the effective prepotential to be
expressible in terms of functions on the resulting
surface of lower genus. A particularly interesting case
is the behavior of $\F$ near a point on the quantum moduli space
of maximum degeneracy, where all $B_k$ cycles degenerate
simultaneously, and the spectral curve degenerates to
two spheres connected by thin tubes. Physically, this means
that a maximum number of mutually local dyons
become simultaneously massless. As shown in [20], the points
of maximum degeneracy occur at the curves
$y^2=A_0(Q)^2-4\Lambda^{2N_c}$, where $A_0(Q)$ is given by the $N$-th Chebyshev
polynomial
$$A_0=2\Lambda^{N_c}C_0({Q\over 2\Lambda}),\ C_0(z)={\rm cos}(N{\rm arccos}(z))
\eqno(6.18)
$$
A neighborhood of the maximum degeneracy point on the quantum moduli space
is parametrized by polynomials $P(Q)$ of degree $N-2$, with the spectral curve
$y^2=A(Q)^2-4\Lambda^{2N_c}$, $A(Q)=A_0(Q)+2\Lambda^{N_c}P({Q\over 2\Lambda})$.
Since it is the $B_k$ cycle which degenerates this time, it is more convenient
to express the prepotential $\F(a_{Dk})$ and the beta function
in terms of the dual variables $a_{Dk}$, which can then be evaluated
to an arbitrary order of accuracy by residue methods. The renormalization 
group equation remains the same under interchange of the dual variables
$a_k\leftrightarrow a_{Dk}$
$$
\sum_{k=1}^{N_c}a_{Dk}{\p \F\over\p a_{Dk}}-2\F
={2N_c\over 2\pi i}u(a_D)\eqno(6.19)
$$
where $u$ is the coefficient of $Q^{N_c-2}$ in $A(Q)$. Residue calculations
show next $a_{Dk}$ is of first order in $P$ and that  
$$a_{Dk}=i(-)^k{s_k\over N_c}P(c_k)
+\sum_{m=1}^{\infty}a_{Dk}^{(m)},\ s_k={\rm sin}({k\pi\over N_c}),
\ c_k={\rm cos}({k\pi\over N_c}),\ k=1,\cdots,N_c\eqno(6.20)
$$
where the $a_{Dk}^{(m)}$ are of order $O^m(a_D)$ and can be evaluated
explicitly. To third order in $a_{Dk}$, we find for $u$ 
$$
u=2i\Lambda\sum_{k=1}^{N_c-1}s_ka_{Dk}+{1\over 4N_c}\sum_{k=1}^{N_c-1}a_{Dk}^2
+{i\over 32N_c^2\Lambda}\big[\sum_{k=1}^{N_c-1}{a_{Dk}^2\over s_k^3}
-4\sum_{k\not=l}^{N_c-1}{a_{Dk}^2a_{Dl}s_l\over(c_k-c_l)^2}\big]\eqno(6.21)
$$
Solving the renormalization group equation, we obtain [14] 
 
\medskip
\noindent
{\bf Theorem 21.} {\it Near the point of maximum degeneracy on the quantum moduli
space given by (6.19), the prepotential $\F$ is given by the
following expression
$$
\eqalignno{\F(a_D)
=&-{2N_c\Lambda\over\pi}\sum_{k=1}^{N_c-1}s_ka_{Dk}
-{i\over 4\pi}\sum_{k=1}^{N_c-1}a_{Dk}^2{\rm log}{a_{Dk}\over\Lambda_k}
+{1\over 32\pi N_c\Lambda}\big[\sum_{k=1}^{N_c-1}{a_{Dk}^3\over s_k^3}\cr
&\qquad-4\sum_{k\not=l}^{N_c-1}{a_{Dk}^2a_{Dl}s_l\over (c_k-c_l)^2}\big]
&(6.22)\cr}
$$
up to third order in the order parameters $a_{Dk}$. Here $\Lambda_k$
is determined by ${\rm log}{\Lambda_k\over\Lambda}={3\over 2}+{\rm log}s_k$.}

\bigskip
We should mention that there is by now an extensive literature
on Seiberg-Witten theories, and we refer to [40]
for a description of other recent advances and for
a more complete list of references.

\vfill\break

\centerline{\bf REFERENCES}

\bigskip

\item{[1]} Adler, M., ``On a trace functional
for formal pseudo-differential operators and the
symplectic structure for Korteweg-de Vries type equations",
Invent. Math. 50 (1979) 219-248.

\item{[2]} Amati, D., Konishi, K., Y. Meurice, G.C. Rossi,
and G. Veneziano, ``Non-perturbative aspects of supersymmetric
gauge theories", Physics Reports 162 (1988) 169-248.

\item{[3]} Arbarello, E. and C. DeConcini, ``On a set of equations
characterizing Riemann matrices", Ann. of Math. 120 (1984) 119-140.

\item{[4]} Argyres, P. and A.P. Shapere, ``The vacuum structure
of N=2 super-QCD with classical gauge groups", Nucl. Phys. B 461 (1996)
437-459, hep-th/9508175.

\item{[5]} Bogoyavlenski, O. and S.P. Novikov, ``On the connection
between Hamiltonian formalisms of stationary and non-stationary
problems", Funct. Anal. and Appl. 10 (1976) 9-13.

\item{[6]} G. Bonelli and M. Matone, Non-perturbative relations
in N=2 SUSY Yang-Mills and WDVV equations, hep-th/9605090.

\item{[7]} Cherednik, I., ``Differential equations for the
Baker-Akhiezer functions of algebraic curves", Funct. Anal. and Appl.
12 (1978) 

\item{[8]} Date, E., M. Jimbo, M. Kashiwara, and T. Miwa,
``Transformation groups for soliton equations", in
M. Jimbo and T. Miwa (eds) ``{\it Non-linear
integrable systems- classical and quantum theory}",
Proc. RIMS Symposium, Singapore, 1983.

\item{[9]} de Wit, B., P.G. Lauwers, and A. van Proeyen, ``Lagrangians
of N=2 supergravity-matter systems", Nucl. Phys. B 255 (1985) 569-608. 

\item{[10]} Dickey, L.A., ``{\it Soliton equations and Hamiltonian systems}",
Advanced Series in Mathematical Physics, Vol. 12 (1991) World Scientific,
Singapore.

\item{[11]} Dijkgraaf, R., E. Verlinde, and H. Verlinde, ``Topological strings in
$d<1$", Nucl. Phys. B 352 (1991) 59-86.

\item{[12]} Dijkgraaf, R., E. Verlinde, and H. Verlinde, ``Notes on topological 
string theory and 2D quantum gravity", Nucl. Phys. B (1991)

\item{[13]} Dijkgraaf, R. and E. Witten, Nucl. Phys. B 342 (1990) 486.

\item{[14]} D'Hoker, E. and D.H. Phong, ``Strong coupling expansions
in SU($N_c$) Seiberg-Witten theory", Phys. Lett. B 397 (1997) 94-103.

\item{[15]} D'Hoker, E., I.M. Krichever, and D.H. Phong, ``The effective 
prepotential for N=2 supersymmetric SU(N) gauge theories",
Nucl. Phys. B 489 (1997) 179-210. 

\item{[16]} D'Hoker, E., I.M. Krichever, and D.H. Phong, ``The effective
prepotential for N=2 supersymmetric SO(N) and Sp(2N) gauge
theories", Nucl. Phys. B 489 (1997) 211-222.

\item{[17]} D'Hoker, E., I.M. Krichever, and D.H. Phong, ``The 
renormalization group equation for N=2 supersymmetric gauge
theories", Nucl. Phys. B (1997), hep-th/9610156.

\item{[18]} Donagi, R. and E. Witten, ``Supersymmetric Yang-Mills
theory and integrable systems", Nucl. Phys. B 460 (1996) 299-334,
hep-th/9610101.

\item{[19]} Douglas, M. and S. Shenker, ``Strings in less than one
dimension", Nucl. Phys. B 335 (1990) 635.

\item{[20]} Douglas, M. and S. Shenker, ``Dynamics of SU(3) gauge theories",
Nucl. Phys. B 447 (1995) 271, hep-th/9503163.

\item{[21]} Dubrovin, B.A., ``Integrable systems in topological field theory",
Nucl. Phys. B 379 (1992) 627-689.

\item{[22]} Dubrovin, B.A., ``Hamiltonian formalism of Whitham-type
hierarchies and topological Landau-Ginzburg models", Comm.
Math. Phys. 145 (1992) 195-207.

\item{[23]} Dubrovin, B.A. and S.P. Novikov, ``The Hamiltonian formalism
of one-dimensional systems of the hydrodynamic type,
and the Bogoliubov-Whitham averaging method",
Sov. Math. Doklady 27 (1983) 665-669.

\item{[24]} Eguchi, T. and S.K. Yang, ``Prepotentials of N=2 supersymmetric
gauge theories and soliton equations", hep-th/9510183.

\item{[25]} Faddeev, L. and L. Takhtajan, {\it ``Hamiltonian approach
in soliton theory"}, Springer-Verlag. 

\item{[26]} Flaschka, H., Forrest, M.G., and D. McLaughlin, ``Multiphase averaging
and the inverse spectral solution of the Korteweg-de Vries
equation", Comm. Pure Appl. Math. 33 (1980) 739.

\item{[27]} Gelfand, I.M. and L.A. Dickey, ``Fractional powers
of operators and Hamiltonian systems", Funct. Anal. and Appl. 10 (1976) 13-29.

\item{[28]} Gorsky, A., I.M. Krichever, A. Marshakov, A. Mironov, 
and A. Morozov, Phys. Lett. B 355 (1995) 466-474.

\item{[29]} Gorsky, A. and N. Nekrasov, ``Elliptic Calogero-Moser system
from 2d current algebra", hep-th/9401021.

\item{[30]} Greene, B., C. Vafa, and N. Warner, ``Calabi-Yau
manifolds and renormalization group
flows", Nucl. Phys. B 324 (1989) 371.

\item{[31]} Greene, B. and S.T. Yau, eds., {\it Mirror Symmetry II}, Studies in
Advanced Mathematics, American Mathematical Society-International
Press (1997), Rhode Island and Hong Kong.

\item{[32]} Klemm, A., W. Lerche, and S. Theisen, ``Non-perturbative
actions of N=2 supersymmetric gauge theories", Int. J. Mod. Phys. A 11 (1996)
1929-1974, hep-th/9505150.

\item{[33]} Krichever, I.M., ``The algebraic-geometric construction of
Zakharov-Shabat equations and their solutions", Doklady Akad. Nauka
USSR 227 (1976) 291-294.

\item{[34]} Krichever, I.M., ``Methods of algebraic geometry
in the theory of non-linear equations", Russian Math Surveys 32 (1977) 185-213.

\item{[35]} Krichever, I.M., ``Elliptic solutions of the Kadomtsev-Petviashvili
equation and integrable systems of particles", Funct. Anal. Appl. 14 (1980)
282-290.

\item{[36]} Krichever, I.M., ``Averaging method for two-dimensional
integrable equations", Funct. Anal. Appl. 22 (1988) 37-52.

\item{[37]} Krichever, I.M., ``Spectral theory of two-dimensional
periodic operators and its applications", Russian Math. Surveys (1989)
44 (1989) 145-225.

\item{[38]} Krichever, I.M., ``The $\tau$-function of the universal
Whitham hierarchy, matrix models, and topological field theories",
Comm. Pure Appl. Math. 47 (1994) 437-475.

\item{[39]} Krichever, I.M. and D.H. Phong, ``On the integrable geometry
of soliton equations and N=2 supersymmetric gauge theories",
J. Differential Geometry 45 (1997) 349-389.

\item{[40]} Lerche, W., ``Introduction to Seiberg-Witten theory
and its stringy origins", Proceedings of the
{\it Spring School and Workshop in String Theory},
ICTP, Trieste (1996), hep-th/9611190.

\item{[41]} Magri, F., ``A simple model of the integrable Hamiltonian
equation", J. Math. Phys. 19 (1978) 1156-1162.

\item{[42]} Martinec, E., ``Integrable structures in supersymmetric
gauge and string theory", hep-th/9510204.

\item{[43]} Martinec, E. and N. Warner, ``Integrable models and supersymmetric
gauge theory", Nucl. Phys. B 459 (1996) 97-112, hep-th/9609161.

\item{[44]} Marshakov, A., ``Non-perturbative quantum theories
and integrable equations", hep-th/9610242.

\item{[45]} Marshakov, A., A. Mironov, and A. Morozov,
``WDVV-like equations in N=2 SUSY Yang-Mills theory", hep-th/9607109.

\item{[46]} Matone, M., Phys. Lett. B 357 (1995) 342.

\item{[47]} Minahan, J. and D. Nemeschansky, Nucl. Phys. B 468 (1996) 73-84,
hep-th/9601059.

\item{[48]} Mumford, D., {\it ``Theta Functions I, II"}, Birkh\"auser
(1980) Boston.

\item{[49]} Nakatsu, T. and K. Takasaki, ``Whitham-Toda
hierarchy and N=2 supersymmetric Yang-Mills theory", Mod.
Phys. Lett. A 11 (1996) 157-168, hep-th/9509162.

\item{[50]} Novikov, S.P. and A. Veselov, ``On Poisson brackets
compatible with algebraic geometry and Korteweg-deVries dynamics on the
space of finite-zone potentials", Soviet Math. Doklady 26 (1982) 357-362.

\item{[51]} Periwal, V. and A. Strominger, ``K\"ahler geometry of the space
of N=2 supersonformal field theories", Phys. Lett. B 235 (1990) 261-267. 

\item{[52]} Seiberg, N. and E. Witten, ``Electric-magnetic duality, 
Monopole Condensation, and Confinement in N=2 Supersymmetric
Yang-Mills Theory", Nucl. Phys. B 426 (1994) 19, hep-th/9407087.

\item{[53]} Seiberg, N. and E. Witten, ``Monopoles, Duality and Chiral
Symmetry Breaking in N=2 Supersymmetric QCD", Nucl. Phys. B 431 (1994)
494, hep-th/9410167.

\item{[54]} Sato, M., ``Soliton equations and the universal
Grassmann manifold, Math. Notes. Series 18 (1984) Sophia University, Tokyo.

\item{[55]} Shiota, T., ``Characterization of Jacobian varieties
in terms of soliton equations", Invent. Math. 83 (1986) 333-382.

\item{[56]} Sonnenschein, J., S. Theisen, and S. Yankielowicz,
Phys. Lett. B 367 (1996) 145-150, hep-th/9510129.

\item{[57]} Strominger, A., ``Special Geometry", Comm. Math. Phys. 133 (1990)
163-180.

\item{[58]} Takebe, T., ``Representation theoretic meaning for
the initial-value problem for the Toda lattice hirearchy",
Lett. Math. Phys. 21 (1991) 

\item{[59]} Terng, C.L., ``Soliton equations and differential geometry",
J. Differential Geometry 45 (1997) 407-445.

\item{[60]} Terng, C.L. and K. Uhlenbeck, to appear.

\item{[61]} Vafa, C., ``Topological Landau-Ginzburg models", Mod. Phys. Lett.
A 6 (1991) 337.

\item{[62]} Whitham, G., {\it ``Linear and non-linear waves"}, Wiley (1974)
New York.

\item{[63]} Witten, E., ``Topological Quantum Field Theory", Comm. Math. Phys.
117 (1988) 353-386.

\item{[64]} Witten, E., ``Topological sigma models", Comm. Math. Phys. 118
(1988) 411-449.

\item{[65]} Witten, E., ``Mirror manifolds and topological field theory",
in {\it ``Essays on Mirror Manifolds"}, ed. by S.T. Yau, International
Press (1992), Hong-Kong, 120-159.

\item{[66]} Witten, E., ``Two-dimensional gravity and intersection
theory on moduli space", Surveys in Differential Geometry 1 (1991) 281-332.

\item{[67]} Witten, E., ``Phases of N=2 theories in two dimensions",
Nucl. Phys. B 403 (1993) 159-222; also reprinted in [31].

\item{[68]} Yau, S.T., ed., {\it ``Essays on Mirror Manifolds"}, International
Press (1992) Hong-Kong.

\end